\documentclass[twocolumn]{aastex701}

\usepackage{amsmath} 
\usepackage{subcaption}
\usepackage{bm}
\usepackage{float}

\begin{document}

\title{Probing the large-scale structure with 21cm-galaxy cross-bispectrum: Estimates from simulations and forecasts for upcoming cosmological surveys}

\author[]{Leon Noble}
\affiliation{Department of Astronomy, Astrophysics \& Space Engineering, Indian Institute of Technology Indore, Indore 453552, India}
\email[show]{leonnoblek@gmail.com}  

\author[]{Suman Majumdar}
\affiliation{Department of Astronomy, Astrophysics \& Space Engineering, Indian Institute of Technology Indore, Indore 453552, India}
\email{}  

\author[]{Matteo Viel}
\affiliation{SISSA - International School for Advanced Studies, Via Bonomea 265, I-34136 Trieste, Italy}
\affiliation{IFPU - Institute for Fundamental Physics of the Universe, Via Beirut 2, I-34151 Trieste, Italy}
\affiliation{INAF - Osservatorio Astronomico di Trieste, Via G. B. Tiepolo 11, I-34143 Trieste, Italy}
\affiliation{INFN - Sezione di Trieste, Via Valerio 2, I-34127 Trieste, Italy}
\email{}  

\author[]{Fabio Fontanot}
\affiliation{INAF - Osservatorio Astronomico di Trieste, Via G. B. Tiepolo 11, I-34143 Trieste, Italy}
\affiliation{IFPU - Institute for Fundamental Physics of the Universe, Via Beirut 2, I-34151 Trieste, Italy}
\email{}  

\author[]{Gabriella De Lucia}
\affiliation{INAF - Osservatorio Astronomico di Trieste, Via G. B. Tiepolo 11, I-34143 Trieste, Italy}
\affiliation{IFPU - Institute for Fundamental Physics of the Universe, Via Beirut 2, I-34151 Trieste, Italy}
\email{}  

\author[]{ Abinash Kumar Shaw}
\affiliation{Max-planck-Institut f$\ddot{u}$r Astrophysik, Karl-Schwarzschild-Strasse 1, D-85741, Garching, Germany}
\email{}

\author[]{Marta Spinelli}
\affiliation{Observatoire de la C\^{o}te d’Azur, Laboratoire Lagrange, Bd de l’Observatoire, CS 34229, 06304 Nice cedex 4, France}
\affiliation{Department of Physics and Astronomy, University of the Western Cape, Robert Sobukwe Road, Cape Town 7535, South Africa}
\email{}

\author[]{Mohd Kamran}
\affiliation{INAF - Osservatorio Astronomico di Trieste, Via G. B. Tiepolo 11, I-34143 Trieste, Italy}
\affiliation{IFPU - Institute for Fundamental Physics of the Universe, Via Beirut 2, I-34151 Trieste, Italy}
\email{}

\author[]{Lizhi Xie}
\affiliation{Astrophysics Center, Tianjin Normal University, Binshuixidao 393, Tianjin, China}
\email{}

\author[]{Michaela Hirschmann}
\affiliation{ EPFL - Institute for Physics, Laboratory for Galaxy Evolution, Observatoire de Sauverny, Chemin Pegasi 51, 1290 Versoix, Switzerland}
\affiliation{INAF - Osservatorio Astronomico di Trieste, Via G. B. Tiepolo 11, I-34143 Trieste, Italy}
\email{}

\begin{abstract}
    The redshifted 21cm signal from the post-reionization epoch is highly non-Gaussian; thus, higher-order statistics, such as the bispectrum, are required to extract this non-Gaussian information. However, high signal-to-noise ratio (SNR) detection of the 21cm auto-bispectrum will be hindered by the presence of residual systematics. Cross-correlating the 21cm signal with galaxies offers a promising path to suppress this uncertainty from residual systematics and potentially increase the SNR. We present a comprehensive analysis of the HI-galaxy cross-bispectrum using the predictions of theoretical galaxy evolution models defined on large cosmological volumes. Our analysis includes the cross-bispectrum for different triangle sizes and shapes, as well as for different combinations of the HI and galaxy fields. We forecast the detectability of the 21cm-galaxy cross-bispectrum at redshift $z\approx1$ with the Euclid-like galaxy survey and SKA-Mid observations in both interferometric and single-dish modes of the survey. We find that the 21cm-galaxy cross-bispectrum shows enhanced detectability compared to the 21cm auto-bispectrum for all unique triangles in the interferometric mode of observations. We forecast a 10$\sigma$ detection of the cross-bispectrum for squeezed-limit triangles and a 100$\sigma$ detection for all shapes combined for scales $0.2~\text{Mpc}^{-1}\leq k_1 \leq 0.9~\text{Mpc}^{-1}$ with 100 hours of SKA-Mid observations per pointing. However, the detectability of the cross-bispectrum for large scales ($k_1 < 0.1~\text{Mpc}^{-1}$), which is accessible with the single-dish mode of the survey, is limited by cosmic variance. Additionally, the signal loss due to foreground removal further suppresses the detectability. Our analysis presents a first step toward an end-to-end analysis pipeline for the future 21cm-galaxy cross-bispectrum observations.
\end{abstract}
\section{Introduction}
Tracing the density fluctuations in the large-scale structure of the Universe  across as large an observational volume as possible is the cornerstone of precision cosmology. Observing wide areas of the sky and reaching deep redshifts provides a robust dataset for testing the standard LCDM model, constraining the nature of dark matter and
dark energy, measuring neutrino masses, constraining primordial non-Gaussianity, and hunting for any new physics. Various completed and ongoing surveys in optical and near-infrared wavelengths have yielded significant progress in this direction. Expanding these efforts to a wider wavelength range is highly beneficial. A multiwavelength approach not only increases the quantity of available data for cosmological inference but also provides robust checks on systematic uncertainties.
\par
Mapping the distribution of neutral hydrogen in the post-reionization epoch offers a powerful probe of the large-scale structure, complementary to galaxy surveys. After the ionization of the intergalactic medium, neutral hydrogen (HI) is  expected to be mostly confined to highly dense, self-shielded regions within the galaxies~\citep{Navarro_2018}. This distribution of HI can be mapped through 21cm line intensity mapping~\citep{Bharadwaj_2001, Bharadwaj_sethi_2001,Battye_2004,Wyithe:2007rq,Chang_2008,Santos:2015gra}. The 21cm line intensity mapping observations measure aggregate 21cm flux from  coarse patches of the sky, with information about the line of sight  encoded in the different frequency channels of the instrument. Even though these observations have low spatial resolution, they have a very fine line-of-sight resolution, enabling the mapping of large cosmic volumes in relatively short observational time.
\par
Various observational efforts have been successful in making the statistical detection of the 21cm signal both in autocorrelation and in cross correlation with other tracers of the large-scale structure.
Recently, the \citet{Chime_detection_2025} reported a detection of the 21cm auto-power spectrum at redshift $z\approx 1$ for scales of $0.4~h~\text{Mpc}^{-1}< k < 1.5~h~\text{Mpc}^{-1}$   with high significance. Furthermore, \citet{Paul_2023} have also  reported a detection of the 21cm auto-power spectrum
at $z \approx 0.32$ and $z \approx 0.44$ for scales of  $0.3~\text{Mpc}^{-1} < k < 8~\text{Mpc}^{-1}$. In addition to these measurements, several observational efforts successfully detected the 21cm signal by cross-correlating with another tracer of the large-scale structure \citep{Chang_2010, Masui_2013, Switzer_2013, Anderson_2018, Li_2021, Cunnington_Li_2023,MeerKLASS:2025,Amiri_2023, Amiri_2024, Carucci_2025}.
A large number of ongoing radio experiments, including MeerKAT~\citep{Meerkat_2017}, the upgraded Giant Metrewave Radio Telescope (uGMRT)~\citep{GMRT_2017}, the Tianlai array ~\citep{Zuo_2021}, and the Five-hundred-meter Aperture Spherical radio Telescope (FAST)~\citep{fast}, are also attempting to make a statistical detection of the 21 cm signal in auto-correlation across various redshifts and scales. In the near future, additional experiments such as the Square Kilometer Array Observatory (SKAO)\footnote{\url{https://www.skao.int/}}~\citep{Braun_2019}, the Canadian Hydrogen Observatory, and the Radio transient Detector (CHORD)~\citep{chord}, the Hydrogen Intensity
and Real-Time Analysis Experiment (HIRAX) \citep{hirax} and the Baryon Acoustic Oscillations from Integrated
Neutral Gas Observations telescope (BINGO) \citep{bingo} will join these ongoing observational efforts.
\par
The redshifted 21cm signal from the post-reionization Universe  is non-Gaussian due to the nonlinear clustering of the matter~\citep{Peebles_1980}, complex astrophysics, and primordial non-Gaussianity~\citep{Bartolo_2004} sourced by inflation. To quantify this non-Gaussianity in the signal, one has to consider a statistic other than the power spectrum. Various summary statistics, including voxel intensity distribution~\citep{Breysse:2016szq, Bernal:2023ovz}, marked power spectrum~\citep{2025JCAP...07..054K, massara_cosmological_2023}, $k$-nearest-neighbour~\citep{Chand_2025}, Minkowski functionals~\citep{bag_shape_2018, pathak_distinguishing_2022}, largest cluster statistics~\citep{bag_shape_2018, dosibhatla_2025_lss-morphology} and machine learning techniques~\citep{mishra26} can be utilized for this purpose. The 21cm bispectrum is a promising higher-order statistic that can quantify the non-Gaussianity present in the 21cm signal~\citep{Ali_2006, Sarkar_2013, Claude_2019, Sarkar_2019,Durrer_2020, Jolicouer_2021, Cunnington_2020, Karagiannis_2020,Chhabra_2025}. Additionally, the inference with the 21cm bispectrum can provide improved constraints on the astrophysics~\citep{Chhabra_2025, Sarkar_2019} and cosmology~\citep{Claude_2019, Karagiannis_2020b, Karagiannis_2022, Randrianjanahary_2024,Joshi:2025swr,Pal:2026hkq,Pinheiro:2026mcm}.  
\par 
Upcoming experiments, such as SKA-Mid and HIRAX, are expected to detect the 21cm auto-bispectrum across a range of redshifts and scales. However, the 21cm line intensity mapping observations are affected by strong astrophysical foregrounds that are several orders of magnitude higher than the cosmological signal. The spectral smoothness of the foreground signal in frequency is used to remove the foreground contribution from the cosmological signal \citep{Wang_2006,2012MNRAS.423.2518C,Chapman_2013,Switzer_2013, Alonso_2015,Zuo_2019,Olivari_2016,Carucci_2020, Carucci_2025, Spinelli_2021}. However, various systematic errors associated with radio instruments and calibration errors restrict the perfect foreground removal. The uncertainty due to this residual foreground will hinder a high-significance detection of the 21cm auto-bispectrum. Looking for a higher-order statistic signal by cross-correlating the 21cm signal with another tracer of large-scale structure~\citep{Guandalin_2022, Moodely,Tristan_2025} offers a promising strategy to suppress the uncertainty due to residual foreground systematics, since the foregrounds of the two signals are unlikely to be correlated. Various ongoing and upcoming photometric and spectroscopic galaxy surveys, such as Euclid, the Nancy Grace Roman Space Telescope,  the Dark Energy Spectroscopic Instrument (DESI), the Rubin Observatory, and 4MOST~\citep{Euclid:2019clj,Spergel_2015, DESI, LSSTDarkEnergyScience:2018jkl, 2019Msngr.175....3D} have an overlap in sky area and redshift with the upcoming SKA-Mid. Given the success in detecting the 21cm-galaxy cross-power spectrum~\cite{Cunnington_Li_2023,MeerKLASS:2024ypg, Carucci_2025}, extending this approach to the cross-bispectrum of the 21cm signal with galaxies using data from SKAO and galaxy surveys offers a promising path toward achieving a high-significance detection and perform cosmological investigations (see e.g.~\citet{villa15,Cunnington_2019,Squarotti:2023nzy,berti24,Karagiannis_2024,Kopana:2024qqq,autieri26}).
\par
In this article, we present a comprehensive analysis of the HI-galaxy cross-bispectrum and quantify its detectability with SKA-Mid and a Euclid-like~\citep{Euclid:2019clj, Spergel_2015} galaxy survey. To do this, we simulate mock HI line intensity maps and galaxy catalogs using the GAEA~\citep{DeLucia_2014,Hirschmann_2016, Fontanot_2025}  semi-analytical galaxy formation model. We examine the HI-galaxy cross-bispectrum for $k$-triangles of different sizes and shapes in both real and redshift space. Additionally, we investigate all possible combinations of the cross-bispectrum involving HI and galaxies. Furthermore, we perform an analysis to identify the $k$-range where the cross-bispectrum can be adequately modeled  using predictions of standard perturbation theory, and we then use this to extract the linear and quadratic HI bias parameters. We quantify the expected signal-to-noise ratio (SNR) of the 21cm-galaxy cross-bispectrum for different combinations of the 21cm and galaxy fields, as a function of triangle size and shape, and observational time for SKA-Mid operating in both interferometric and single-dish modes. This article is a first step toward  an end-to-end analysis pipeline for future observations of the 21cm-galaxy cross-bispectrum.
\par
This article is organized as follows: in Section~\ref{sec:simulations}, we discuss the suite of simulations used to generate HI line intensity maps and mock galaxy catalogs. Section~\ref{sec:bispectrum estimation} describes the details of auto and cross-bispectrum estimation from the simulations. In Section~\ref{sec:bispectrum results}, we discuss the HI and galaxy auto and cross-bispectrum both in real and redshift space, followed by a discussion on modelling the HI-galaxy cross-bispectrum from perturbation theory. In Section~\ref{sec:detectability}, we present forecasts for the detectability of the 21cm-galaxy cross-bispectrum with SKA-Mid and Euclid-like galaxy survey. Finally, in Section~\ref{sec:summary}, we summarize our results.
\section{Simulations of HI line intensity maps and mock galaxy Catalogs}
\label{sec:simulations}
\label{sec:simulation_signal} 
In this section, we describe the details of the simulation of HI (and 21cm) line intensity maps and mock galaxy catalogs  used in this study. We employ two distinct sets of simulations. The first is based on the state-of-the-art semi-analytic galaxy formation model GAEA~\citep{DeLucia_2014,Hirschmann_2016, Fontanot_2025} and the other is based on halo occupation distribution (HOD) modeling. The former set of simulations is used to study the nature of the HI-galaxy cross-bispectrum and to forecast its detectability with upcoming observations. The latter, which contains 50 independent realizations of the signal, is used to compare HI-galaxy cross-bispectrum predictions from standard perturbation theory with estimates from simulations. 
\par 
Galaxy survey observations deliver the number density of galaxies $n(\bm{x})$ at each position $\bm{x}$ and at each redshift $z$. We can then estimate the overdensity of the galaxies, $\delta_{\rm Gal} = \frac{n(\bm{x}) - \bar n}{\bar n}$, where $\bar n $ is the average number density of the galaxies. Similarly, for the  HI line intensity maps, we estimate the HI mass overdensity $\delta_{\rm HI}(\bm{x})$ by interpolating the HI mass from each galaxy/halo into a voxel of the intensity map. The differential brightness temperature ($\delta T_b$), which is the observable from 21cm radio experiments, can be written as~\citep{Furlanetto:2006jb}
\begin{multline}
    \delta T_b(\bm{x}) = 23.88~x_{\rm HI} (1 + \delta_{\rm HI}(\bm{x}))\bigg( \frac{\Omega_b h^2}{0.02}\bigg) \\\sqrt{\frac{0.15}{\Omega_m h^2} \frac{(1+z)}{10}}~ \rm mK.
\end{multline}
Here, $x_{\rm HI} = \Omega_{\rm HI}/\Omega_{\rm H}$ is the  neutral atomic hydrogen fraction. We estimate the hydrogen fraction as $\Omega_{\rm H} = 0.74 \Omega_b$. We will now describe in detail the two sets of simulations we used to generate galaxy catalogs and HI line intensity maps.

\subsection{GAEA Simulations}
\label{sec:GAEA_simulations} 
GAEA is a state-of-the-art semi-analytic model (SAM)  of galaxy formation and evolution. It traces the evolution of various baryonic components, incorporating treatments for star formation, chemical enrichment, stellar feedback, gas accretion onto supermassive black holes, and AGN feedback. Additionally, the GAEA model explicitly partitions cold gas into its atomic and molecular components, tuning the relevant model parameters against the observed  HI and H$_2$ galaxy mass function. More details on the latest GAEA implementation used in this work can be found in \citet{Xie_2017,Xie_2020,Fontanot_2020,DeLucia_2024b} and references therein.
\par
For our study, we use the latest GAEA realization run on dark matter merger trees extracted from the Planck Millennium Simulation \citep{Fontanot_2025}. The P-Millennium~\citep{Baugh_2019} simulates the evolution of dark matter particles with a mass of  $1.56\times 10^{8} M_{\odot}$ in a cubic comoving volume of $800^{3} \rm Mpc^{3}$. At each snapshot, the halos and subhalos in the volume are identified using a friends-of-friends (FoF) algorithm and the substructure finding code SUBFIND~\citep{Springel_2001}. The most bound part of the FoF group hosts the central galaxy, while all other bound subhalos are associated with satellite galaxies. GAEA predictions based on the  P-Millennium outputs show excellent agreement with observational data for the galaxy two-point correlation function at both lower and higher redshifts. Moreover, they accurately reproduce the primary dependencies of the two-point correlation function as a function of stellar mass, star formation activity, HI content, and redshift~\citep{Fontanot_2025}. 
This large simulation volume is ideal for our study, as it is comparable to the observing volumes expected from the SKA-Mid surveys.
\par 
To generate the HI intensity maps, we interpolate the HI mass extracted from the GAEA galaxy catalog  to a $600^3$ grid using the cloud-in-cell (CIC) algorithm. This results in a HI intensity map with a grid resolution of 1.33 Mpc. To generate a mock spectroscopic H$\alpha$ sample of galaxies for a Euclid-like galaxy survey, we selected only galaxies that have a stellar mass greater than $10^{10.6 } M_{\odot}$ from the GAEA catalog. The stellar mass cut is chosen such that it reproduces the  number densities of H$\alpha$-emitting galaxies galaxies that  match Model 3 described in \cite{Pozzetti_2016} and the HOD of the Euclid Flagship simulation~\citep{Pezzotta_2024,castander_2025}. For estimating summary statistics, we interpolate galaxy positions onto a grid of the same size as the HI line intensity maps using CIC.
\subsection{HOD model simulations}
\label{sec:HOD_simulations}
We simulate 50 independent realizations of the HI line intensity maps and mock galaxy catalogs using HOD models.
First, we simulate the dark matter distribution at redshift $z=1$ using a dark matter only Particle-Mesh (PM)~\citep{Bharadwaj_Srikant_2004, Mondal_2015} $N$-body simulation\footnote{\url{https://github.com/rajeshmondal18/N-body}}. We run the simulation with a comoving volume of $215^3~\text{Mpc}^3$ with a $3072^3$ grid using $1536^3$ dark matter particles. Next, we employed an FoF algorithm\footnote{\url{https://github.com/rajeshmondal18/FoF-Halo-finder}} to identify the dark matter halos from the outputs of the $N$-body simulation. We use a linking length of 0.2 times the mean inter-particle separation, and halos with at least 10 dark matter particles are selected. This results in a minimum halo mass of $10^9~M_{\odot}$. Later, these halo catalogs were used as inputs for the HI and galaxy HOD models to generate HI line intensity maps and mock galaxy catalogs.
\par
At lower redshifts, the intergalactic medium is highly ionized, with most of the HI mass confined to dense, self-shielded regions within galaxies~\citep{Navarro_2018}. Additionally, the HI line intensity mapping experiments have poor spatial resolution, and each voxel in the line intensity map will contain many galaxies. Hence, one can use prescriptions that assign HI mass ($M_{\rm HI}$) to halo mass ($M_{\rm h}$) to populate halos with HI. Here, we use the $M_{\rm HI}-M_{\rm h}$ relation from \cite{Spinelli_2020}, which is derived from an earlier realization of GAEA. According to this model, the HI mass inside the halo follows the relation 
\begin{multline}
        M_{\rm HI}(M_{\rm h}) = \\ M_{\rm h}\bigg[a_1\bigg(\frac{M_{\rm h}}{10^{10}}\bigg)^{\beta}e^{-\big(\frac{M_{\rm h}}{M_{\rm break}}\big)^{\alpha}} + a_2\bigg]e^{-\big(\frac{M_{\rm min}}{M_{\rm h}}\big)^{0.5}},
\end{multline}
where $a_1,a_2,\alpha, \beta, M_{\rm break},$ and $M_{\rm min}$ are free parameters. Later, we interpolate the HI mass to a $384^3$ grid using CIC, which results in a HI line intensity map with a grid resolution of 0.56 Mpc.
\par
To populate galaxies within halos and create a mock spectroscopic H$\alpha$ sample of galaxies for a Euclid-like galaxy survey, we utilize the HOD algorithm. Each halo will contain at most one central galaxy situated at the center of the halo and can contain many satellite galaxies. Following ~\cite{Zheng_2007}, the mean occupation number of central galaxies depends on the host halo mass,
\begin{align}
    \big<N_{\rm cen}(M_{\rm h})\big> = \frac{f_{\rm max}}{2}\bigg[1 + \text{erf} \big(\frac{\log M_{\rm h} - \log M_{\rm min} }{\sigma_{\log M}}\big)\bigg].
\end{align}
Here, $M_{\rm min}$ is the halo mass below which we do not expect a halo to contain a central galaxy. The parameter $\sigma_{\log M}$ determines the transition of the mean occupation number between 0 and $f_{\rm max}$. The transition of $ \big<N_{\rm cen}(M_{\rm h})\big>=0$ to $ \big<N_{\rm cen}(M_{\rm h})\big>=f_{\rm max}$ occurs quickly if the value of $\sigma_{\log M}$ is lower, and the transition is slower for larger values of $\sigma_{\log M}$. The mean occupation of the satellite galaxy, which depends on the host halo mass, is given by 
\begin{align}
    \big<N_{\rm sat}(M_{\rm h})\big> = \bigg(\frac{M_{\rm h} - M_{\rm cut}}{M_1}\bigg)^{\alpha}.
\end{align}
Here, $M_{\rm cut}$ is the minimum halo mass to host a satellite galaxy, and $M_1$ is the typical halo mass at which a halo hosts one satellite galaxy. The parameter $\alpha$ is the power-law index. The position and velocity of the central galaxy are set to be the same as the halo center. To determine the position of satellite galaxies within the halo virial radius, we use the Navarro–Frenk–White (NFW) proﬁle~\citep{Navarro_1996}. The values of the free parameters we adopt to generate the mock galaxy catalogs are $\{f_{\rm max},\log M_{\rm min}, \sigma_{\log M}, \log M_{\rm cut}, \log M_1, \alpha\}=\{0.3362,12.11,0.4765,12.06,13.72,1.175 \}$. These values are chosen such that the HOD applied to our simulation box reproduces the number densities of H$\alpha$-emitting galaxies that match Model 3 described in \cite{Pozzetti_2016} and the HOD of the Euclid Flagship simulation~\citep{Pezzotta_2024,castander_2025}.
\section{Power spectrum and bispectrum estimation}
\label{sec:bispectrum estimation}
\subsection{Power Spectrum}
The power spectrum of a signal $S$, which has spatial fluctuation $\delta_{S}(\bm{x})$ is defined as 
\begin{equation}
\big<\Delta_{S}(\bm{k}) \Delta_{S}^{*}(\bm{k}^{'})\big> = V \delta^{K}_{\bm{k}-\bm{k'},0}~P_{S}(\bm{k}),
\end{equation}
where $\Delta_{S}(\bm{k})$ represents the 3D Fourier transform of the signal fluctuation $\delta_S(\bm{x})$  and $V$ is the volume under consideration. The $\delta^{K}_{\bm{k}-\bm{k'},0}$ ensures that the power spectrum is nonzero only when $\bm{k}=\bm{k'}$, whereas $\big<\cdots\big>$ denotes the ensemble average. The cross-power spectrum of two signals $S_1$ and $S_2$ can be analogously defined as 
\begin{equation}
    \big<\Delta_{S_1}(\bm{k}) \Delta_{S_2}^{*}(\bm{k^{'}})\big> = V \delta^{K}_{\bm{k}-\bm{k'},0}~P_{S_1 \times S_2}(\bm{k}).
\end{equation}
\subsection{Bispectrum}
The bispectrum of a signal $S$, which has spatial fluctuation $\delta_{S}(\bm{x})$ is defined as 
\begin{multline}
    \langle \Delta_{S}(\bm{k_1}) \Delta_{S}(\bm{k_2}) \Delta_{S}(\bm{k_3}) \rangle = 
        \\ V \delta_{\bm{k_1} + \bm{k_2} + \bm{k_3},0}~B_{S}(\bm{k_1},\bm{k_2},\bm{k_3}),
\end{multline}
    where $\delta_{\bm{k_1} + \bm{k_2} + \bm{k_3},0}$ is the Kronecker delta function, which equals unity when the condition  $\bm{k_1} + \bm{k_2} + \bm{k_3}=0$ is satisfied and zero otherwise. This ensures that only the closed $k$-triangles contribute to the bispectrum. Similarly, the cross-bispectrum for three fields $\delta_{S_{1}}(\bm{x})$, $\delta_{S_{2}}(\bm{x})$ and $\delta_{S_{3}}(\bm{x})$ can be defined as 
    \begin{multline}
        \langle \Delta_{S_1}(\bm{k_1}) \Delta_{S_2}(\bm{k_2}) \Delta_{S_3}(\bm{k_3}) \rangle = \\ V \delta_{\bm{k_1} + \bm{k_2} + \bm{k_3},0}~B_{S_1,S_2,S_3}(\bm{k_1},\bm{k_2},\bm{k_3}).
    \end{multline}
    The cross-bispectrum of the signals can be estimated for various cross-combinations.  In this work, we only consider the cross-bispectrum of the two signals (denoted by $S_1$ and $S_2$). This results in six different cross-combinations and they are $B_{S_1, S_2, S_2}$, $B_{S_2, S_1, S_2}$, $B_{S_2, S_2, S_1}$, $B_{S_1, S_1, S_2}$, $B_{S_1, S_2, S_1}$ and $B_{S_2, S_1, S_1}$.
    \par
    We estimate the binned auto-bispectrum $\bar B_{S}(k_1,k_2,k_3)$ and the cross-bispectrum $\bar B_{S_1 \times S_2 \times S_3}(k_1,k_2,k_3)$ from the simulated data using an updated version of the fast bispectrum estimator presented in \citet{Shaw_2021}, which closely follows the algorithm of \citet{Scoccimarro_2015} and \citet{Sefusatti_2015}. The binned auto-bispectrum is given by
    \begin{multline} \label{eq_binned_auto_bispec}
        \bar B_{S}(k_1,k_2,k_3)=\\ \frac{1}{V}\frac{\sum_{\bm{x}}D_S(k_1,\bm{x})D_S(k_2,\bm{x})D_S(k_3,\bm{x})}{\sum_{\bm{x}}I_S(k_1,\bm{x})I_S(k_2,\bm{x})I_S(k_3,\bm{x})}.
    \end{multline}
     Here, the terms $D_S(k,\bm{x})$ and $I_S(k,\bm{x})$ are the inverse Fourier transform of the masked $\delta_{S}(\bm{x})$ field and unit field, respectively,
    \begin{align}
        D_s(k, \bm{x}) = \sum_{||\bm{k}|-k|<\Delta k/2} \Delta_S(\bm{k}) \exp(-i\bm{k}\cdot \bm{x}),
    \end{align}
        \begin{align}
        I_s(k, \bm{x}) = \sum_{||\bm{k}|-k|<\Delta k/2}  \exp(-i\bm{k}\cdot \bm{x}),
    \end{align}
    where all $\bm{k}-\text{modes}$ within a $k-\text{shell}$ of width $\Delta k$ centered on $k$ are summed. We refer the reader to \citet{Shaw_2021} for further details of the method.
    Similarly, the binned cross-bispectrum is given by
    \begin{multline} \label{eq_binned_cross_bispec}
        \bar B_{S_1 \times S_2 \times S_3}(k_1,k_2,k_3)= \\
        \frac{1}{V}\frac{\sum_{\bm{x}}D_{S_1}(k_1,\bm{x})D_{S_2}(k_2,\bm{x})D_{S_3}(k_3,\bm{x})}{\sum_{\bm{x}}I_{S_1}(k_1,\bm{x})I_{S_2}(k_2,\bm{x})I_{S_3}(k_3,\bm{x})},
    \end{multline}
    where $D_{S_1}(k,\bm{x})$, $D_{S_2}(k,\bm{x})$ and $D_{S_2}(k,\bm{x})$ are inverse Fourier transforms of the masked $\delta_{S_1}(\bm{x})$, $\delta_{S_2}(\bm{x})$ and $\delta_{S_3}(\bm{x})$ fields, respectively. 

    \subsubsection{The unique triangle configurations in the triangle parameter space}
\begin{figure}
    \centering
    \includegraphics[width=0.9\linewidth]{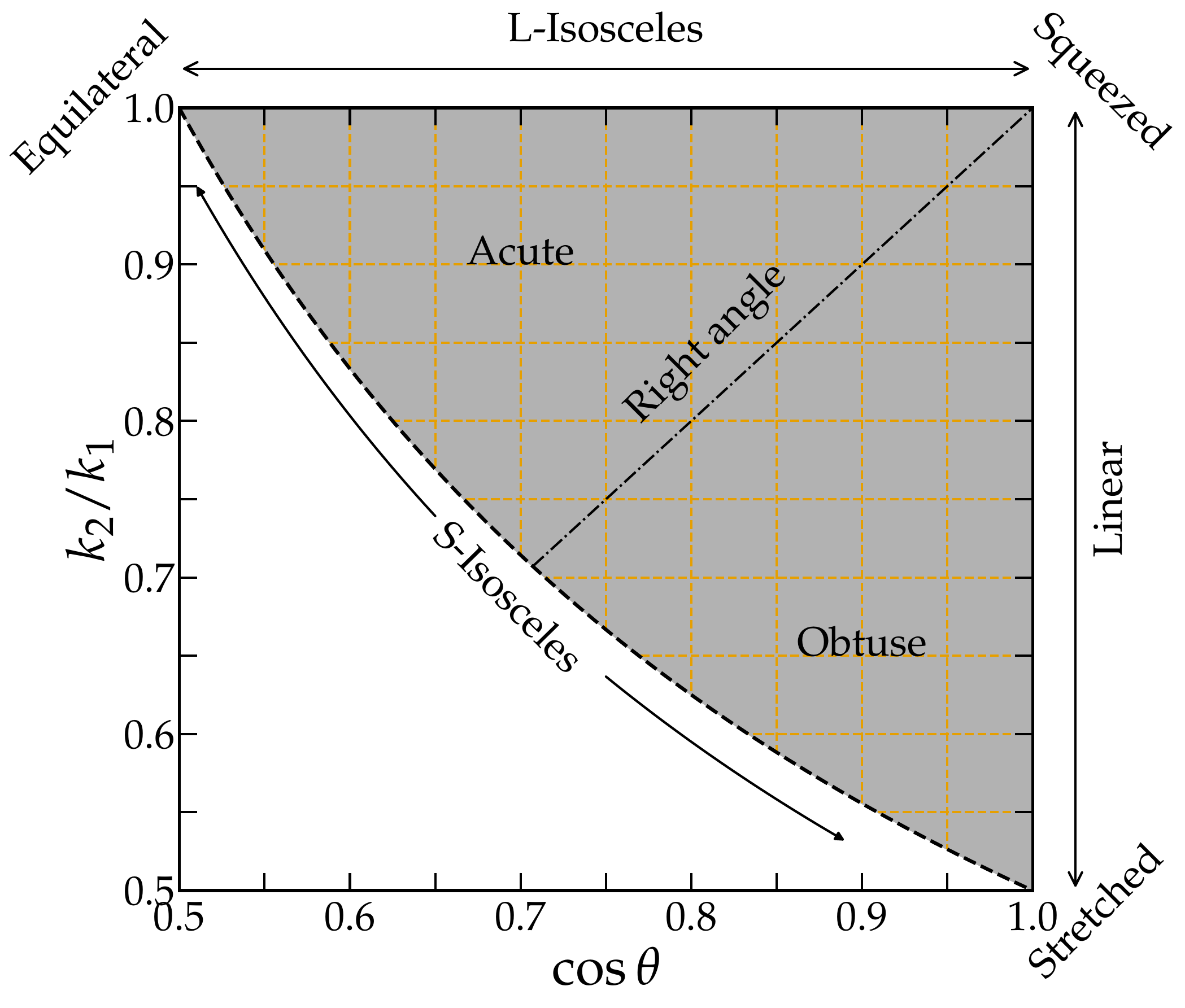}
    \caption{The unique shapes of the $k-\text{triangles}$ distributed over the $k_2/k_1-\cos \theta$ plane. Unique shapes are confined to the region where $\frac{k_2}{k_1}\cos\theta \geq 0.5$, shown in gray. We divided the entire $k_2/k_1 -\cos\theta$ plane with a grid size of $\Delta~k_2/k_1 = 0.05$ and $\Delta \cos \theta = 0.05$, represented by the orange color  grids.}
    \label{fig:n_cos_theta_plane}
\end{figure}

    The auto and cross-bispectrum can be estimated for different shapes and sizes of triangles in the Fourier space. To find all unique shapes of $k-\text{triangles}$ in Fourier space, we followed the bispectrum parameterization introduced in~\citet{Bharadwaj_2020} and \citet{Majumdar_2020}. For a triangle in Fourier space with $k_1\geq k_2 \geq k_3$, its size is determined by $k_1$ and the shape is determined by $k_2/k_1$ and the cosine of the acute angle between $\bm{k_1}$ and $\bm{k_2}$ ($\cos \theta = \bm{k_1}\cdot \bm{k_2}/k_1k_2$) along with following additional conditions:
    \begin{align}
        0.5\leq \cos \theta \leq 1.0, \\
        0.5 \leq k_2/k_1 \leq 1.0.
    \end{align}
    The triangles that satisfy the above conditions, along with $\frac{k_2}{k_1}\cos\theta \geq 0.5$, are unique in shape and are confined within the gray shaded region in Figure~\ref{fig:n_cos_theta_plane} in the $k_2/k_1-\cos \theta$ plane. We divided the entire $k_2/k_1 -\cos\theta$ plane with grid size $\Delta~k_2/k_1 = 0.05$ and $\Delta \cos \theta = 0.05$, represented by the orange color  grids. Various unique $k-\text{triangle}$ configurations include the L-isosceles ($k_2/k_1 = 1 ~\text{and} \cos\theta ~\epsilon~[0.5,1]$), S-isosceles ($k_2/k_1 \cos \theta = 0.5$), linear ($\cos \theta \rightarrow 1~\text{and}~k_2/k_1~\epsilon~[0.5,1]$), right angle  ($k_2/k_1 = \cos \theta$), acute angle ($\cos \theta<k_2/k_1$) and obtuse angle ($\cos \theta > k_2/k_1$), as shown in Figure~\ref{fig:n_cos_theta_plane}. Among these, the squeezed-limit ($k_2 / k_1 = 1, \cos \theta \rightarrow 1$), stretched  ($k_2/k_1 = 0.5, \cos \theta \rightarrow1$) and equilateral ($k_2/k_1 = 1, \cos \theta=0.5$) are $k-\text{triangles}$, which we explore in detail in this article. We present our results in terms of the normalized bispectrum defined as 
    \begin{align}
        \Delta^3(k_1, k_2/k_1,\cos \theta) = \frac{k_{1}^{3} k_{2}^{3} B(k_1, k_2/k_1,\cos \theta)}{(2\pi^{2})^2}.
    \end{align}
\par We estimated the binned bispectrum for $k_1$ in the range $[k_f, \frac{N}{3} k_f]$ with a bin width of $\Delta k_1=2k_f$, where $k_f = 2\pi / \text{box size}$ is the fundamental wavenumber. We divided the entire  $k_2 / k_1 - \cos{\theta}$ plane with grid size $\Delta k_2/k_1=0.05$ and $\Delta\cos\theta$ = 0.05.
\section{Bispectrum estimates from simulations}
\label{sec:bispectrum results}
\begin{figure*}
    \centering
    \begin{subfigure}{1\textwidth}
        \centering
        \includegraphics[width=\linewidth]{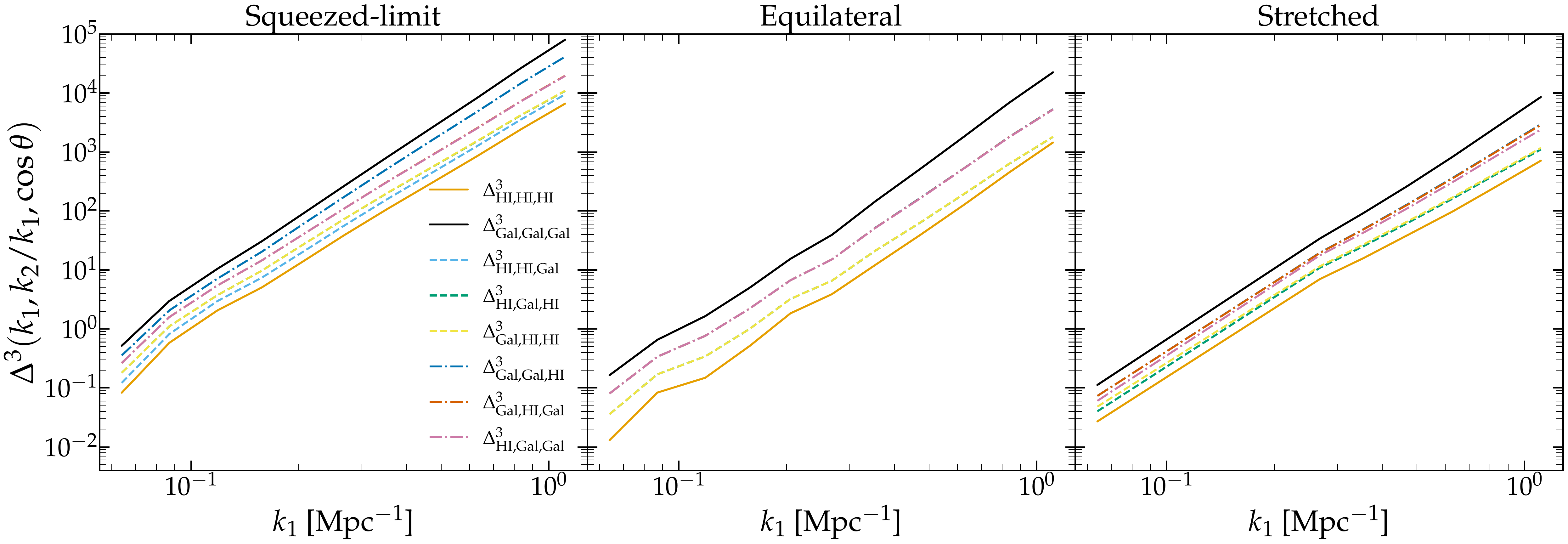}
        \caption{The left, middle, and right panels show the results for squeezed-limit, equilateral, and stretched $k-\text{triangles}$. The solid orange and black lines represent the HI and galaxy auto-bispectrum, respectively. The dashed lines correspond to cross-bispectrum combinations containing two HI fields $\Delta^3_{\rm HI,HI,Gal}, \Delta^3_{\rm HI,Gal,HI}$ and $\Delta^3_{\rm Gal, HI,HI}$), while the dash-dotted lines indicate combinations with a single HI field ($\Delta^3_{\rm Gal,Gal,HI}, \Delta^3_{\rm Gal,HI,Gal}$ and $\Delta^3_{\rm HI,Gal,Gal}$).}
        \label{sq_eq_st}
    \end{subfigure}
    \begin{subfigure}{1\textwidth}
        \centering
        \includegraphics[width=0.8\linewidth]{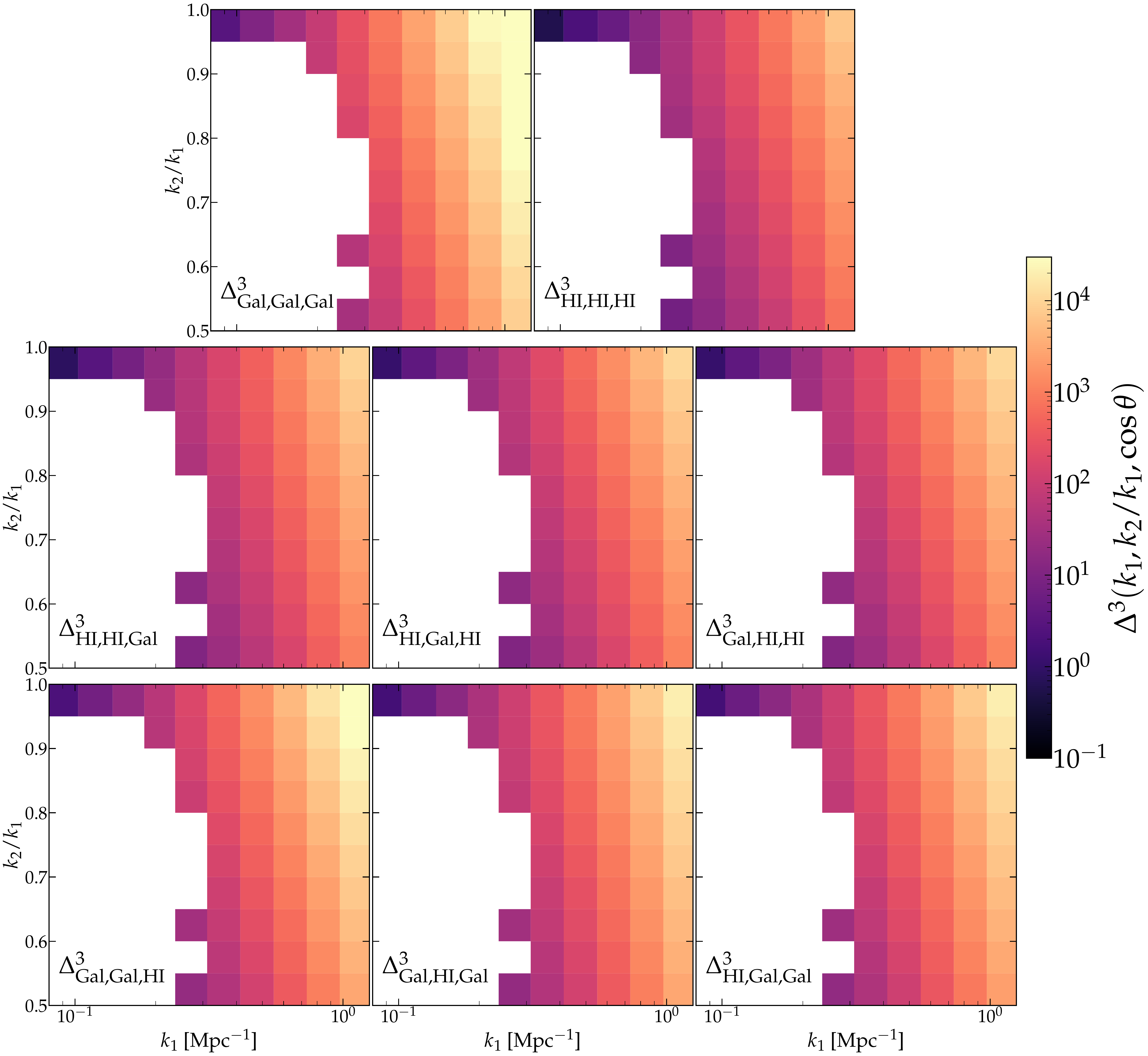}
        \caption{The bispectra for linear $k-{\rm triangles}$ ($k_2 / k_1 ~\epsilon ~[0.5, 1]$ and $\cos \theta \rightarrow 1$). The top panel presents the HI and galaxy cross-bispectrum. The middle and bottom panels show HI-galaxy cross-bispectrum with two HI fields ($\Delta^3_{\rm HI,HI,Gal}, \Delta^3_{\rm HI,Gal,HI}$ and $\Delta^3_{\rm Gal, HI,HI}$) and with a single HI field ($\Delta^3_{\rm Gal,Gal,HI}, \Delta^3_{\rm Gal,HI,Gal}$ and $\Delta^3_{\rm HI,Gal,Gal}$), respectively.}
        \label{linear_bispectrum}
    \end{subfigure}
    \caption{The HI and galaxy auto and cross-bispectrum as a function of $k_1$ at $z=0.99$ estimated from GAEA simulations~( see Section~\ref{sec:GAEA_simulations}).}
    \label{}
\end{figure*}
In this section, we discuss the HI and galaxy auto- and cross-bispectrum estimates from HI line intensity maps and mock galaxy catalogs generated using GAEA SAMs. We begin by presenting the results for the HI and galaxy auto-bispectrum (Section~\ref{sec:auto_bispectrum}), followed by the cross-bispectrum for all possible combinations of HI and galaxy fields (Section~\ref{sec:HI_galaxy_cross_bispectrum}). In Section~\ref{sec:impact_of_RSD}, we examine the impact of redshift space distortions (RSD) on the HI-galaxy cross-bispectrum. Finally, Section~\ref{sec:modelliing_cross_bispec_from_PT} discusses the modeling of the HI-galaxy cross-bispectrum using perturbation theory.
\subsection{Auto-bispectrum}
\label{sec:auto_bispectrum}
\subsubsection{Squeezed-limit, equilateral, and  stretched $k-\text{triangles}$}
In Figure~\ref{sq_eq_st}, we present the normalized bispectrum for squeezed-limit, equilateral, and stretched $\text{triangles}$ as a function of $k_{1}$ at redshift $z=0.99$. The solid orange and black lines represent the HI auto-bispectrum ($\Delta^3_{\rm HI,HI,HI}$) and the galaxy auto-bispectrum ($\Delta^{3}_{\rm Gal,Gal,Gal}$), respectively. Both $\Delta^{3}_{\rm HI,HI,HI}$ and $\Delta^3_{\rm Gal,Gal,Gal}$ increase monotonically going from large scales (small $k_{1}$) to small scales (large $k_1$) across all triangle shapes. This trend arises because the large-scale structure becomes increasingly nonlinear for small scales due to the gravitational clustering of matter, making these scales highly non-Gaussian. As a result, the magnitude of the bispectrum increases toward smaller scales. The magnitude of $\Delta^3_{\rm Gal, Gal, Gal}$ is higher than that of $\Delta^3_{\rm HI,HI,HI}$ across all $k_1$ bins. This is due to the higher galaxy bias compared to the HI line intensity maps, which trace the HI distribution across all galaxies. Note that the magnitude of $\Delta^3_{\rm Gal,Gal,Gal}$ depends on the number density of the galaxies, which in turn varies with the stellar mass cut we applied to generate the mock galaxy catalog. However, irrespective of this selection criterion, $\Delta^3_{\rm Gal,Gal,Gal}$ will always have a higher magnitude than $\Delta^3_{\rm HI,HI,HI}$ across all $k_1$ bins. The squeezed-limit auto-bispectrum for both HI and galaxies exhibits higher magnitudes than the equilateral and stretched $\text{triangles}$ across all $k_1$ bins. The auto-bispectrum shows a power-law behavior  $\Delta^3\approx k_1^{n}$ and the spectral index ($n$) varies with the shape of the triangle. The spectral index values for $ \Delta^3_{\rm HI,HI,HI}$ and $\Delta^3_{\rm Gal,Gal,Gal}$ for the squeezed-limit $\text{triangles}$ are $n\approx3.6$ and $n\approx4$, respectively.
The $\Delta^3_{\rm HI,HI,HI}$ have $n\approx3.9$ and $n\approx3.6$ for equilateral and stretched, where the corresponding values for $\Delta^3_{\rm Gal,Gal,Gal}$ are $n\approx4.3$ and $n\approx3.1$. These results are consistent with \cite{Sarkar_2019}, who investigated the HI auto-bispectrum using a set of semi-numerical simulations of the HI distribution.
\subsubsection{Linear $k-\text{triangles}$}
In the upper panel of Figure \ref{linear_bispectrum}, we present $\Delta^3_{\rm HI,HI,HI}$ and $\Delta^3_{\rm Gal,Gal,Gal}$ for linear $\text{triangles}$ as a function of $k_1$ at redshift $z=0.99$. For linear $\text{triangles}$ $\cos \theta \rightarrow 1$ and $k_2/k_1$ vary from 0.5 (stretched) to 1 (squeezed). First, we focus on $\Delta^3_{\rm HI,HI,HI}$. Considering any fixed $k_2/k_1$ bin, the magnitude of the bispectrum increases from large scales (small $k_1$ bin) to small scales (large $k_1$ bin) as the HI field becomes highly non-Gaussian for small scales due to the gravitational clustering of matter. For any fixed $k_1$ bin, the magnitude of the bispectrum increases going from stretched ($k_2/k_1 = 0.5 $) to squeezed-limit ($k_2/k_1 = 1 $) $\text{triangles}$. The galaxy bispectrum exhibits a similar nature as the HI bispectrum, with an increase in magnitude in every $k_1$ and $k_2/k_1$ bin.  
\subsection{HI-galaxy cross-bispectrum}
\label{sec:HI_galaxy_cross_bispectrum}
\subsubsection{Squeezed-limit, equilateral, and  stretched $k-\text{triangles}$}
In Figure \ref{sq_eq_st}, we present six different combinations of the normalized HI-galaxy cross-bispectrum for the squeezed-limit, equilateral and stretched $\text{triangles}$ as a function of $k_{1}$ at redshift $z=0.99$. The dashed line represents the cross-bispectrum combinations that contain two HI fields and a single galaxy field, namely $\Delta^3_{\rm HI, HI, Gal},~\Delta^3_{\rm HI, Gal, HI}~\text{and}~ \Delta^3_{\rm Gal, HI, HI}$. Furthermore, the dashed-dotted line corresponds to the combination with a single HI field and two galaxy fields, $\Delta^3_{\rm Gal,Gal,HI},~\Delta^3_{\rm Gal,HI,Gal}~\text{and}~\Delta^3_{\rm HI,Gal,Gal}$.
\par
The magnitude of all the cross-bispectrum combinations increases monotonically, going from large scales (small $k_1$ bin)  to small scales (large $k_1$ bin) across all $\text{triangles}$. This behavior is similar to that of HI and the galaxy auto-bispectrum. However, the magnitudes of different cross-bispectrum combinations for a fixed $k_1$ bin vary depending on the field combinations. The HI-galaxy cross-bispectrum that contains two HI fields ($\Delta^3_{\rm HI,HI,Gal}$, $\Delta^3_{\rm HI,Gal,HI}$, $\Delta^3_{\rm Gal,HI,HI}$) shows a lower bispectrum value than the combination with a single HI field ($\Delta^3_{\rm Gal,Gal,HI}$, $\Delta^3_{\rm Gal,HI,Gal}$, $\Delta^3_{\rm HI,Gal,Gal}$) across all $\text{triangles}$. Additionally, depending on the shape ($k_2/k_1$ and $\cos \theta$) of the $\text{triangle} $, each cross-bispectrum combination shows variation in magnitude for every $k_1$ bin.
\par
First, we consider the cross-bispectrum for squeezed-limit $\text{triangles}$. Considering cross-bispectrum combination with two HI fields, the magnitude of $\Delta^3_{\rm HI,Gal,HI}$ and $\Delta^3_{\rm Gal,HI,HI}$ are the same. Recall that for the squeezed-limit bispectrum, the two sides of the $\text{triangle}$ are equal, $k_1=k_{2} $, and the third side $k_3 \rightarrow 0$. Thus, interchanging the HI and galaxy field between $k_{1}$ and $k_{2}$ does not change the magnitude of the bispectrum. A similar trend is observed for combinations $\Delta^3_{\rm Gal,HI,Gal}$ and $\Delta^3_{\rm Gal,HI,Gal}$, which exhibit the same magnitude across all scales. Note that cross-bispectrum estimates do not capture the bispectrum exactly for $k_2/k=1$ and $\cos\theta\rightarrow1$; rather, we obtain binned estimates for $k_2/k_1 = 0.975$  and $\cos\theta=0.975$ bins. This slight deviation from perfect symmetry results in a minor variation in the magnitude of $\Delta^3_{\rm HI,Gal,HI}$ and $\Delta^3_{\rm Gal,HI,HI}$ (as well as between $\Delta^3_{\rm Gal,HI,Gal}$ and $\Delta^3_{\rm Gal,HI,Gal}$ ) for large $k_1$ bins. However, this difference remains below 2\%. 
\par 
Next, we look at the cross-bispectrum for the equilateral and stretched $\text{triangles}$, which is presented in the middle and right panels of Figure~\ref{sq_eq_st}. For equilateral $\text{triangles}$, all sides are equal, $k_1 = k_2 = k_3$, which results in the same cross-bispectrum magnitude across all $k_1$ bins for all combinations with two HI fields ($\Delta^3_{\rm HI,HI,Gal}$, $\Delta^3_{\rm HI,Gal,HI}$ and $\Delta^3_{\rm Gal,HI,HI}$). This trend is also true for cross-bispectrum combinations with a single HI field and two galaxy fields ($\Delta^3_{\rm Gal,Gal,HI}$, $\Delta^3_{\rm Gal,HI,Gal}$ and $\Delta^3_{\rm HI,Gal,Gal}$). This happens due to the same reason we pointed out for the squeezed-limit $k-\text{triangles}$. The behavior of the cross-bispectrum for stretched $\text{triangles}$ as a function of $k_1$ is similar to that of the squeezed-limit and equilateral. However, the magnitude of each cross-bispectrum combination is lower compared to the corresponding cross-bispectrum for squeezed-limit and equilateral.
\par
Similarly to the auto-bispectrum, the cross-bispectrum for all combinations shows a power-law behavior, where the spectral index varies depending on the shape of the $\text{triangles}$ and cross-bispectrum combination. The cross-bispectrum for the squeezed-limit with two HI fields and a single HI field has spectral indices $n\approx3.3$ and $n\approx3.4$, respectively. The corresponding values for stretched are $n\approx3.2$ and $n\approx3.4$. All cross-bispectrum combinations of equilateral $k-\text{triangles}$ have $n\approx3.7$.
\subsubsection{Linear $k-\text{triangles}$}
In Figure~\ref{linear_bispectrum}, we present the HI-galaxy cross-bispectrum for linear $k-\text{triangles}$ as a function of $k_1$ at $z=0.99$. For any fixed $k_2/k_1$, the magnitude of cross-﻿bispectrum for all combinations increases from the small $k_1$ bin to the large $k_1$ bin. Considering any fixed $k_1$ bin, the magnitude of the cross-bispectrum for all combinations increases going from the stretched to the squeezed-limit. The cross-bispectrum combinations that contain two HI fields ($\Delta^3_{\rm HI,HI,Gal}$, $\Delta^3_{\rm HI,Gal,HI}$, $\Delta^3_{\rm Gal, HI,HI}$) have lower bispectrum magnitude than with a single HI field ($\Delta^3_{\rm Gal, Gal, HI}$, $\Delta^3_{\rm Gal, HI, Gal}$, $\Delta^3_{\rm HI, Gal, Gal}$). 
\begin{figure*}
\centering
\includegraphics[width=0.85\linewidth]{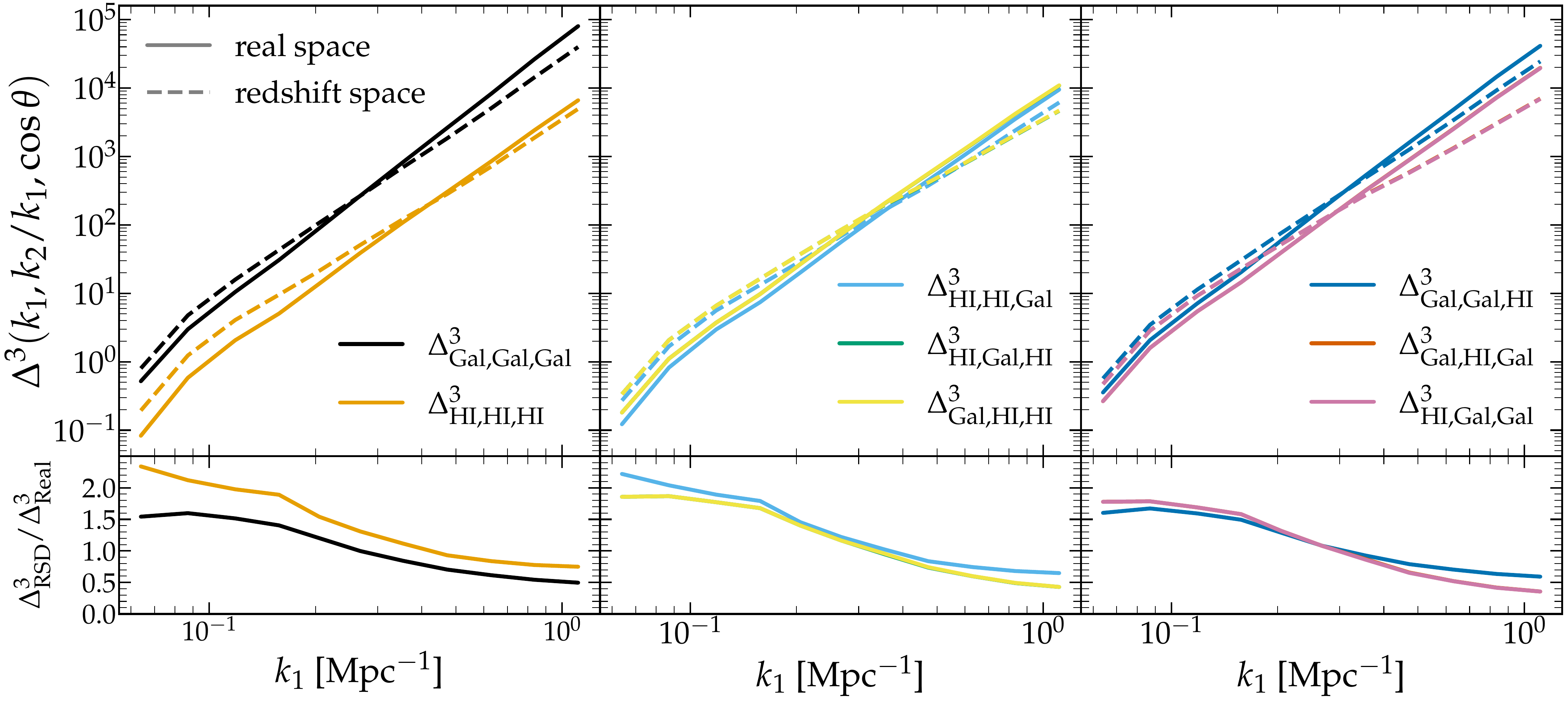}
\caption{Impact of RSD on the HI-galaxy auto and cross-bispectrum for squeezed-limit $\text{triangles}$ at $z=0.99$. The solid lines show auto and cross-bispectrum in the real space, while the dashed line corresponds to the redshift space bispectrum. The left panel presents the HI and galaxy auto-bispectrum. The middle and right panels show HI-galaxy cross-bispectrum with two HI fields ($\Delta^3_{\rm HI,HI, Gal},\Delta^3_{\rm HI,Gal,HI}~\text{and}~\Delta^3_{\rm Gal,HI,HI}$) and a single HI field ($\Delta^3_{\rm Gal,Gal, HI},\Delta^3_{\rm Gal,HI,Gal}~\text{and}~\Delta^3_{\rm HI,Gal,Gal}$), respectively.}
\label{fig:impact_of_rsd_squeezed}
\end{figure*}
\subsection{Impact of Redshift Space Distortions}
\label{sec:impact_of_RSD}
Redshift Space Distortions (RSD), arising from the peculiar velocities of galaxies, will distort the observed cross-bispectrum. Hence, incorporating these effects in the modeling  of the cross-bispectrum is crucial for the correct interpretation of the signal. In this section, we discuss  the impact of the RSD on the HI-galaxy cross-bispectrum estimated from the GAEA simulations.
For implementing the RSD, we displace the real space position $\bm{x}$ of the galaxy along the $z$-axis, which we assume as  the line of sight, following the plane parallel approximation 
\begin{align}
        \bm{s} = \bm{x} + \frac{1+z}{H(z)}\big[ \bm{v}(\bm{x})\cdot \bm{\hat z}\big]\bm{\hat z} .
\end{align}
Here, $\bm{s}$ is the redshift space position of the galaxy and $\bm{v}(\bm{x})$ is its peculiar velocity. 
In the top panel of  Figure~\ref{fig:impact_of_rsd_squeezed}, the solid and dashed lines show the HI and galaxy auto and cross-bispectrum for the squeezed-limit $\text{triangles}$ in real space and redshift space, respectively. In the bottom panel, we show the ratio of the redshift space bispectrum to the real space bispectrum. The left, middle and right panels correspond to the HI/galaxy auto-bispectrum, HI-galaxy cross-bispectrum with two HI fields and with a single HI field, respectively. First, we discuss the impact of RSD on the auto-bispectrum. RSD enhances the signal for large scales (small $k_1$) and suppresses it for small scales (large $k_1$). The enhancement in the magnitude for large scales is due to the fact that there is more clustering at these scales due to the Kaiser effect~\citep{Kaiser_1987}, while the suppression in the magnitude for small scales is due to the randomness of peculiar velocities of galaxies~\citep{Scoccimarro:1999ed,Taruya:2010mx, Scoccimarro:2004tg, Zheng:2016zxc}. For $k_1 = 0.1~\text{Mpc}^{-1}$ bin, we observe an increase in the signal of approximately 100\% for $\Delta^3_{\rm HI,HI,HI}$ and 50\% for $\Delta^3_{\rm Gal,Gal,Gal}$. Considering $k_1 = 1~\text{Mpc}^{-1}$ bin, the signal is suppressed by 20\% and 80\% for $\Delta^3_{\rm HI,HI,HI}$ and $\Delta^3_{\rm Gal,Gal,Gal}$, respectively.
\par
The redshift space HI-galaxy cross-bispectrum as a function of $k_1$ for different combinations follows a similar trend as the HI auto-bispectrum. For $k_1 = 0.1~\text{Mpc}^{-1}$ bin, the cross-bispectrum combinations with two HI field, $\Delta^3_{\rm HI,HI,Gal}$ and $\Delta^3_{\rm Gal,HI,HI}$, show an approximately 95\%  and 80\% increase in the signal in comparison to the corresponding real space cross-bispectrum. For combinations with a single HI field, the increase in signal magnitude is approximately 90\% and 80\% for $\Delta^3_{\rm Gal,Gal,HI}$ and $\Delta^3_{\rm HI,Gal,Gal}$, respectively. Considering $k_1=1~\text{Mpc}^{-1}$ bin, the suppression in signal magnitude of redshift space cross-bispectrum in comparison with real space for $\Delta^3_{\rm HI,HI,Gal}$, $\Delta^3_{\rm Gal,HI,HI}$, $\Delta^3_{\rm Gal,Gal,HI}$ and $\Delta^3_{\rm HI,Gal,Gal}$ are approximately  30\%, 70\%, 70\%, and 70\%, respectively. This analysis shows that including RSD is essential for the correct interpretation of the auto and cross-bispectrum.
\subsection{Modelling the HI-galaxy cross-bispectrum from perturbation theory}
\label{sec:modelliing_cross_bispec_from_PT}
   \begin{table*}[]
    \centering
    \renewcommand{\arraystretch}{1.5} 
    \begin{tabular}{l c c c c}
        \hline
        Data Set & $b_{1,\rm HI}$ & $b_{2,\rm HI}$ & $b_{1,\rm Gal}$ & $b_{2,\rm Gal}$ \\
        \hline
        Base$^\dagger$ + $B_{\rm HI,HI,Gal} + B_{\rm HI,Gal,HI} + B_{\rm Gal,HI,HI}$ & $1.212^{+0.013}_{-0.013}$ & $-0.341^{+0.041}_{-0.045}$ & $1.888^{+0.021}_{-0.022}$ & $0.156^{+0.087}_{-0.090}$ \\
        Base + $B_{\rm HI,HI,Gal}$ & $1.251^{+0.013}_{-0.013}$ & $-0.371^{+0.041}_{-0.045}$ & $1.884^{+0.023}_{-0.023}$ & $0.301^{+0.104}_{-0.102}$ \\
        Base + $B_{\rm HI,Gal,HI}$ & $1.245^{+0.013}_{-0.013}$ & $-0.315^{+0.048}_{-0.048}$ & $1.904^{+0.022}_{-0.022}$ & $0.209^{+0.087}_{-0.098}$ \\
        Base + $B_{\rm Gal,HI,HI}$ & $1.234^{+0.013}_{-0.013}$ & $-0.246^{+0.048}_{-0.048}$ & $1.925^{+0.023}_{-0.023}$ & $0.085^{+0.094}_{-0.092}$ \\
        Base: $P_{\rm HI} + P_{\rm Gal} + B_{\rm HI,HI,HI} + B_{\rm Gal,Gal,Gal}$ & $1.265^{+0.013}_{-0.013}$ & $-0.264^{+0.049}_{-0.049}$ & $1.916^{+0.024}_{-0.024}$ & $0.256^{+0.104}_{-0.105}$ \\
        $P_{\rm HI}+B_{\rm HI,HI,HI}$ &$1.266^{+0.013}_{-0.012}$ &$-0.268^{+0.050}_{-0.054}$ &$-$ &$-$ \\
        $B_{\rm HI,HI,HI}$ &$1.103^{+0.053}_{-0.045}$ &$0.239^{+0.151}_{-0.203}$ & & \\
        $P_{\rm HI} + P_{\rm Gal} + P_{\rm HI,Gal}$ &$1.234^{+0.014}_{-0.014}$ &$-$ &$1.948^{+0.024}_{-0.023}$ &$-$ \\
        $P_{\rm HI}$ &$1.286^{+0.015}_{-0.014}$ & $-$ & $-$ & $-$ \\
        \hline
    \end{tabular}
    \caption{Marginalized constraints on the linear ($b_1$) and quadratic ($b_2$) bias. $^\dagger$Base includes $P_{\rm HI} + P_{\rm Gal} + B_{\rm HI,HI,HI} + B_{\rm Gal,Gal,Gal}$.}
    \label{tab:bias_parameters}
\end{table*}
\begin{figure}
        \centering
        \includegraphics[width=1\linewidth]{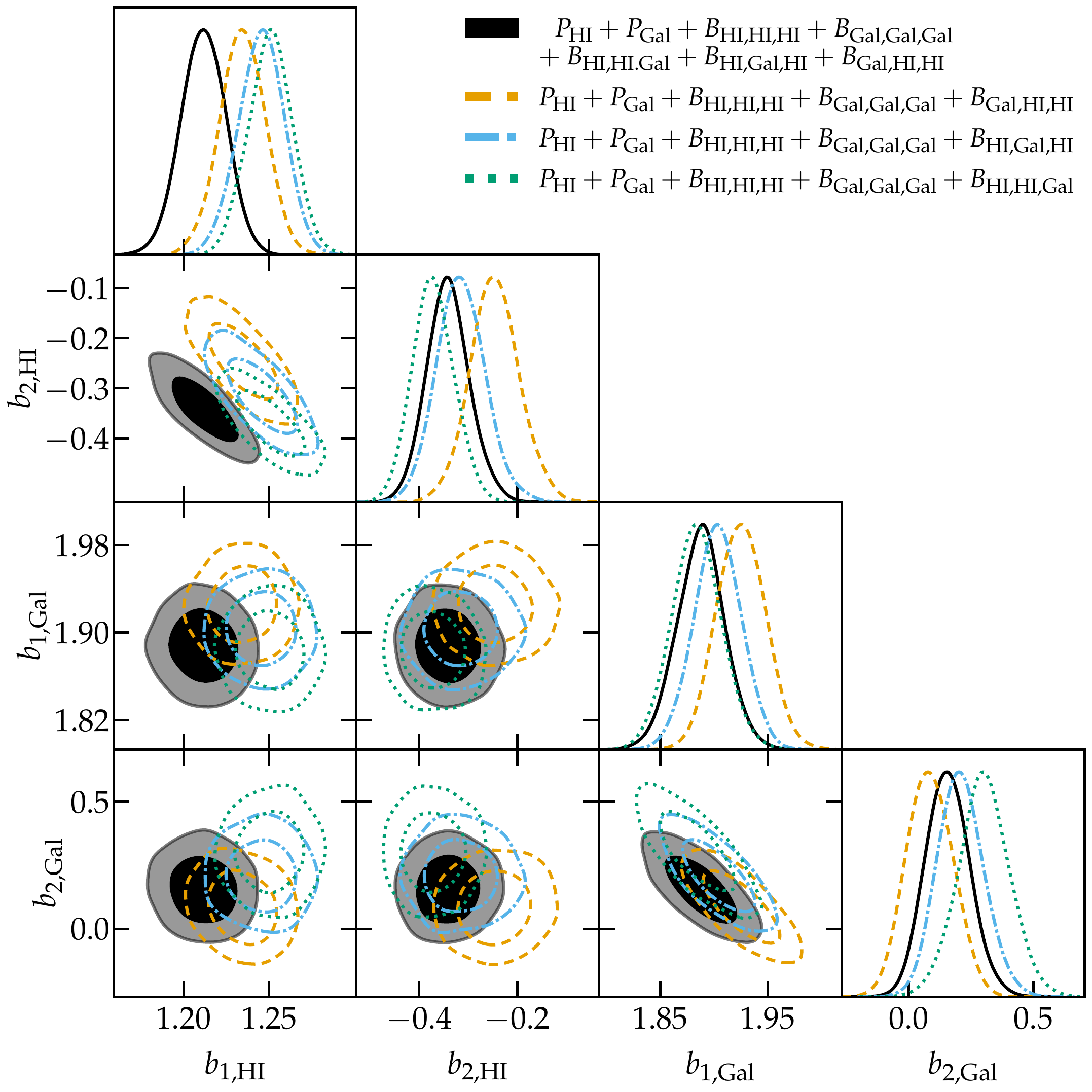}
        \caption{Joint constraints ($1\sigma$ and $2\sigma$ confidence regions) and marginalized posterior distributions for the linear ($b_{1,\rm HI}~\text{and}~b_{1,\rm Gal}$) and quadratic ($b_{2,\rm HI}~\text{and}~b_{2,\rm Gal}$) bias parameters of the HI and galaxy fields. The results correspond to the various dataset combinations listed in Table~\ref{tab:bias_parameters}. The black contours represent the constraints obtained from the combined dataset: $P_{\rm HI} + P_{\rm Gal} + B_{\rm HI,HI,HI} + B_{\rm Gal,Gal,Gal} + B_{\rm HI,HI,Gal}$ $ + B_{\rm HI,Gal,HI} + B_{\rm Gal,HI,HI}$. We utilize the best-fit bias parameters from this specific dataset for the HI-galaxy cross-bispectrum model predictions presented in Figure~\ref{fig:bs_SPT_fit}.}
        \label{fig:bias_fit_corner_plot}
\end{figure}
\begin{figure}
    \centering
    \includegraphics[width=0.7\linewidth]{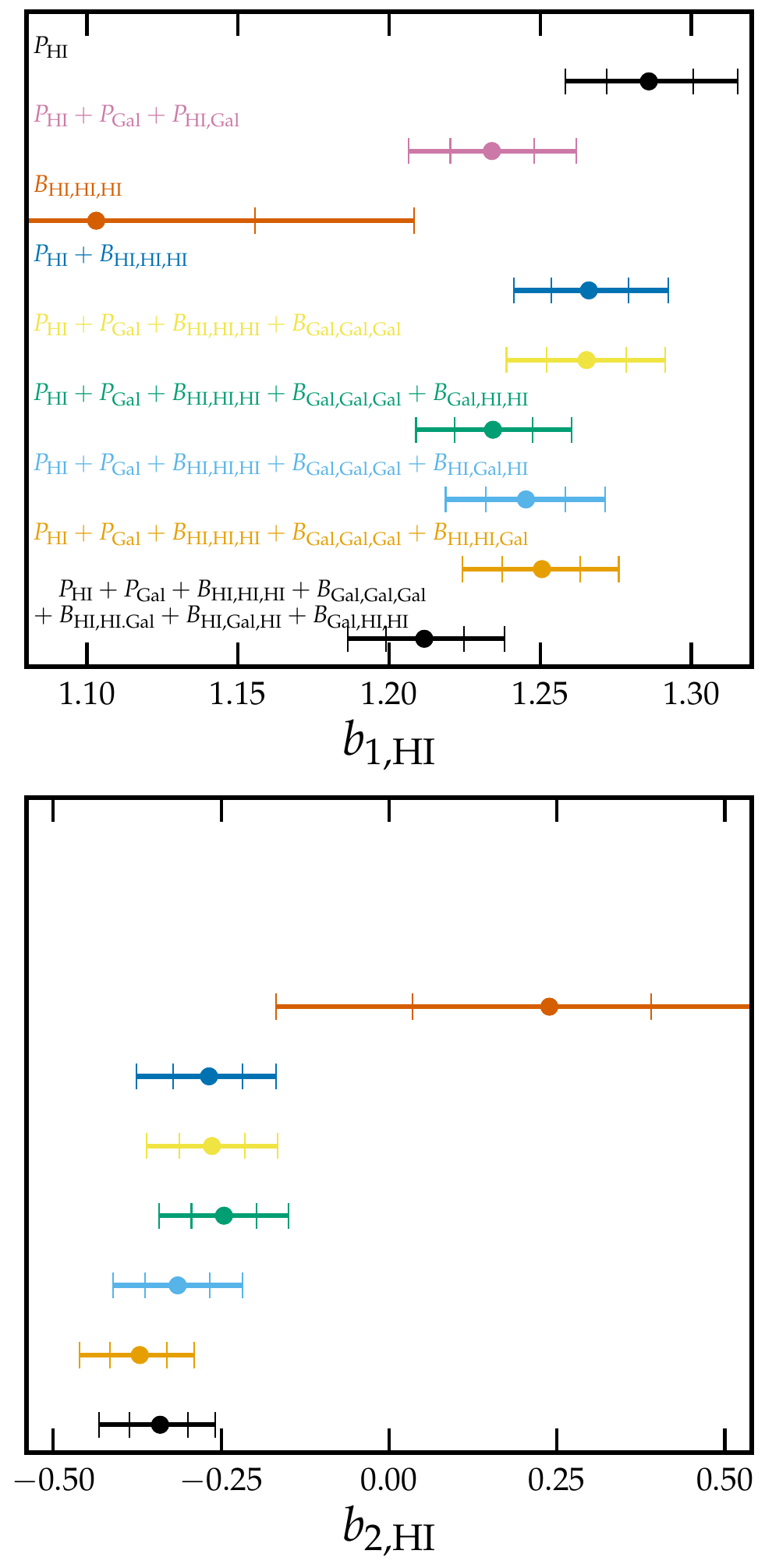}
    \caption{ID marginalized constraints on HI linear ($b_{1,\rm HI}$) and quadratic ($b_{2,\rm HI}$) bias parameters obtained from the different datasets listed in Table~\ref{tab:bias_parameters}.}
    \label{fig:b1_b2_bias_for_diff_dataset}
\end{figure}
In this section, we compare the predictions of the HI-galaxy cross-bispectrum from standard perturbation theory with bispectrum estimates from simulations. This analysis aims to identify the range of $k_1$ values for which the cross-bispectrum can be adequately modeled using perturbation theory and then use this to extract the linear and quadratic HI bias parameters. We utilize bispectra estimated from 50 independent realizations of HI line intensity maps and mock galaxy catalogs generated using the HOD (see section \ref{sec:HOD_simulations}) for this analysis. Here, our analysis is limited to real space, and a more comprehensive examination in redshift space will be addressed in future work.
\par
The HI and galaxies are biased tracers of the large-scale structure. The distribution of the HI and galaxy follows the dark matter distribution with a bias. We model this relationship by expanding the density contrasts  $\delta_{\rm HI}(\bm{x})$ and $\delta_{\rm Gal}(\bm{x})$ perturbatively up to quadratic order~\citep{Bernardeau_2002,Desjacques_2018}
\begin{equation}
        \delta_{r}(\bm{x}) = b_{1, r} \delta(\bm{x}) + \frac{b_{2,r}}{2} \delta(\bm{x})^2.
\end{equation}
Here, the subscript $r$ denotes the biased tracer probe of the dark matter, which can be HI or a galaxy field, and $\delta(\bm{x})$ represents the dark matter density contrast. The coefficients $b_{1, r}$ and $b_{2, r}$ denote the linear and quadratic bias of the tracer $r$, respectively. The linear bias of the probe $r$ can be constrained using the auto-power spectrum of that probe, given by
\begin{equation}
        P_r(k) = b_{1,r}^2 P(k)~, 
\end{equation}
where $P(k)$ is the matter power spectrum.
\par 
Following ~\cite{Guandalin_2022}, we model the cross-bispectrum $ B_{r,s,t}(\bm{k_1, k_2,k_3})$  of three biased tracers ($r,s,t$) in the real space as
\begin{multline}
        B_{r,s,t}(\bm{k_1, k_2,k_3}) = \mathcal{L}_{r,s,t} B(\bm{k_1}, \bm{k_2}, \bm{k_3}) + \\ \mathcal{Q}^{112}_{r,s,t} P(k_1) P(k_2) + \mathcal{Q}^{121}_{r,s,t} P(k_1) P(k_3) + \mathcal{Q}^{211}_{r,s,t} P(k_2) P(k_3),
\end{multline}
where the kernel $\mathcal{L}_{r,s,t} = b_{1,r}~b_{1,s}~b_{1,t}$ and kernel $\mathcal{Q}^{112}_{r,s,t} = b_{1,r} b_{1,s} b_{2,t}$ and so on. The matter bispectrum $B(\bm{k_1}, \bm{k_2}, \bm{k_3})$, is modelled as \citep{Fry_1984, Matarrese_1997, Scoccimarro_2000}
\begin{multline}
            B(\bm{k_1}, \bm{k_2}, \bm{k_3}) = 2F(\bm{k_1}, \bm{k_2}) P(k_1) P(k_2) + \text{cyc},\\
\end{multline}
\begin{multline}
\text{where }\\
F(\bm{k_1}, \bm{k_2}) = \frac{5}{7} + \frac{\bm{k_1} \bm{k_2}}{2} \bigg(\frac{1}{k_1^2} + \frac{1}{k_2^2}\bigg) + \frac{2}{7} \bigg(\frac{\bm{k_1}\bm{k_2}}{k_1 k_2}\bigg)^2.
\end{multline}
By fitting this model to the HI-galaxy cross-bispectrum estimates from a simulated set of HI line intensity maps and mock galaxy catalogs, we obtain best-fit linear and quadratic bias parameters for both HI and galaxies.
    \par 
    We employed a Bayesian framework to get the constraints on the bias parameters. Our goal is to estimate the posterior  distribution $\mathcal{P}(\theta|\mathcal{D},\mathcal{M})$ of the model parameters $\theta$
    of the model $\mathcal{M}$ conditioned on observations $\mathcal{D}$ (in our case, the mock signal). Under the Bayesian statistical framework, the posterior distribution is given by 
    \begin{align}
        \mathcal{P}(\theta|\mathcal{D},\mathcal{M}) = \frac{\mathcal{L}(\mathcal{D}|\theta,\mathcal{M}) \Pi(\theta,\mathcal{M})}{\mathcal{P}(\mathcal{D},\mathcal{M})},
    \end{align}
    where $\mathcal{L}(\mathcal{D}|\theta,\mathcal{M})$ $\Pi(\theta,\mathcal{M})$ and $\mathcal{P}(\mathcal{D},\mathcal{M})$ are the likelihood, prior, and Bayesian evidence, respectively. Here, we use a  multivariate
    Gaussian likelihood. The log-likelihood is given by  
    \begin{multline}
        \ln \mathcal{L}(\mathcal{D}|\theta,\mathcal{M}) = -\frac{1}{2}\bigg[(\mathcal{D}-\mu)^{T} \Sigma^{-1}(\mathcal{D}-\mu)\bigg] - \\ \frac{1}{2}\ln(2\pi|\Sigma|),
    \end{multline}
    where $\mu$ is the model prediction and $\Sigma$ is the covariance. The covariance is estimated from the data using the 50 independent realizations. We use a uniform prior, and the prior ranges are $b_{1,\rm HI}\epsilon[0,10]$, $b_{1,\rm Gal}\epsilon[0,10]$, $b_{2,\rm HI}\epsilon[-10,10]$ and $b_{2,\rm Gal}\epsilon[-10,10]$.  We restrict the model fitting upto $k_1 = 0.31~\text{Mpc}^{-1}$ as the standard perturbation theory is only expected to hold for weakly nonlinear scales. We estimate the bias parameters for different datasets quoted in the Table~\ref{tab:bias_parameters}. To sample the posterior distribution, we utilized the Markov Chain Monte Carlo (MCMC) approach. We employed the publicly available affine-invariant MCMC sampler \texttt{emcee}\footnote{\url{https://emcee.readthedocs.io/en/stable/}}~\citep{emcee_2013}.
\par
In Figure~\ref{fig:bias_fit_corner_plot}, we show the estimates of the parameters $\{b_{1,\rm HI}, b_{2,\rm HI}, b_{1, \rm Gal}, b_{2,\rm Gal}\}$ obtained from the first five datasets listed in Table~\ref{tab:bias_parameters}. In addition, we also show 1D marginalized constraints on $b_{1,\rm HI}$ and $b_{2,\rm HI}$ for all datasets in Figure~\ref{fig:b1_b2_bias_for_diff_dataset}. The parameters $b_{1,\rm HI}$ and $b_{2,\rm HI}$ are degenerate, and the bispectrum alone cannot constrain these parameters. The $B_{\rm HI,HI,HI}$ is able to provide tighter constraints on $b_{1,\rm HI}$ and $b_{2,\rm HI}$ only in combination with $P_{\rm HI}$ (see \citet{Yankelevich_2018} for galaxy bispectrum). Different datasets contain different combinations of the cross-bispectrum, and all combinations show slightly different mean estimates for the bias parameters, but these estimates are within their error bars.
\par
\begin{figure*}
        \centering
        \includegraphics[width=0.85\linewidth]{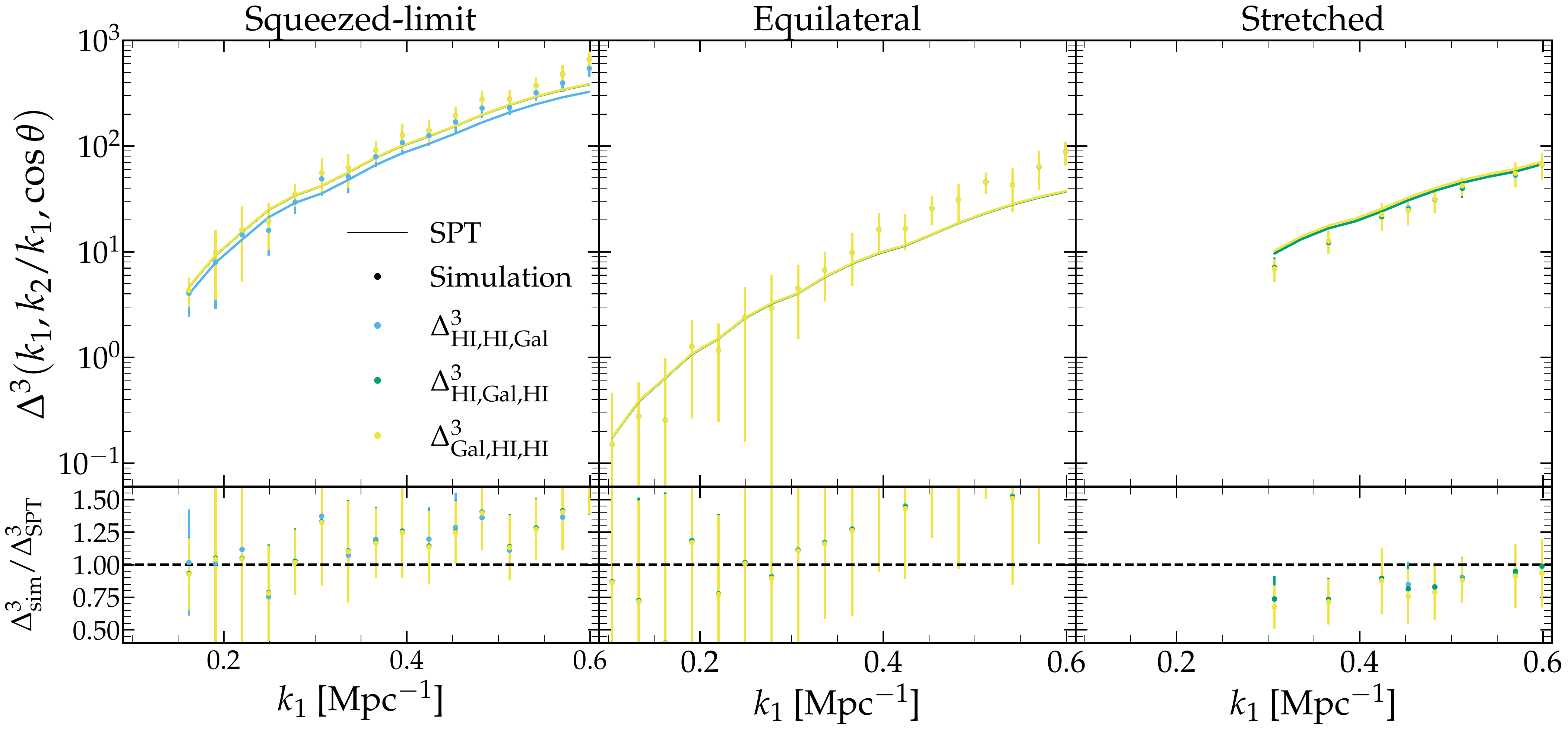}
        \caption{Comparing the perturbation theory predictions of the HI-galaxy cross-bispectrum with simulations at $z=0.99$. The solid line represents the perturbation theory predictions, while the points correspond to bispectrum estimates from simulations. Left, middle, and right panels present comparisons for squeezed-limit, equilateral, and stretched $k-\text{triangles}$ as a function of $k_1$. Different colors correspond to different combinations of the HI-galaxy cross-bispectrum.}
        \label{fig:bs_SPT_fit}
    \end{figure*}

In Figure~\ref{fig:bs_SPT_fit}, we present the comparison of the HI-galaxy cross-bispectrum predictions from the perturbation theory with the bispectrum estimates from the simulations as a function of $k_1$. The left, middle, and right panels show results for squeezed-limit, equilateral, and stretched $k-\text{triangles}$ respectively. The solid lines represent the perturbation theory predictions of the cross-bispectrum, while dotted points are the estimates from the simulations. Different colors denote different cross-combinations of the HI-galaxy cross-bispectrum. Here, we choose to show the cross-bispectrum combinations that contain two HI and one galaxy field. The error bars represent the $1\sigma$ uncertainty in the estimated cross-bispectrum measured across different realizations of the simulations. In the bottom panels,  we show the relative deviation of the perturbation theory predictions from the simulation estimates.
   \par 
    For squeezed-limit $k-\text{triangles}$, the deviation between the perturbation theory predictions and simulations remains below  10\% for the scales $k_1={0.35}~\text{Mpc}^{-1}$ across all cross-bispectrum combinations. Beyond this scale, the difference exceeds 20\%. However, the perturbation theory predictions of the cross-bispectrum for equilateral $k-\text{triangles}$ deviations are higher than 20\% even at larger scales. This might arise because the three sides of the equilateral triangles enter the nonlinear scales at the same time, and the effect of nonlinearities might be more substantial in comparison with the squeezed-limit~\citep{Marin_2012}. Our analysis shows that the cross-bispectrum can be modeled  from the standard perturbation theory for scales less than $k_1 = 0.35~\text{Mpc}^{-1}$. Our modeling can be improved by incorporating RSD and using the effective field theory of large-scale structure~\citep{Senatore:2014vja, Ivanov:2022mrd}, which we plan to pursue in a future study.
\section{Detectability of the 21cm-galaxy cross-bispectrum}
\label{sec:detectability}
 In this section, we present the forecast for the detectability of the 21cm-galaxy cross-bispectrum with SKA-Mid and a Euclid-like galaxy redshift survey. The SKA-Mid AA$^{*}$ array configuration~\citep{seethapuram_sridhar_2025_16951088} consists of 144 dishes. Among them, 64 are MeerKAT dishes and 80 are SKA-Mid dishes. SKA-Mid can operate in both interferometric and single-dish modes of the survey. SKA-Mid in interferometric mode can probe very small scales ($k_1 > 1~\text{Mpc}^{-1}$), which allow us to study complex astrophysics and probe physics at small scales, including the impact of warm dark matter and primordial magnetic fields on the large-scale structure. However, SKA-Mid in the interferometric mode does not provide short baselines~\citep{Bull_2015} (which is limited by the field of view (FoV) of SKA-Mid) to probe large scales ($k_1 < 0.1~\text{Mpc}^{-1}$), which is essential to study BAOs, ultra-large-scale effects, and the impact of neutrino masses~\citep{SKA:2018_red_book}. An alternative is to use SKA-Mid in single-dish mode to scan wide areas of the sky, which is pioneered by MeerKLASS~\citep{Meerkat_2017,Wang:2020lkn, MeerKLASS:2024ypg}. Here, we present forecasts for both SKA-Mid in single-dish and interferometric survey modes.
 \par
 The variance in the 21cm-galaxy cross-bispectrum will have contributions from cosmic variance, thermal noise of the radio telescope, any residual foreground left in the 21cm maps after foreground cleaning, and various instrumental systematics. For our SNR forecast, we consider variance due to thermal noise ($\sigma_{B, \rm  th}(k_1, k_2/k_1, \cos \theta)$) and a Gaussian approximation of the cosmic variance ($\sigma_{B,\rm cv}(k_1, k_2/k_1, \cos \theta)$). Additionally, we also include the impact of the SKA-Mid telescopic beam and the signal loss due to foreground cleaning. To estimate the variance in the 21cm-auto bispectrum and  21cm-galaxy cross-bispectrum due to thermal noise, we simulate 100 statistically independent realizations of  21cm thermal noise maps with the same volume as the cosmological signal and add them to the signal. The mean of the 21cm auto-bispectrum/21cm-galaxy cross-bispectrum ($\bar B(k_1, k_2/k_1, \cos \theta)$) and variance ($\sigma_{B,\rm th}(k_1,k_2/k_1, \cos \theta)$) is estimated from this ensemble. To get an estimate of the cosmic variance, following~\citet{Scoccimarro_2000} we add a Gaussian analytic prediction given by 
 \begin{equation}
     \sigma^2_{B,\rm cv}(k_1,k_2/k_1, \cos \theta) = \frac{s_b k_f^3 (2\pi)^3 P(k_1)P(k_2)P(k_3)}{N_{\rm tri}},
 \end{equation}
 where the symmetry factor is $s_b = 6,2,1$ for equilateral, isosceles, and general triangles, respectively. Here, $N_{\rm tri}$ represents the number of $\text{triangles}$ used to estimate the bispectrum. 
 \par
 The SNR for the bispectrum for each ($k_1,k_2/k_1,\cos \theta$) bin is estimated as  follows:
 \begin{equation}
      \text{SNR}(k_1, k_2/k_1, \cos \theta)=\frac{\bar B(k_1, k_2/k_1,\cos \theta)}{\sigma_{\rm tot}(k_1,k_2/k_1, \cos \theta)},
 \end{equation}
 where the total variance in the bispectrum is given by $\sigma^2_{B,\rm tot} = \sigma^2_{B,\rm th}(k_1, k_2/k_1, \cos \theta) + \sigma^2_{B,\rm cv}(k_1, k_2/k_1, \cos \theta)$. We discuss the details of generating the thermal noise maps for the interferometric mode and the single-dish mode of the survey in Section~\ref{sec:interfero_thermal noise} and Section~\ref{sec:single_dish_thermal_noise}, respectively.
\begin{figure*}
    \centering
    \begin{subfigure}{1\textwidth}
        \centering
        \includegraphics[width=0.85\linewidth]{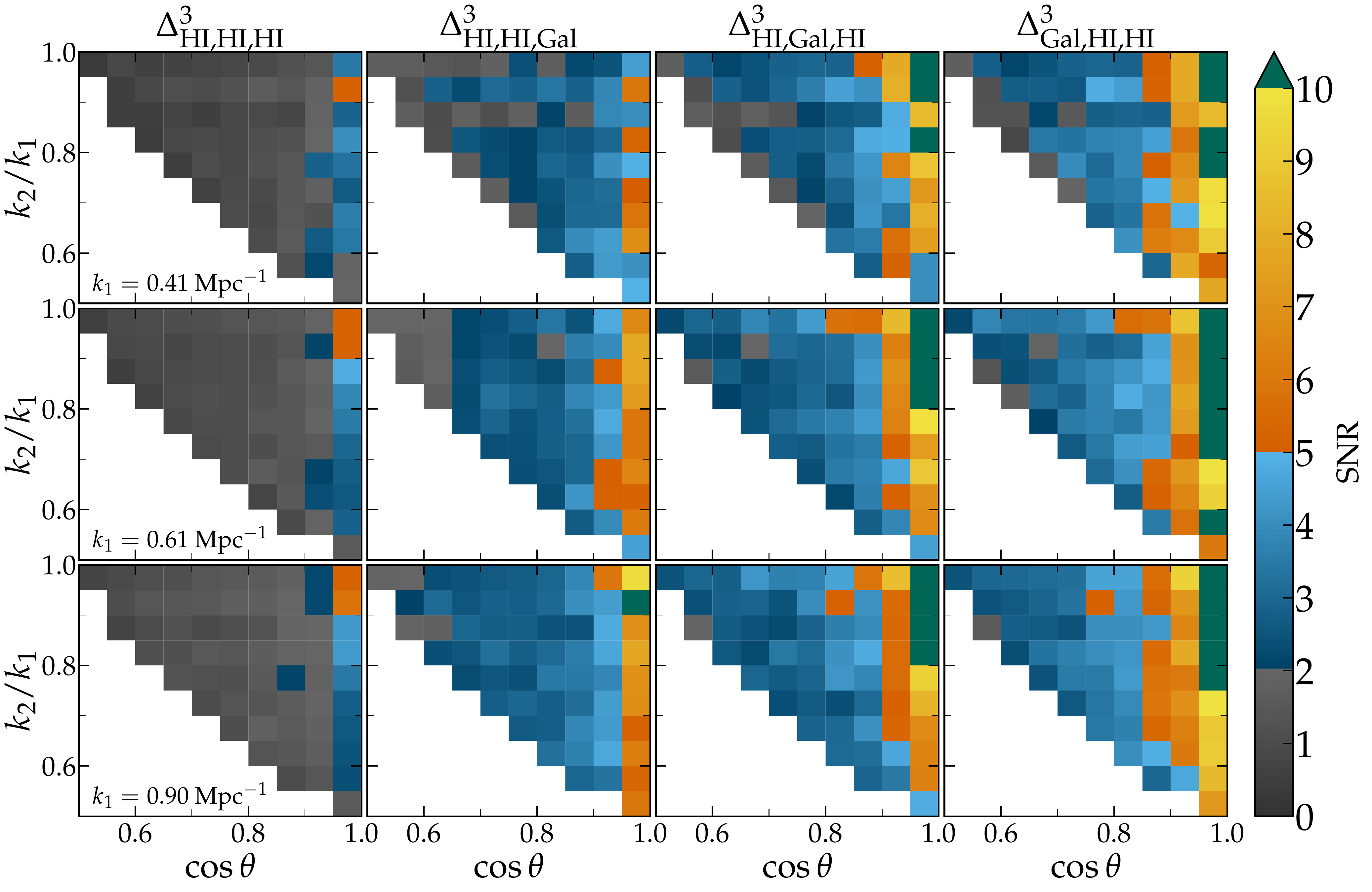}
        \caption{The SNR for detecting the bispectrum for $t_{\rm p}=100$ hours.}
        \label{fig:t_p_100}
    \end{subfigure}
    \begin{subfigure}{1\textwidth}
        \centering
        \includegraphics[width=0.85\linewidth]{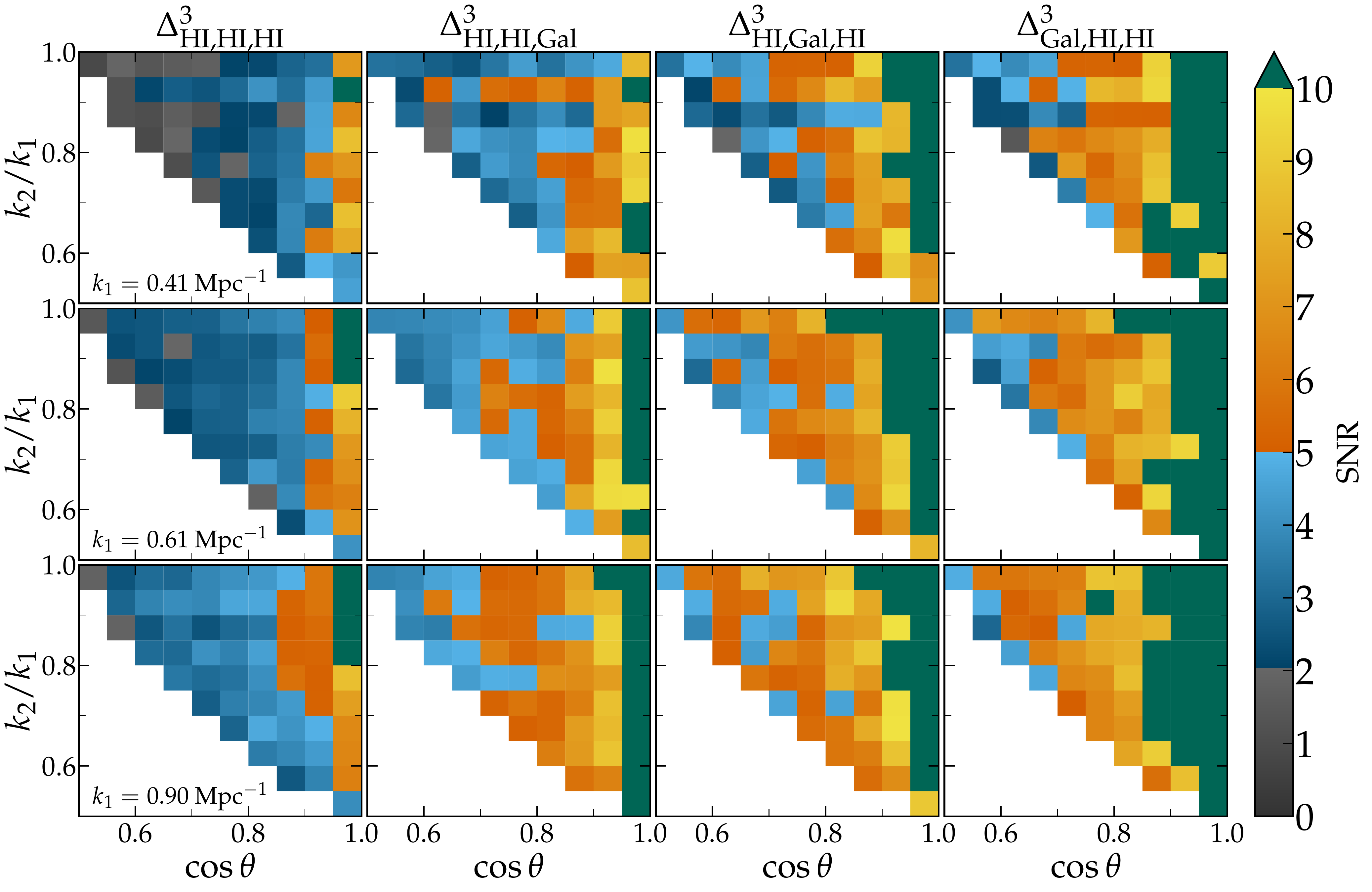}
        \caption{The SNR for detecting the bispectrum for $t_{\rm p}=200$ hours.}
        \label{fig:t_p_200}
    \end{subfigure}
    
    \caption{The SNR for detecting 21-cm auto bispectrum and 21-cm-galaxy cross-bispectrum for all unique $\text{triangles}$ at $z=0.99$ with SKA-Mid (interferometric mode) and Euclid-like galaxy survey. Gray, blue, orange and green colors indicate SNR $<2 \sigma$,   $2\sigma \leq$ SNR $<5\sigma$,  $5\leq$ SNR $<10\sigma$ and SNR $>10\sigma$ respectively. Figure \ref{fig:t_p_100} and \ref{fig:t_p_200} correspond to SNR estimates for 100 and 200 hours of SKA-Mid observations per pointing ($t_{\rm p}$), respectively.  In each Figure, the first, second, and third rows correspond to SNR estimates for $k_1 = 0.41, 0.61$ and $0.90~\text{Mpc}^{-1}$.}
    \label{fig:cross_bispec_SNR_interfero}
\end{figure*}
\subsection{Interferometer thermal noise}
\label{sec:interfero_thermal noise}
We assume that noise maps are Gaussian with thermal noise power spectrum given by~\citep{Bull_2015}:
\begin{align}
    P_N(k_{\perp},z) = \frac{D^2(z) r_{\nu}(z) \lambda^4(z)}{\bigg(\frac{A_{\rm eff}}{T_{\rm sys}}\bigg)^2 t_{\rm obs} n_{\rm pol} n(\bm{u},z)\nu_{\rm 21cm}} \frac{S_{\rm area}}{\Omega_{\rm FOV}}.
\end{align}
Here, $D(z)$ is the comoving distance to redshift $z$, $A_{\rm eff}$ is the effective collecting area of  SKA-Mid dish, $T_{\rm sys}$ is the system temperature, and $t_{\rm obs}$ is the total observation time. The numerical value of $A_{\rm eff} / T_{\rm sys}$ is taken from the Anticipated
SKA1 Science Performance document
\footnote{\url{https://www.skao.int/sites/default/files/documents/SKAO-TEL-0000818-V2_SKA1_Science_Performance.pdf}}. $\lambda = 21 \times (1+z) ~\rm cm$ and $v_{\rm 21cm} = 1420~\text{MHz}$ represent the redshifted wavelength and rest-frame frequency of the 21cm radiation, respectively. We assume the number of polarization ($n_\text{pol}$) to be 2.  The  baseline number density, $n(\bm{u}, z)$ , is estimated using the \texttt{ska\_ost\_array\_config}
\footnote{\url{https://gitlab.com/ska-telescope/ost/ska-ost-array-config/-/tree/master/src/ska_ost_array_config?ref_type=heads}} package assuming the  SKA-Mid AA$^{*}$ array configuration~\citep{seethapuram_sridhar_2025_16951088}. 
 We choose a subvolume of $400^3$ Mpc$^3$ from the full coeval box at $z=0.99$ from the GAEA simulations (see Section~\ref{sec:GAEA_simulations}) for our forecast. This corresponds to a survey area of 46 deg$^2$ and a bandwidth of 56 MHz with a frequency resolution of 120 KHz. Considering the central redshift of $z = 0.99$, FoV, $\Omega_{\rm FOV}$, of SKA-Mid in the interferometric survey mode is 2.57 deg$^2$. To cover a survey area of  46 deg$^2$, one has to consider a total of 18 pointings.
\subsection{Single-dish  mode of operation}
\subsubsection{Impact of single-dish instrumental beam}
We also include the impact of the radio telescope beam when the observations are carried out in single-dish mode of operation. The telescope beam will smooth the 21cm fluctuations in all directions perpendicular to the line of sight. To include this effect, we smooth the 21cm field with a Gaussian beam with standard deviation $R = D(z)\theta_{\rm FWHM}/(2\sqrt{2\ln2})$. The full width at half maximum (FWHM) of the radio telescope with a diameter of the dish $D_{\rm dish}$ and observing frequency $\nu$ is given by $\theta_{\rm FWHM} = c / \nu D_{\rm dish}$. For our forecast at $z=0.99$ with SKA-Mid, which has a dish diameter of $D_{\rm dish} = 15~\rm m$, the values are $\theta_{\rm FWHM} = 1.596^{0}$ and $R= 20.034~\text{Mpc}$.
\subsubsection{Single-dish thermal noise}
\label{sec:single_dish_thermal_noise}
We modeled the instrumental noise to be Gaussian and white. The standard deviation of the thermal noise at each pixel is given by~\citep{Matshawule_2020, Spinelli_2021}
\begin{align}
    \sigma_{\rm N} = \frac{T_{\rm sys}(\nu)}{\sqrt{2t_{\rm pix}\Delta \nu}},
\end{align}
where $T_{\rm sys}$, $t_{\rm pix}$ and $\Delta \nu$ represent the system temperature, observational time per pixel and frequency resolution, respectively. The system temperature is given by 
\begin{align}
    T_{\rm sys}(\nu) = T_{\rm rcv}(\nu) + T_{\rm spill} + T_{\rm CMB} + T_{\rm gal}(\nu),
\end{align}
where $T_{\rm rcv}(\nu)$, $T_{\rm CMB}$, $T_{\rm gal}$ and $T_{\rm gal}$ are the receiver temperature, spillover temperature, cosmic microwave background (CMB) temperature, and contribution from our own galaxy, respectively. 
The total observational time per pixel is given by $t_{\rm pix} = t_{\rm obs}N_{\rm dish}\frac{\Omega_{\rm pix}}{\Omega_{\rm survey}}$, where $N_{\rm dish}$ is the total number of dishes scanning  the sky, and $\Omega_{\rm survey}$ is the total survey area. We assume the pixel area to be $\Omega_{\rm pix}=(\theta_{\rm FWHM}/3)^2$~\citep{Cunnington_2022, Wang_2020}.
Following the anticipated SKA1 Science Performance document\footnote{\url{https://www.skao.int/sites/default/files/documents/SKAO-TEL-0000818-V2_SKA1_Science_Performance.pdf}} , we choose $T_{\rm rcv}(\nu) = 15 + 30 \big(\frac{\nu}{\rm GHz} - 0.75\big)^2 \rm K $, $T_{\rm spill} = 3 \rm K$, $T_{\rm CMB} = 2.73 \rm K$ and $T_{\rm gal}=25\big(\frac{408 \rm MHz}{\nu}\big)^{2.75}~\rm K$. We choose $N_{\rm dish} = 144$, which will be the number of dishes for the SKA-Mid AA$^*$ array configuration. We estimated the $\sigma_{\rm N}$ assuming the SKA-Mid will scan a survey area of  $\Omega_{\rm survey} = 185~\rm deg^2$ with frequency resolution of $\Delta \nu = 191~\rm KHz$  for a total observation time of $t_{\rm obs}=200~\rm hours$.
\subsubsection{Impact of foreground removal}
\label{sec:foreground_removal}
The astrophysical foregrounds pose a major challenge in detecting the 21cm signal. Blind cleaning techniques~\citep{Wang_2006,Switzer_2013, Alonso_2015,Carucci_2020, Spinelli_2021} are used to remove these foregrounds. However, these blind foreground cleaning methods cause a reduction in the amplitude of the 21cm signal, which affects the estimated summary statistics. To mimic this signal loss due to foreground cleaning, following~\citet{Chand_2025, Bernal:2019jdo, Soares:2020zaq, Cunnington_2020} we smooth the $\delta T_{\rm b}(\bm{x})$ with a Gaussian filter in Fourier space given by
\begin{equation}
G_{\rm FG} = 1-\exp{\bigg(-\frac{k_{\parallel}^2}{2k_{\parallel,\rm FG}^2}\bigg)},    
\end{equation}
where $k_{\parallel, \rm FG}$ controls the extent of signal loss. A higher value of $k_{\parallel, \rm FG}$  means that the damping in the signal amplitude extends toward a higher $k_{\parallel}$ and the signal loss is severe for lower $k_{\parallel}$. We choose $k_{\parallel, \rm FG} = 0.008~\text{and}~ 0.016~\text{Mpc}^{-1}.$
\subsection{Signal-to-noise ratio}
\subsubsection{Interferometric mode of operation}
In Figure~\ref{fig:cross_bispec_SNR_interfero}, we present the SNR estimates for the 21cm auto-bispectrum and 21cm-galaxy cross-bispectrum for all unique triangles at $z=0.99$. Figures \ref{fig:t_p_100} and \ref{fig:t_p_200} show SNR estimates for observational time per pointing $t_{\rm p}= 100$ and 200 hours. The corresponding total observational time  for 18 pointings to cover the survey area is $t_{\rm obs} = 1800$ and 3600 hours, respectively. In each Figure, the first, second, and third rows show the SNR estimate for $k_1 = 0.41, 0.61$, and $0.90~\text{Mpc}^{-1}$. The results are color-coded by significance level. Gray, blue, orange, and green colors indicate SNR $<2 \sigma$,   $2\sigma \leq$ SNR $<5\sigma$,  $5\leq$ SNR $<10\sigma$ and SNR $>10\sigma$ respectively.
\par 
First, we discuss the SNR estimates for 21cm auto-bispectrum, which are presented in the first column of Figure \ref{fig:t_p_100} and Figure~\ref{fig:t_p_200}. For $t_{\rm p}=100$ hours, we find that, except for a few linear $\text{triangles}$, all unique $\text{triangles}$ have an SNR below 2$\sigma$ for all the $k_1$ bins. Increasing $t_{\rm p}$ to 200 hours results in a significant boost in detectability. The squeezed-limit $k-\text{triangles}$ achieve an SNR exceeding $10\sigma$ for $k_1 = 0.61, 0.90~\text{Mpc}^{-1}$. Additionally, a $2\sigma$ detection is possible across almost the entire space of unique triangle configurations. For all $k_1$ bins presented here, the linear $k-\text{triangles}$ have the highest detectability. This is due to the higher magnitude of the bispectrum for these configurations.
\par
The second, third, and fourth columns in Figure~\ref{fig:t_p_100} and Figure~\ref{fig:t_p_200} present the SNR estimates for $\Delta^3_{\rm HI,HI,Gal}$, $\Delta^3_{\rm HI,Gal,HI}$ and $\Delta^3_{\rm Gal,HI,HI}$, respectively. First, we discuss results for $t_{\rm p} = 100\rm~hours$. For the  $k_1 = 0.41~\text{Mpc}^{-1}$ bin, we find that almost the entire unique triangle configuration space of the cross-bispectra has detectability higher than $2\sigma$. Among the cross-bispectrum combinations considered here, $\Delta^3_{\rm Gal, HI, HI}$ shows the highest detectability. For $\Delta^3_{\rm Gal, HI,HI}$, linear triangles achieve an SNR higher than $10\sigma$. This is due to the higher signal strength observed for linear triangles. Furthermore, sensitivity improves going form $k_1= 0.41~\text{Mpc}^{-1}$ to $k_1=0.90~\text{Mpc}^{-1}$ across all the unique triangle configurations.
Increasing $t_{\rm p}$ to 200 hours results in improved detectability across the unique triangle configurations. For $k_1 =0.61~\text{Mpc}^{-1}$ and $k_1=0.9~{\rm Mpc}^{-1}$, the SNR  exceeds $5\sigma$ across almost all unique triangle configurations. Noticeably, linear triangles and triangles in their vicinity achieve an SNR higher than $10\sigma$ for the cross-combinations $\Delta^3_{\rm HI, Gal,HI}$ and $\Delta^3_{\rm Gal, HI,HI}$.
\par
In comparison with the SNR estimates of the 21cm auto-bispectrum, the 21cm-galaxy cross-bispectrum for all combinations considered  here exhibits a significant boost in SNR. The detectability improves by more than a factor of three across the entire configuration space for $\Delta^3_{\rm HI,Gal,HI}$ and $\Delta^3_{\rm Gal,HI,HI}$, thanks to the high SNR of the galaxy surveys. This is true for $t_{\rm p}=100$ and 200 hours. These results highlight which triangle configurations and 21cm-galaxy cross-bispectrum combinations one should target with future observations with SKA-Mid. 
\par
\begin{figure*}
\centering 
\includegraphics[width=0.7\linewidth]{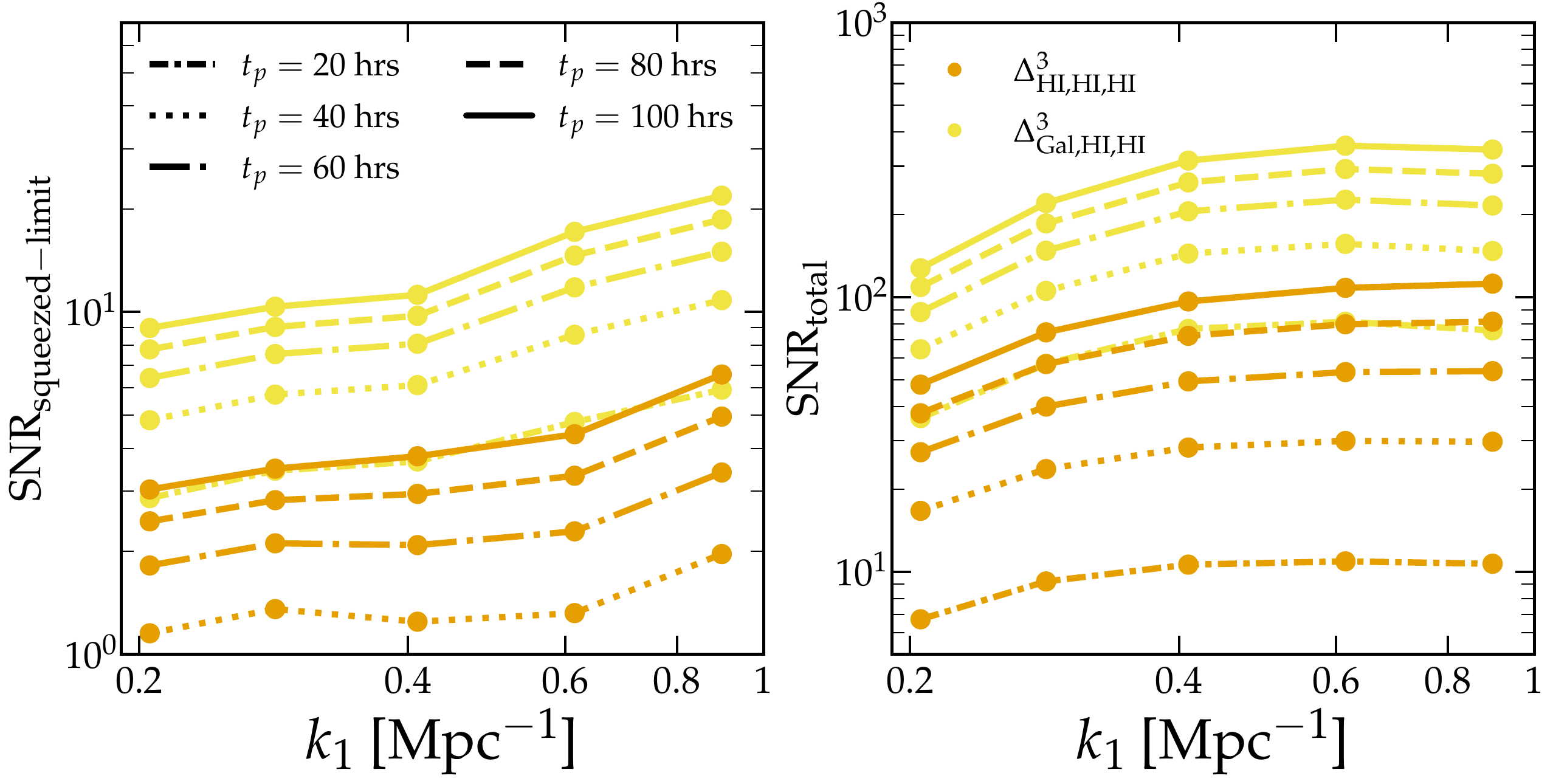}
\caption{The SNR for detecting the  21cm auto-bispectrum (orange) and 21cm-galaxy cross-bispectrum (yellow) as a function of $k_1$ for different observation time per pointing ($t_{\rm p}$). The left panel presents results for squeezed-limit $k-\text{triangles}$ and the right panel corresponds to all shapes ($k_2/k_1,\cos \theta$) combined.}
\label{fig:t_p_diff_SN}
\end{figure*}

In Figure~{\ref{fig:t_p_diff_SN}}, we present SNR estimates for squeezed-limit $\text{triangles}$ and for all shapes combined as a function of $k_1$ at $z=0.99$ for different choices of $t_{\rm p}$. Here, we show estimates for $\Delta^3_{\rm HI,HI,HI}$ and cross-bispectrum combination $\Delta^3_{\rm Gal,HI,HI}$, which show the highest SNR. We vary $t_{\rm p}$ from 20 to 100 hours in steps of 20 hours. The orange and yellow colors in Figure~\ref{fig:t_p_diff_SN} represent $\Delta^3_{\rm HI,HI,HI}$ and $\Delta^3_{\rm Gal,HI,HI}$, respectively. Different line styles show SNR estimates for different choices of $t_{\rm p}$.  Considering $\Delta^3_{\rm HI,HI,HI}$, increasing $t_{\rm p}$ from 20 hours to 100 hours results in an increase in SNR across all $k_1$ bins. The $\Delta^3_{\rm Gal,HI,HI}$ shows significant boost in SNR for every $k_1$ bin for each choice of $t_{\rm p}$ compared to $\Delta^3_{\rm HI,HI,HI}$. A $t_{\rm p}=40~\rm hours$ is enough to achieve a $5\sigma$ detection for $\Delta^3_{\rm Gal, HI,HI}$.  For $t_p=20~\rm hours$, $\Delta^3_{\rm HI,HI,HI}$ is not detectable for every $k_1$ bin. However, even for $t_{\rm p}=20~\rm hours$, $\Delta^3_{\rm Gal,HI,HI}$ shows an SNR higher than that of $\Delta^3_{\rm HI,HI,HI}$ with $t_{\rm p}=100~\rm hours$.
\par
The total SNR as a function of $k_1$ presented in the right panel of Figure~\ref{fig:t_p_diff_SN} is estimated by summing the SNR across all unique shapes ( $k_2/ k_1$ and $\cos \theta$ bin). Considering $\Delta^3_{\rm HI,HI,HI}$, with $t_{\rm p}=40~\rm hours$ the combined SNR reaches above $10\sigma$ for every $k_1$ bin. Similarly to the squeezed-limit $k-\text{triangles}$, $\Delta^3_{\rm Gal,HI,HI}$ shows an enhanced SNR compared to $\Delta^3_{\rm HI,HI,HI}$. Higher than $50\sigma$ detection is possible across every $k_1$ bin for $\Delta^3_{\rm Gal,HI,HI}$ with $t_{\rm p} =40\rm~hours $.
\subsubsection{Single-dish mode of operation}
\begin{figure*}
\centering
\includegraphics[width=1\linewidth]{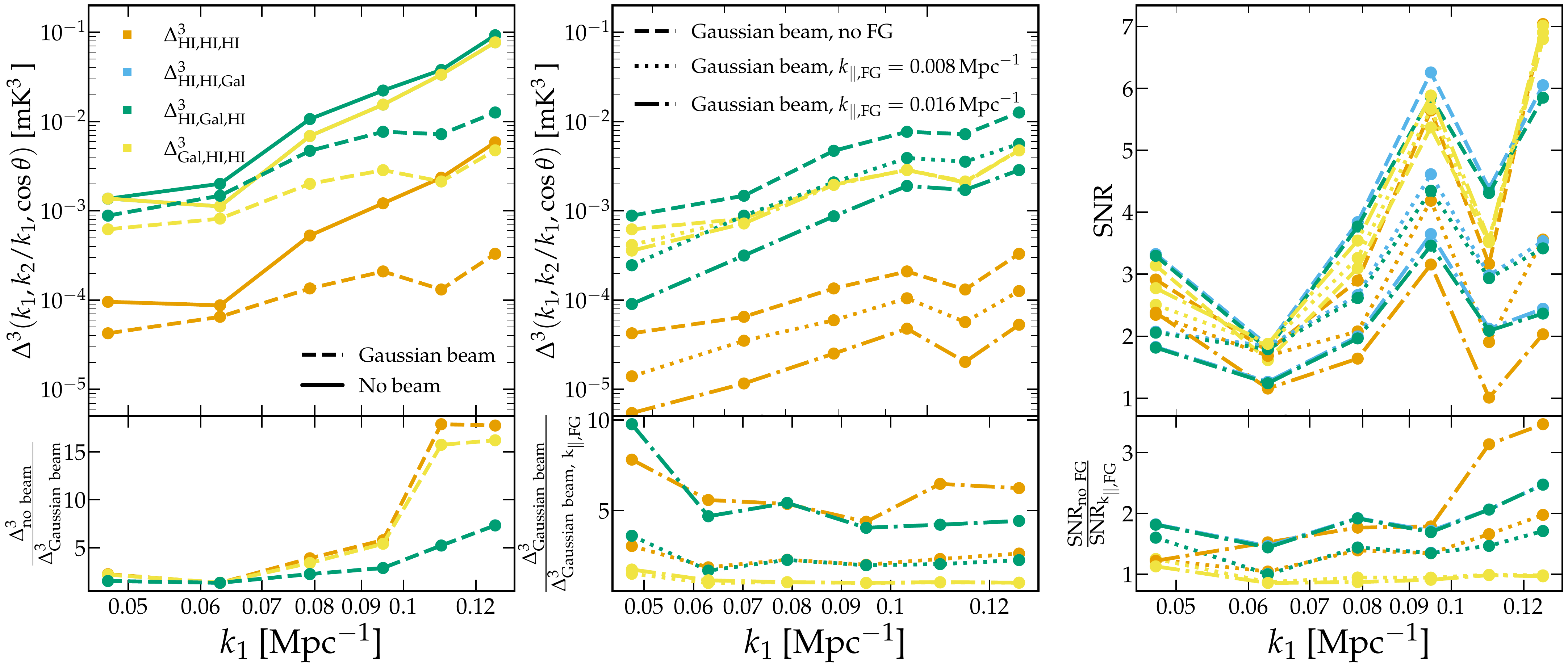}
\caption{\textbf{Left:} Impact of instrumental beam on the 21cm-galaxy cross-bispectrum for squeezed-limit $k-\text{triangles}$. The solid lines represent the bispectrum estimates without any telescopic beam, and dashed lines correspond to bispectrum estimates from maps where the impact of a Gaussian beam is included. \textbf{Middle:} Impact of signal loss in the bispectrum due to foreground removal. The dotted and dashed–dotted lines correspond to different levels of signal loss, where signal loss is severe for $k_{\parallel, \rm FG} = 0.016~\rm Mpc^{-1}$ (see Section~\ref{sec:foreground_removal} for the details). \textbf{Right:} The SNR for detecting 21-cm auto bispectrum for squeezed-limit $k-\text{triangles}$ as a function of $k_1$ at $z=0.99$ with SKA-Mid (single-dish mode) and Euclid-like galaxy survey. Different colors correspond to different combinations of the 21cm-galaxy cross-bispectrum. }
\label{fig:SKA_mid_single_dish}
\end{figure*}

Before we discuss the SNR estimates, we present the impact of the SKA-Mid telescopic beam and foreground removal on the 21cm auto-bispectrum and 21cm-galaxy cross-bispectrum, which is shown in the upper left panel of Figure~\ref{fig:SKA_mid_single_dish}. Here, we only show the estimates for the squeezed-limit $k-\text{triangles}$, which have the highest SNR. Additionally, the choice of $k-\text{triangles}$ is constrained by the limitation of our bispectrum estimator in sampling other unique $k-\text{triangles}$ for the low $k_1$ bin. The solid lines represent the bispectrum estimates without any telescopic beam, while the  dashed lines correspond to bispectrum estimates from maps that include the effects of a Gaussian beam. The dotted and dashed–dotted lines show estimates which include foreground removal effects with $k_{\parallel, \rm FG} = 0.008~\rm Mpc^{-1}$ and $0.016~\text{Mpc}^{-1}$, in addition to the Gaussian beam. Different colors distinguish between estimates for the 21cm auto-bispectrum and various combinations of the 21cm-galaxy cross-bispectrum. The inclusion of the telescopic beam suppresses the magnitude of both the auto- and cross-bispectrum. This suppression is severe for $k_1> 0.1~\text{Mpc}^{-1}$, resulting in a reduction of more than 15 times in the magnitude. Even for scales $k_1 < 0.1~\text{Mpc}^{-1}$, the inclusion of the beam reduces the bispectrum magnitude by more than a factor of two. Including the foreground removal effects further suppresses the magnitudes of the auto- and cross-bispectrum. We observe a reduction in the magnitude of $\Delta^3_{\rm HI,HI,HI}$ by a factor of two and five for $k_{\parallel, \rm FG}=0.008~\rm Mpc^{-1}$ and  $0.016~\text{Mpc}^{-1}$, respectively. However, the effect is minimal for $\Delta^3_{\rm HI, HI, Gal}$. This is because a galaxy field is placed at $k_3$, which probes the largest scales where the foreground removal effect in the 21cm map is severe.
\par
In the right panel of the Figure~\ref{fig:SKA_mid_single_dish}, we present the SNR estimates for the 21cm auto-bispectrum and 21cm-galaxy cross-bispectrum as a function of $k_1$, with and without foreground removal effects. The SNR for 21cm auto-bispectrum and for different cross-combinations of the 21cm-galaxy cross-bispectrum without signal loss due to foreground removal varies between $1\sigma$ and $7\sigma$, where the highest SNR is observed for $k_1 = 0.13~\text{Mpc}^{-1}$. In contrast with SNR estimates for SKA-Mid in interferometric mode, the 21cm-cross-bispectrum does not show an enhancement in the SNR in comparison with the 21cm auto-bispectrum. This is because the SNR estimates for the single-dish mode of the survey presented here are limited by cosmic variance rather than thermal noise. This cosmic variance can be suppressed by observing a larger survey area,  thereby increasing the number of triangles sampled in each $k_1,k_2/k_1~\text{and}~\cos \theta$ bin. The signal loss due to the foreground removal suppresses the SNR across all the scales, and this suppression is higher for $\Delta^3_{\rm HI,HI,HI}$ and minimal for $\Delta^3_{\rm HI, HI, Gal}$. These results are consistent with \citet{Cunnington_2020}, who investigated the impact of foreground on the redshift space 21cm auto-bispectrum. Our analysis also indicates that one must correct for this signal loss in the bispectrum using similar reconstruction techniques used for power spectrum analysis~\citep{Cunnington:2023jpq, Cunnington_Li_2023, MeerKLASS:2024ypg}.
\section{Summary and discussion}
\label{sec:summary}
Upcoming 21cm line intensity mapping experiments, such as SKA-Mid and HIRAX, are expected to detect the 21cm auto-bispectrum across a wide range of redshifts and scales. However, the residual foregrounds leftover due to the imperfect foreground cleaning pose a significant barrier to achieving high-significance detections. Given the success in detecting the 21cm-galaxy cross-power spectrum, extending this approach to the cross-bispectrum of the 21cm signal with galaxies using data from SKA-Mid and galaxy surveys offers a promising path to suppress uncertainty from residual systematics and achieve a high-significance detection. In this article, we explore the modeling of the 21cm–galaxy cross-bispectrum using simulations and present forecasts for its detectability with SKA-Mid and a Euclid-like galaxy survey.  
\par
We present a detailed analysis of the HI-galaxy cross-bispectrum in real and redshift space for different sizes, shapes, and combinations of the HI and galaxy fields. This is achieved by HI line intensity maps and mock galaxy catalogs generated by postprocessing the GAEA SAM simulations. Furthermore, we have modeled the HI-galaxy cross-bispectrum using standard perturbation theory and compared the theoretical predictions with bispectrum estimates from a suite of HI line intensity maps and mock galaxy catalogs generated using HOD models. This analysis aims to identify the $k$-range where the cross-bispectrum can be adequately modeled using perturbation theory, and then use this to extract the linear and quadratic HI bias parameters. Additionally, we present a forecast for the detectability of the 21cm-galaxy cross-bispectrum for all the unique $k-\text{triangles}$ with a Euclid-like galaxy survey and SKA-Mid in both interferometric and single-dish survey modes.
\par
We summarize our main findings below:
\begin{itemize}
    \item Both the HI and galaxy auto-bispectrum show similar behavior in terms of variation in magnitude as a function of size ($k_1$) and shape ($k_2/k_1$ and $\cos \theta$). The magnitude of the auto-bispectrum increases monotonically with $k_1$. Considering the shape dependence, the  bispectrum peaks for linear $k-\text{triangles}$. The galaxy auto-bispectrum has a higher magnitude across all $k_1,k_2/k_1~\text{and}~\cos\theta$ bins compared to the HI auto-bispectrum,  which is attributed to the higher galaxy bias than the HI line intensity maps. 
    \item The variation in magnitude for the HI-galaxy cross-bispectrum as a function of size ($k_1$) and shape ($k_1$ and $\cos \theta$) of the triangles across all combinations shows a similar trend as observed with the auto-bispectra. The HI-galaxy cross-bispectrum for all combinations of the HI and galaxy field increases monotonically with $k_1$ and reaches peak magnitudes for linear triangles across all combinations. However, for each ($k_1, k_2/k_1, \cos \theta$) bin, we find variation in the magnitude across different cross-bispectrum combinations. 
    \item Inclusion of RSD enhances the magnitude of the HI-galaxy cross-bispectrum for large scales (small $k_1$ bins) and suppresses the magnitude for the small scales (large $k_1$ bins) across all the cross-bispectrum combinations. The level of suppression at each $k_1$ bin varies depending on the cross-bispectrum combination. Including the effect of RSD is essential for the correct interpretation of the signal cross-bispectrum. 

    \item We find that the HI-galaxy cross-bispectrum can be modeled from the standard perturbation theory for scales less than $k_1 = 0.35~\text{Mpc}^{-1}$, where the deviation of the perturbation theory predictions compared to the cross-bispectrum estimates from the simulations remains less than 10\%.
    \item  Our forecast on the detectability of the 21cm-galaxy cross-bispectrum with SKA-Mid and a  Euclid-like galaxy survey shows that cross-correlation achieves a higher SNR for scales of  $k_1 > 0.2~\text{Mpc}^{-1}$(interferometric mode of the survey) than the 21cm auto-bispectrum across every $k_1, k_2/k_1$ and $\cos \theta$ bin. Among all the unique $k-\text{triangles}$, we find that  linear  $k-\text{triangles}$ show the highest SNR (higher than 10$\sigma$) for $\Delta^3_{\rm HI,Gal,HI}$ and $\Delta^3_{\rm Gal,HI,HI}$ (see Figure~\ref{fig:cross_bispec_SNR_interfero}). The squeezed-limit $k-\text{triangles}$ and all shapes combined  for $\Delta^3_{\rm Gal,HI,HI}$ achieve an   SNR $>10\sigma$ and an SNR $>100\sigma$ respectively, across the scales $0.2~\text{Mpc}^{-1}\leq k_1 \leq 0.9~\text{Mpc}^{-1}$  with 100 hours of observations per pointing (see Figure~\ref{fig:t_p_diff_SN}). 
    \item 
     The detectability of the 21cm-auto bispectrum and 21cm-galaxy cross-bispectrum for large scales,  which can be measured with SKA-Mid in single-dish survey mode, is primarily limited by cosmic variance rather than uncertainty due to thermal noise. One has to observe a larger number of triangles in each $k_1, k_2/k_1~\text{and}~\cos \theta$ bin to suppress cosmic variance, which can be achieved by observing a large survey area. Additionally, signal loss due to foreground removal will further decrease the SNR.
    \end{itemize}
Our analysis presents a first step toward an end-to-end analysis pipeline for observations of the 21cm-galaxy cross-bispectrum with future cosmological surveys. The forecast for the detectability of the 21cm-galaxy cross-bispectrum presented here is optimistic, as we do not account for uncertainties arising from residual systematics and foreground filtering in the 21cm line intensity maps, which we will address in follow-up work. Additionally, we lack a realistic estimate of uncertainty due to cosmic variance and do not include the effect of line-of-sight anisotropy arising due to the light-cone effect, which is crucial for the correct interpretation of the signal.
\section*{Acknowledgements}
We are grateful to the anonymous reviewer for very useful comments
and suggestions that improved the quality of this paper.
The authors thank Samit Kumar Pal,  Emiliano Sefusatti, and Manas Mohit Dosibhatla for the helpful discussions.
L.N acknowledges the financial support by the Department of Science and Technology, Government of India, through the INSPIRE Fellowship [IF210392]. L.N acknowledges the support from the Abdus Salam International Centre for Theoretical Physics (ICTP) under the `ICTP Sandwich Training Educational Programme (STEP)’  SMR.3991 and SMR.4129. S.M, M.V and L.N acknowledge financial support through the project titled “Illuminating the Dark Sector of the Cosmos in the SKA Era” (Project No. P3497) funded under the “Scheme for Promotion of Academic and Research Collaboration (SPARC)” from the Ministry of Education, India. M.V is supported by the INFN INDARK and SISSA IDEAS grants.
L.N and S.M acknowledge the use of computing infrastructure for this work, which is hosted at the DAASE, IIT Indore, and was procured through funding via the Department of Science and Technology, Government of India sponsored DST-FIST grant No. SR/FST/PSII/2021/162 (C) awarded to the DAASE, IIT Indore.   
An introduction to GAEA, a list of our recent work, as well as  datafile containing published model predictions, can be found at  \url{https://sites.google.com/inaf.it/gaea/home}. We acknowledge the use of INAF-OATs computational resources within the framework of the  CHIPP project \citep{Taffoni_2020} and the INAF PLEIADI program
  (\url{http://www.pleiadi.inaf.it}).

\bibliography{REFERENCES}{}

@article{Senatore:2014vja,
    author = "Senatore, Leonardo and Zaldarriaga, Matias",
    title = "{Redshift Space Distortions in the Effective Field Theory of Large Scale Structures}",
    eprint = "1409.1225",
    archivePrefix = "arXiv",
    primaryClass = "astro-ph.CO",
    month = "9",
    year = "2014"
}

@inbook{Ivanov:2022mrd,
    author = "Ivanov, Mikhail M.",
    title = "{Effective Field Theory for Large-Scale Structure}",
    eprint = "2212.08488",
    archivePrefix = "arXiv",
    primaryClass = "astro-ph.CO",
    doi = "10.1007/978-981-19-3079-9_5-1",
    year = "2023"
}

@article{Cunnington:2023jpq,
    author = "Cunnington, Steven and others",
    title = "{The foreground transfer function for H{\,}i intensity mapping signal reconstruction: MeerKLASS and precision cosmology applications}",
    eprint = "2302.07034",
    archivePrefix = "arXiv",
    primaryClass = "astro-ph.CO",
    doi = "10.1093/mnras/stad1567",
    journal = "Mon. Not. Roy. Astron. Soc.",
    volume = "523",
    number = "2",
    pages = "2453--2477",
    year = "2023"
}

@article{Scoccimarro:1999ed,
    author = "Scoccimarro, Roman and Couchman, H. M. P. and Frieman, Joshua A.",
    title = "{The Bispectrum as a Signature of Gravitational Instability in Redshift-Space}",
    eprint = "astro-ph/9808305",
    archivePrefix = "arXiv",
    reportNumber = "FERMILAB-PUB-98-254-A, CITA-98-16",
    doi = "10.1086/307220",
    journal = "Astrophys. J.",
    volume = "517",
    pages = "531--540",
    year = "1999"
}

@article{Zheng:2016zxc,
    author = "Zheng, Yi and Song, Yong-Seon",
    title = "{Study on the mapping of dark matter clustering from real space to redshift space}",
    eprint = "1603.00101",
    archivePrefix = "arXiv",
    primaryClass = "astro-ph.CO",
    doi = "10.1088/1475-7516/2016/08/050",
    journal = "JCAP",
    volume = "08",
    pages = "050",
    year = "2016"
}

@article{Taruya:2010mx,
    author = "Taruya, Atsushi and Nishimichi, Takahiro and Saito, Shun",
    title = "{Baryon Acoustic Oscillations in 2D: Modeling Redshift-space Power Spectrum from Perturbation Theory}",
    eprint = "1006.0699",
    archivePrefix = "arXiv",
    primaryClass = "astro-ph.CO",
    doi = "10.1103/PhysRevD.82.063522",
    journal = "Phys. Rev. D",
    volume = "82",
    pages = "063522",
    year = "2010"
}

@article{Scoccimarro:2004tg,
    author = "Scoccimarro, Roman",
    title = "{Redshift-space distortions, pairwise velocities and nonlinearities}",
    eprint = "astro-ph/0407214",
    archivePrefix = "arXiv",
    doi = "10.1103/PhysRevD.70.083007",
    journal = "Phys. Rev. D",
    volume = "70",
    pages = "083007",
    year = "2004"
}

@article{Bernal:2019jdo,
    author = "Bernal, Jos{\'e} Luis and Breysse, Patrick C. and Gil-Mar{\'\i}n, H{\'e}ctor and Kovetz, Ely D.",
    title = "{User{\textquoteright}s guide to extracting cosmological information from line-intensity maps}",
    eprint = "1907.10067",
    archivePrefix = "arXiv",
    primaryClass = "astro-ph.CO",
    doi = "10.1103/PhysRevD.100.123522",
    journal = "Phys. Rev. D",
    volume = "100",
    number = "12",
    pages = "123522",
    year = "2019"
}

@article{Soares:2020zaq,
    author = "Soares, Paula S. and Cunnington, Steven and Pourtsidou, Alkistis and Blake, Chris",
    title = "{Power spectrum multipole expansion for HI intensity mapping experiments: unbiased parameter estimation}",
    eprint = "2008.12102",
    archivePrefix = "arXiv",
    primaryClass = "astro-ph.CO",
    doi = "10.1093/mnras/stab027",
    journal = "Mon. Not. Roy. Astron. Soc.",
    volume = "502",
    number = "2",
    pages = "2549--2564",
    year = "2021"
}

@ARTICLE{Kaiser_1987,
       author = {{Kaiser}, Nick},
        title = "{Clustering in real space and in redshift space}",
      journal = {\mnras},
         year = 1987,
        month = jul,
       volume = {227},
        pages = {1-21},
          doi = {10.1093/mnras/227.1.1},
       adsurl = {https://ui.adsabs.harvard.edu/abs/1987MNRAS.227....1K}
}

@ARTICLE{Spergel_2015,
       author = {{Spergel}, D. and {Gehrels}, N. and {Baltay}, C. and {Bennett}, D. and {Breckinridge}, J. and {Donahue}, M. and {Dressler}, A. and {Gaudi}, B.~S. and {Greene}, T. and {Guyon}, O. and {Hirata}, C. and {Kalirai}, J. and {Kasdin}, N.~J. and {Macintosh}, B. and {Moos}, W. and {Perlmutter}, S. and {Postman}, M. and {Rauscher}, B. and {Rhodes}, J. and {Wang}, Y. and {Weinberg}, D. and {Benford}, D. and {Hudson}, M. and {Jeong}, W.-S. and {Mellier}, Y. and {Traub}, W. and {Yamada}, T. and {Capak}, P. and {Colbert}, J. and {Masters}, D. and {Penny}, M. and {Savransky}, D. and {Stern}, D. and {Zimmerman}, N. and {Barry}, R. and {Bartusek}, L. and {Carpenter}, K. and {Cheng}, E. and {Content}, D. and {Dekens}, F. and {Demers}, R. and {Grady}, K. and {Jackson}, C. and {Kuan}, G. and {Kruk}, J. and {Melton}, M. and {Nemati}, B. and {Parvin}, B. and {Poberezhskiy}, I. and {Peddie}, C. and {Ruffa}, J. and {Wallace}, J.~K. and {Whipple}, A. and {Wollack}, E. and {Zhao}, F.},
        title = "{Wide-Field InfrarRed Survey Telescope-Astrophysics Focused Telescope Assets WFIRST-AFTA 2015 Report}",
      journal = {arXiv e-prints},
         year = 2015,
        month = mar,
          eid = {arXiv:1503.03757},
        pages = {arXiv:1503.03757},
          doi = {10.48550/arXiv.1503.03757},
archivePrefix = {arXiv},
       eprint = {1503.03757},
 primaryClass = {astro-ph.IM},
       adsurl = {https://ui.adsabs.harvard.edu/abs/2015arXiv150303757S}
}

@article{Santos:2015gra,
    author = "Santos, Mario G. and others",
    editor = "Bourke, Tyler L. and others",
    title = "{Cosmology from a SKA HI intensity mapping survey}",
    eprint = "1501.03989",
    archivePrefix = "arXiv",
    primaryClass = "astro-ph.CO",
    doi = "10.22323/1.215.0019",
    journal = "PoS",
    volume = "AASKA14",
    pages = "019",
    year = "2015"
}

@article{Wyithe:2007rq,
    author = "Wyithe, Stuart and Loeb, Abraham and Geil, Paul",
    title = "{Baryonic Acoustic Oscillations in 21cm Emission: A Probe of Dark Energy out to High Redshifts}",
    eprint = "0709.2955",
    archivePrefix = "arXiv",
    primaryClass = "astro-ph",
    doi = "10.1111/j.1365-2966.2007.12631.x",
    journal = "Mon. Not. Roy. Astron. Soc.",
    volume = "383",
    pages = "1195",
    year = "2008"
}

@ARTICLE{Chang_2008,
       author = {{Chang}, Tzu-Ching and {Pen}, Ue-Li and {Peterson}, Jeffrey B. and {McDonald}, Patrick},
        title = "{Baryon Acoustic Oscillation Intensity Mapping of Dark Energy}",
      journal = {\prl},
         year = 2008,
        month = mar,
       volume = {100},
       number = {9},
          eid = {091303},
        pages = {091303},
          doi = {10.1103/PhysRevLett.100.091303},
archivePrefix = {arXiv},
       eprint = {0709.3672},
 primaryClass = {astro-ph},
       adsurl = {https://ui.adsabs.harvard.edu/abs/2008PhRvL.100i1303C}
}

@ARTICLE{Battye_2004,
       author = {{Battye}, Richard A. and {Davies}, Rod D. and {Weller}, Jochen},
        title = "{Neutral hydrogen surveys for high-redshift galaxy clusters and protoclusters}",
      journal = {\mnras},
         year = 2004,
        month = dec,
       volume = {355},
       number = {4},
        pages = {1339-1347},
          doi = {10.1111/j.1365-2966.2004.08416.x},
archivePrefix = {arXiv},
       eprint = {astro-ph/0401340},
 primaryClass = {astro-ph},
       adsurl = {https://ui.adsabs.harvard.edu/abs/2004MNRAS.355.1339B}
}

@article{Euclid:2019clj,
    author = "Blanchard, A. and others",
    collaboration = "Euclid",
    title = "{Euclid preparation. VII. Forecast validation for Euclid cosmological probes}",
    eprint = "1910.09273",
    archivePrefix = "arXiv",
    primaryClass = "astro-ph.CO",
    doi = "10.1051/0004-6361/202038071",
    journal = "Astron. Astrophys.",
    volume = "642",
    pages = "A191",
    year = "2020"
}

@article{LSSTDarkEnergyScience:2018jkl,
    author = "Mandelbaum, Rachel and others",
    collaboration = "LSST Dark Energy Science",
    title = "{The LSST Dark Energy Science Collaboration (DESC) Science Requirements Document}",
    eprint = "1809.01669",
    archivePrefix = "arXiv",
    primaryClass = "astro-ph.CO",
    reportNumber = "FERMILAB-PUB-18-465-A",
    doi = "10.2172/1471560",
    month = "9",
    year = "2018"
}

@ARTICLE{2019Msngr.175....3D,
       author = {{de Jong}, R.~S. and {Agertz}, O. and {Berbel}, A.~A. and {Aird}, J. and {Alexander}, D.~A. and {Amarsi}, A. and {Anders}, F. and {Andrae}, R. and {Ansarinejad}, B. and {Ansorge}, W. and {Antilogus}, P. and {Anwand-Heerwart}, H. and {Arentsen}, A. and {Arnadottir}, A. and {Asplund}, M. and {Auger}, M. and {Azais}, N. and {Baade}, D. and {Baker}, G. and {Baker}, S. and {Balbinot}, E. and {Baldry}, I.~K. and {Banerji}, M. and {Barden}, S. and {Barklem}, P. and {Barth{\'e}l{\'e}my-Mazot}, E. and {Battistini}, C. and {Bauer}, S. and {Bell}, C.~P.~M. and {Bellido-Tirado}, O. and {Bellstedt}, S. and {Belokurov}, V. and {Bensby}, T. and {Bergemann}, M. and {Bestenlehner}, J.~M. and {Bielby}, R. and {Bilicki}, M. and {Blake}, C. and {Bland-Hawthorn}, J. and {Boeche}, C. and {Boland}, W. and {Boller}, T. and {Bongard}, S. and {Bongiorno}, A. and {Bonifacio}, P. and {Boudon}, D. and {Brooks}, D. and {Brown}, M.~J.~I. and {Brown}, R. and {Br{\"u}ggen}, M. and {Brynnel}, J. and {Brzeski}, J. and {Buchert}, T. and {Buschkamp}, P. and {Caffau}, E. and {Caillier}, P. and {Carrick}, J. and {Casagrande}, L. and {Case}, S. and {Casey}, A. and {Cesarini}, I. and {Cescutti}, G. and {Chapuis}, D. and {Chiappini}, C. and {Childress}, M. and {Christlieb}, N. and {Church}, R. and {Cioni}, M.-R.~L. and {Cluver}, M. and {Colless}, M. and {Collett}, T. and {Comparat}, J. and {Cooper}, A. and {Couch}, W. and {Courbin}, F. and {Croom}, S. and {Croton}, D. and {Daguis{\'e}}, E. and {Dalton}, G. and {Davies}, L.~J.~M. and {Davis}, T. and {de Laverny}, P. and {Deason}, A. and {Dionies}, F. and {Disseau}, K. and {Doel}, P. and {D{\"o}scher}, D. and {Driver}, S.~P. and {Dwelly}, T. and {Eckert}, D. and {Edge}, A. and {Edvardsson}, B. and {Youssoufi}, D.~E. and {Elhaddad}, A. and {Enke}, H. and {Erfanianfar}, G. and {Farrell}, T. and {Fechner}, T. and {Feiz}, C. and {Feltzing}, S. and {Ferreras}, I. and {Feuerstein}, D. and {Feuillet}, D. and {Finoguenov}, A. and {Ford}, D. and {Fotopoulou}, S. and {Fouesneau}, M. and {Frenk}, C. and {Frey}, S. and {Gaessler}, W. and {Geier}, S. and {Gentile Fusillo}, N. and {Gerhard}, O. and {Giannantonio}, T. and {Giannone}, D. and {Gibson}, B. and {Gillingham}, P. and {Gonz{\'a}lez-Fern{\'a}ndez}, C. and {Gonzalez-Solares}, E. and {Gottloeber}, S. and {Gould}, A. and {Grebel}, E.~K. and {Gueguen}, A. and {Guiglion}, G. and {Haehnelt}, M. and {Hahn}, T. and {Hansen}, C.~J. and {Hartman}, H. and {Hauptner}, K. and {Hawkins}, K. and {Haynes}, D. and {Haynes}, R. and {Heiter}, U. and {Helmi}, A. and {Aguayo}, C.~H. and {Hewett}, P. and {Hinton}, S. and {Hobbs}, D. and {Hoenig}, S. and {Hofman}, D. and {Hook}, I. and {Hopgood}, J. and {Hopkins}, A. and {Hourihane}, A. and {Howes}, L. and {Howlett}, C. and {Huet}, T. and {Irwin}, M. and {Iwert}, O. and {Jablonka}, P. and {Jahn}, T. and {Jahnke}, K. and {Jarno}, A. and {Jin}, S. and {Jofre}, P. and {Johl}, D. and {Jones}, D. and {J{\"o}nsson}, H. and {Jordan}, C. and {Karovicova}, I. and {Khalatyan}, A. and {Kelz}, A. and {Kennicutt}, R. and {King}, D. and {Kitaura}, F. and {Klar}, J. and {Klauser}, U. and {Kneib}, J.-P. and {Koch}, A. and {Koposov}, S. and {Kordopatis}, G. and {Korn}, A. and {Kosmalski}, J. and {Kotak}, R. and {Kovalev}, M. and {Kreckel}, K. and {Kripak}, Y. and {Krumpe}, M. and {Kuijken}, K. and {Kunder}, A. and {Kushniruk}, I. and {Lam}, M.~I. and {Lamer}, G. and {Laurent}, F. and {Lawrence}, J. and {Lehmitz}, M. and {Lemasle}, B. and {Lewis}, J. and {Li}, B. and {Lidman}, C. and {Lind}, K. and {Liske}, J. and {Lizon}, J.-L. and {Loveday}, J. and {Ludwig}, H.-G. and {McDermid}, R.~M. and {Maguire}, K. and {Mainieri}, V. and {Mali}, S. and {Mandel}, H.},
        title = "{4MOST: Project overview and information for the First Call for Proposals}",
      journal = {The Messenger},
         year = 2019,
        month = mar,
       volume = {175},
        pages = {3-11},
          doi = {10.18727/0722-6691/5117},
archivePrefix = {arXiv},
       eprint = {1903.02464},
 primaryClass = {astro-ph.IM},
       adsurl = {https://ui.adsabs.harvard.edu/abs/2019Msngr.175....3D}
}

@article{Joshi:2025swr,
    author = "Joshi, Bhuwan and Kothari, Rahul",
    title = "{Constraining statistical isotropy using 21 cm power spectrum and bispectrum}",
    eprint = "2502.10717",
    archivePrefix = "arXiv",
    primaryClass = "astro-ph.CO",
    doi = "10.1088/1475-7516/2025/08/047",
    journal = "JCAP",
    volume = "08",
    pages = "047",
    year = "2025"
}

@article{Pinheiro:2026mcm,
    author = "Pinheiro, Rodrigo F. and Costa, Andr{\'e} A. and Sang, Yu",
    title = "{Cosmological forecast from the full-sky angular power spectrum and bispectrum of 21cm intensity mapping}",
    eprint = "2603.20160",
    archivePrefix = "arXiv",
    primaryClass = "astro-ph.CO",
    month = "3",
    year = "2026"
}

@article{Squarotti:2023nzy,
    author = "Squarotti, Matilde Barberi and Camera, Stefano and Maartens, Roy",
    title = "{Radio-optical synergies at high redshift to constrain primordial non-Gaussianity}",
    eprint = "2307.00058",
    archivePrefix = "arXiv",
    primaryClass = "astro-ph.CO",
    doi = "10.1088/1475-7516/2024/02/043",
    journal = "JCAP",
    volume = "02",
    pages = "043",
    year = "2024"
}

@article{Kopana:2024qqq,
    author = "Kopana, Mponeng and Jolicoeur, Sheean and Maartens, Roy",
    title = "{Constraining primordial non-Gaussianity by combining photometric galaxy and 21~cm intensity mapping surveys}",
    eprint = "2409.19383",
    archivePrefix = "arXiv",
    primaryClass = "astro-ph.CO",
    doi = "10.1140/epjc/s10052-025-14241-7",
    journal = "Eur. Phys. J. C",
    volume = "85",
    number = "5",
    pages = "538",
    year = "2025"
}

@article{Pal:2026hkq,
    author = "Pal, Sourav and Sarkar, Debanjan",
    title = "{Redshift-space 21-cm bispectrum multipoles as an SKA-era gravity test in the post-reionization Universe}",
    eprint = "2601.18862",
    archivePrefix = "arXiv",
    primaryClass = "astro-ph.CO",
    month = "1",
    year = "2026"
}

@article{Furlanetto:2006jb,
    author = "Furlanetto, Steven and Oh, S. Peng and Briggs, Frank",
    title = "{Cosmology at Low Frequencies: The 21 cm Transition and the High-Redshift Universe}",
    eprint = "astro-ph/0608032",
    archivePrefix = "arXiv",
    doi = "10.1016/j.physrep.2006.08.002",
    journal = "Phys. Rept.",
    volume = "433",
    pages = "181--301",
    year = "2006"
}

@article{MeerKLASS:2024ypg,
    author = "{MeerKLASS Collaboration} and others",
    collaboration = "MeerKLASS",
    title = "{MeerKLASS L-band deep-field intensity maps: entering the H{\,}i dominated regime}",
    eprint = "2407.21626",
    archivePrefix = "arXiv",
    primaryClass = "astro-ph.CO",
    doi = "10.1093/mnras/staf195",
    journal = "Mon. Not. Roy. Astron. Soc.",
    volume = "537",
    number = "4",
    pages = "3632--3661",
    year = "2025"
}

@article{Wang:2020lkn,
    author = "Wang, Jingying and others",
    title = "{H{\,}i intensity mapping with MeerKAT: calibration pipeline for multidish autocorrelation observations}",
    eprint = "2011.13789",
    archivePrefix = "arXiv",
    primaryClass = "astro-ph.CO",
    doi = "10.1093/mnras/stab1365",
    journal = "Mon. Not. Roy. Astron. Soc.",
    volume = "505",
    number = "3",
    pages = "3698--3721",
    year = "2021"
}

@article{SKA:2018_red_book,
    author = "{SKA Cosmology SWG} and others",
    collaboration = "SKA",
    title = "{Cosmology with Phase 1 of the Square Kilometre Array: Red Book 2018: Technical specifications and performance forecasts}",
    eprint = "1811.02743",
    archivePrefix = "arXiv",
    primaryClass = "astro-ph.CO",
    doi = "10.1017/pasa.2019.51",
    journal = "Publ. Astron. Soc. Austral.",
    volume = "37",
    pages = "e007",
    year = "2020"
}

@misc{seethapuram_sridhar_2025_16951088,
  author       = {Seethapuram Sridhar, Sarrvesh and
                  Williams, Wendy and
                  Price, Danny and
                  Breen, shari and
                  Ball, Lewis},
  title        = {SKA Low and Mid subarray templates},
  month        = jun,
  year         = 2025,
  publisher    = {SKAO},
  version      = {SKAO-TEL-0002380, Revision 02},
  doi          = {10.5281/zenodo.16951088},
  url          = {https://doi.org/10.5281/zenodo.16951088},
}

@INPROCEEDINGS{Taffoni_2020,
       author = {{Taffoni}, G. and {Becciani}, U. and {Garilli}, B. and {Maggio}, G. and {Pasian}, F. and {Umana}, G. and {Smareglia}, R. and {Vitello}, F.},
        title = "{CHIPP: INAF Pilot Project for HTC, HPC and HPDA}",
    booktitle = {Astronomical Data Analysis Software and Systems XXIX},
         year = 2020,
       editor = {{Pizzo}, R. and {Deul}, E.~R. and {Mol}, J.~D. and {de Plaa}, J. and {Verkouter}, H.},
       series = {Astronomical Society of the Pacific Conference Series},
       volume = {527},
        month = jan,
        pages = {307},
          doi = {10.48550/arXiv.2002.01283},
archivePrefix = {arXiv},
       eprint = {2002.01283},
 primaryClass = {astro-ph.IM},
       adsurl = {https://ui.adsabs.harvard.edu/abs/2020ASPC..527..307T}
}

@article{Yankelevich_2018,
    author = "Yankelevich, Victoria and Porciani, Cristiano",
    title = "{Cosmological information in the redshift-space bispectrum}",
    eprint = "1807.07076",
    archivePrefix = "arXiv",
    primaryClass = "astro-ph.CO",
    doi = "10.1093/mnras/sty3143",
    journal = "Mon. Not. Roy. Astron. Soc.",
    volume = "483",
    number = "2",
    pages = "2078--2099",
    year = "2019"
}

@article{MeerKLASS:2025,
    author = "Barberi-Squarotti, Matilde and others",
    collaboration = "MeerKLASS",
    title = "{MeerKLASS L-band deep-field intensity maps: entering the H{\,}i dominated regime}",
    eprint = "2407.21626",
    archivePrefix = "arXiv",
    primaryClass = "astro-ph.CO",
    doi = "10.1093/mnras/staf195",
    journal = "Mon. Not. Roy. Astron. Soc.",
    volume = "537",
    number = "4",
    pages = "3632--3661",
    year = "2025"
}

@article{Karagiannis_2024,
    author = "Karagiannis, Dionysios and Maartens, Roy and Fonseca, Jos{\'e} and Camera, Stefano and Clarkson, Chris",
    title = "{Multi-tracer power spectra and bispectra: formalism}",
    eprint = "2305.04028",
    archivePrefix = "arXiv",
    primaryClass = "astro-ph.CO",
    doi = "10.1088/1475-7516/2024/03/034",
    journal = "JCAP",
    volume = "03",
    pages = "034",
    year = "2024"
}

@article{Cunnington_2019,
    author = "Cunnington, Steven and Wolz, Laura and Pourtsidou, Alkistis and Bacon, David",
    title = "{Impact of foregrounds on HI intensity mapping cross-correlations with optical surveys}",
    eprint = "1904.01479",
    archivePrefix = "arXiv",
    primaryClass = "astro-ph.CO",
    doi = "10.1093/mnras/stz1916",
    journal = "Mon. Not. Roy. Astron. Soc.",
    volume = "488",
    number = "4",
    pages = "5452--5472",
    year = "2019"
}

@INPROCEEDINGS{DESI,
       author = {{Levi}, Michael and {Allen}, Lori E. and {Raichoor}, Anand and {Baltay}, Charles and {BenZvi}, Segev and {Beutler}, Florian and {Bolton}, Adam and {Castander}, Francisco J. and {Chuang}, Chia-Hsun and {Cooper}, Andrew and {Cuby}, Jean-Gabriel and {Dey}, Arjun and {Eisenstein}, Daniel and {Fan}, Xiaohui and {Flaugher}, Brenna and {Frenk}, Carlos and {Gonzalez-Morales}, Alma X. and {Graur}, Or and {Guy}, Julien and {Habib}, Salman and {Honscheid}, Klaus and {Juneau}, Stephanie and {Kneib}, Jean-Paul and {Lahav}, Ofer and {Lang}, Dustin and {Leauthaud}, Alexie and {Lusso}, Betta and {de la Macorra}, Axel and {Manera}, Marc and {Martini}, Paul and {Mao}, Shude and {Newman}, Jeffrey A. and {Palanque-Delabrouille}, Nathalie and {Percival}, Will J. and {Allende Prieto}, Carlos and {Rockosi}, Constance M. and {Ruhlmann-Kleider}, Vanina and {Schlegel}, David and {Seo}, Hee-Jong and {Song}, Yong-Seon and {Tarle}, Greg and {Wechsler}, Risa and {Weinberg}, David and {Yeche}, Christophe and {Zu}, Ying},
        title = "{The Dark Energy Spectroscopic Instrument (DESI)}",
    booktitle = {Bulletin of the American Astronomical Society},
         year = 2019,
       volume = {51},
        month = sep,
          eid = {57},
        pages = {57},
          doi = {10.48550/arXiv.1907.10688},
archivePrefix = {arXiv},
       eprint = {1907.10688},
 primaryClass = {astro-ph.IM},
       adsurl = {https://ui.adsabs.harvard.edu/abs/2019BAAS...51g..57L}
}

@ARTICLE{Moodely,
       author = {{Moodley}, Kavilan and {Naidoo}, Warren and {Prince}, Heather and {Penin}, Aurelie},
        title = "{A cross-bispectrum estimator for CMB-HI intensity mapping correlations}",
      journal = {arXiv e-prints},
         year = 2023,
        month = nov,
          eid = {arXiv:2311.05904},
        pages = {arXiv:2311.05904},
          doi = {10.48550/arXiv.2311.05904},
archivePrefix = {arXiv},
       eprint = {2311.05904},
 primaryClass = {astro-ph.CO},
       adsurl = {https://ui.adsabs.harvard.edu/abs/2023arXiv231105904M}
}

@ARTICLE{Tristan_2025,
       author = {{CHIME Collaboration} and {Chakraborty}, Arnab and {Dobbs}, Matt and {Foreman}, Simon and {Gray}, Liam and {Halpern}, Mark and {Hinshaw}, Gary and {Joseph}, Albin and {MacEachern}, Joshua and {Masui}, Kiyoshi W. and {Mena-Parra}, Juan and {Newburgh}, Laura and {Pinsonneault-Marotte}, Tristan and {Reda}, Alex and {Shaikh}, Shabbir and {Siegel}, Seth and {Wang}, Haochen and {Wulf}, Dallas and {Ahmed}, Zeeshan and {Kokron}, Nickolas and {Schaan}, Emmanuel},
        title = "{The Squeezed Bispectrum from CHIME HI Emission and Planck CMB Lensing: Current Sensitivity and Forecasts}",
      journal = {arXiv e-prints},
         year = 2026,
        month = jan,
          eid = {arXiv:2601.03240},
        pages = {arXiv:2601.03240},
          doi = {10.48550/arXiv.2601.03240},
archivePrefix = {arXiv},
       eprint = {2601.03240},
 primaryClass = {astro-ph.CO},
       adsurl = {https://ui.adsabs.harvard.edu/abs/2026arXiv260103240C}
}

@ARTICLE{Chand_2025,
       author = {{Chand}, Eishica and {Banerjee}, Arka and {Foreman}, Simon and {Villaescusa-Navarro}, Francisco},
        title = "{Boosting H I-galaxy cross-clustering signal through higher order cross-correlations}",
      journal = {\mnras},
         year = 2025,
        month = apr,
       volume = {538},
       number = {3},
        pages = {2204-2218},
          doi = {10.1093/mnras/staf433},
archivePrefix = {arXiv},
       eprint = {2410.21225},
 primaryClass = {astro-ph.CO},
       adsurl = {https://ui.adsabs.harvard.edu/abs/2025MNRAS.538.2204C}
}

@ARTICLE{Braun_2019,
       author = {{Braun}, Robert and {Bonaldi}, Anna and {Bourke}, Tyler and {Keane}, Evan and {Wagg}, Jeff},
        title = "{Anticipated Performance of the Square Kilometre Array -- Phase 1 (SKA1)}",
      journal = {arXiv e-prints},
         year = 2019,
        month = dec,
          eid = {arXiv:1912.12699},
        pages = {arXiv:1912.12699},
          doi = {10.48550/arXiv.1912.12699},
archivePrefix = {arXiv},
       eprint = {1912.12699},
 primaryClass = {astro-ph.IM},
       adsurl = {https://ui.adsabs.harvard.edu/abs/2019arXiv191212699B}
}

@ARTICLE{Zuo_2021,
       author = {{Zuo}, S. and {Li}, J. and {Li}, Y. and {Santanu}, D. and {Stebbins}, A. and {Masui}, K.~W. and {Shaw}, R. and {Zhang}, J. and {Wu}, F. and {Chen}, X.},
        title = "{Data processing pipeline for Tianlai experiment}",
      journal = {Astronomy and Computing},
         year = 2021,
        month = jan,
       volume = {34},
          eid = {100439},
        pages = {100439},
          doi = {10.1016/j.ascom.2020.100439},
archivePrefix = {arXiv},
       eprint = {2011.10757},
 primaryClass = {astro-ph.IM},
       adsurl = {https://ui.adsabs.harvard.edu/abs/2021A&C....3400439Z}
}

@ARTICLE{Scoccimarro_2015,
       author = {{Scoccimarro}, Rom{\'a}n},
        title = "{Fast estimators for redshift-space clustering}",
      journal = {\prd},
         year = 2015,
        month = oct,
       volume = {92},
       number = {8},
          eid = {083532},
        pages = {083532},
          doi = {10.1103/PhysRevD.92.083532},
archivePrefix = {arXiv},
       eprint = {1506.02729},
 primaryClass = {astro-ph.CO},
       adsurl = {https://ui.adsabs.harvard.edu/abs/2015PhRvD..92h3532S}
}

@article{Sefusatti_2015,
    author = "Sefusatti, Emiliano and Crocce, Martin and Scoccimarro, Roman and Couchman, Hugh",
    title = "{Accurate Estimators of Correlation Functions in Fourier Space}",
    eprint = "1512.07295",
    archivePrefix = "arXiv",
    primaryClass = "astro-ph.CO",
    doi = "10.1093/mnras/stw1229",
    journal = "Mon. Not. Roy. Astron. Soc.",
    volume = "460",
    number = "4",
    pages = "3624--3636",
    year = "2016"
}

@article{Spinelli_2021,
    author = "Spinelli, Marta and Carucci, Isabella P. and Cunnington, Steven and Harper, Stuart E. and Irfan, Melis O. and Fonseca, Jos{\'e} and Pourtsidou, Alkistis and Wolz, Laura",
    title = "{SKAO H{\,}i intensity mapping: blind foreground subtraction challenge}",
    eprint = "2107.10814",
    archivePrefix = "arXiv",
    primaryClass = "astro-ph.CO",
    doi = "10.1093/mnras/stab3064",
    journal = "Mon. Not. Roy. Astron. Soc.",
    volume = "509",
    number = "2",
    pages = "2048--2074",
    year = "2021"
}

@article{Matshawule_2020,
    author = "Matshawule, Siyambonga D. and Spinelli, Marta and Santos, Mario G. and Ngobese, Sibonelo",
    title = "{H{\,}i intensity mapping with MeerKAT: primary beam effects on foreground cleaning}",
    eprint = "2011.10815",
    archivePrefix = "arXiv",
    primaryClass = "astro-ph.CO",
    doi = "10.1093/mnras/stab1688",
    journal = "Mon. Not. Roy. Astron. Soc.",
    volume = "506",
    number = "4",
    pages = "5075--5092",
    year = "2021"
}

@article{Cunnington_2022,
    author = "Cunnington, Steven",
    title = "{Detecting the power spectrum turnover with H~i intensity mapping}",
    eprint = "2202.13828",
    archivePrefix = "arXiv",
    primaryClass = "astro-ph.CO",
    doi = "10.1093/mnras/stac576",
    journal = "Mon. Not. Roy. Astron. Soc.",
    volume = "512",
    number = "2",
    pages = "2408--2425",
    year = "2022"
}

@article{Wang_2020,
    author = "Wang, Jingying and others",
    title = "{H{\,}i intensity mapping with MeerKAT: calibration pipeline for multidish autocorrelation observations}",
    eprint = "2011.13789",
    archivePrefix = "arXiv",
    primaryClass = "astro-ph.CO",
    doi = "10.1093/mnras/stab1365",
    journal = "Mon. Not. Roy. Astron. Soc.",
    volume = "505",
    number = "3",
    pages = "3698--3721",
    year = "2021"
}

@ARTICLE{Bull_2015,
       author = {{Bull}, Philip and {Ferreira}, Pedro G. and {Patel}, Prina and {Santos}, M{\'a}rio G.},
        title = "{Late-time Cosmology with 21 cm Intensity Mapping Experiments}",
      journal = {\apj},
         year = 2015,
        month = apr,
       volume = {803},
       number = {1},
          eid = {21},
        pages = {21},
          doi = {10.1088/0004-637X/803/1/21},
archivePrefix = {arXiv},
       eprint = {1405.1452},
 primaryClass = {astro-ph.CO},
       adsurl = {https://ui.adsabs.harvard.edu/abs/2015ApJ...803...21B}
}

@ARTICLE{Scoccimarro_2000,
       author = {{Scoccimarro}, Rom{\'a}n},
        title = "{The Bispectrum: From Theory to Observations}",
      journal = {\apj},
         year = 2000,
        month = dec,
       volume = {544},
       number = {2},
        pages = {597-615},
          doi = {10.1086/317248},
archivePrefix = {arXiv},
       eprint = {astro-ph/0004086},
 primaryClass = {astro-ph},
       adsurl = {https://ui.adsabs.harvard.edu/abs/2000ApJ...544..597S}
}

@ARTICLE{Navarro_2018,
       author = {{Villaescusa-Navarro}, Francisco and {Genel}, Shy and {Castorina}, Emanuele and {Obuljen}, Andrej and {Spergel}, David N. and {Hernquist}, Lars and {Nelson}, Dylan and {Carucci}, Isabella P. and {Pillepich}, Annalisa and {Marinacci}, Federico and {Diemer}, Benedikt and {Vogelsberger}, Mark and {Weinberger}, Rainer and {Pakmor}, R{\"u}diger},
        title = "{Ingredients for 21 cm Intensity Mapping}",
      journal = {\apj},
         year = 2018,
        month = oct,
       volume = {866},
       number = {2},
          eid = {135},
        pages = {135},
          doi = {10.3847/1538-4357/aadba0},
archivePrefix = {arXiv},
       eprint = {1804.09180},
 primaryClass = {astro-ph.CO},
       adsurl = {https://ui.adsabs.harvard.edu/abs/2018ApJ...866..135V}
}

@ARTICLE{Ali_2006,
       author = {{Saiyad Ali}, SK. and {Bharadwaj}, Somnath and {Pandey}, Sanjay K.},
        title = "{Probing the bispectrum at high redshifts using 21-cm HI observations}",
      journal = {\mnras},
         year = 2006,
        month = feb,
       volume = {366},
       number = {1},
        pages = {213-218},
          doi = {10.1111/j.1365-2966.2005.09847.x},
archivePrefix = {arXiv},
       eprint = {astro-ph/0510118},
 primaryClass = {astro-ph},
       adsurl = {https://ui.adsabs.harvard.edu/abs/2006MNRAS.366..213S}
}

@ARTICLE{Anderson_2018,
       author = {{Anderson}, C.~J. and {Luciw}, N.~J. and {Li}, Y. -C. and {Kuo}, C.~Y. and {Yadav}, J. and {Masui}, K.~W. and {Chang}, T. -C. and {Chen}, X. and {Oppermann}, N. and {Liao}, Y. -W. and {Pen}, U. -L. and {Price}, D.~C. and {Staveley-Smith}, L. and {Switzer}, E.~R. and {Timbie}, P.~T. and {Wolz}, L.},
        title = "{Low-amplitude clustering in low-redshift 21-cm intensity maps cross-correlated with 2dF galaxy densities}",
      journal = {\mnras},
         year = 2018,
        month = may,
       volume = {476},
       number = {3},
        pages = {3382-3392},
          doi = {10.1093/mnras/sty346},
archivePrefix = {arXiv},
       eprint = {1710.00424},
 primaryClass = {astro-ph.CO},
       adsurl = {https://ui.adsabs.harvard.edu/abs/2018MNRAS.476.3382A}
}

@ARTICLE{Amiri_2023,
       author = {{Amiri}, Mandana and {Bandura}, Kevin and {Chen}, Tianyue and {Deng}, Meiling and {Dobbs}, Matt and {Fandino}, Mateus and {Foreman}, Simon and {Halpern}, Mark and {Hill}, Alex S. and {Hinshaw}, Gary and {H{\"o}fer}, Carolin and {Kania}, Joseph and {Landecker}, T.~L. and {MacEachern}, Joshua and {Masui}, Kiyoshi and {Mena-Parra}, Juan and {Milutinovic}, Nikola and {Mirhosseini}, Arash and {Newburgh}, Laura and {Ordog}, Anna and {Pen}, Ue-Li and {Pinsonneault-Marotte}, Tristan and {Polzin}, Ava and {Reda}, Alex and {Renard}, Andre and {Shaw}, J. Richard and {Siegel}, Seth R. and {Singh}, Saurabh and {Vanderlinde}, Keith and {Wang}, Haochen and {Wiebe}, Donald V. and {Wulf}, Dallas and {CHIME Collaboration}},
        title = "{Detection of Cosmological 21 cm Emission with the Canadian Hydrogen Intensity Mapping Experiment}",
      journal = {\apj},
         year = 2023,
        month = apr,
       volume = {947},
       number = {1},
          eid = {16},
        pages = {16},
          doi = {10.3847/1538-4357/acb13f},
archivePrefix = {arXiv},
       eprint = {2202.01242},
 primaryClass = {astro-ph.CO},
       adsurl = {https://ui.adsabs.harvard.edu/abs/2023ApJ...947...16A}
}

@ARTICLE{Amiri_2024,
       author = {{Amiri}, Mandana and {Bandura}, Kevin and {Chakraborty}, Arnab and {Dobbs}, Matt and {Fandino}, Mateus and {Foreman}, Simon and {Gan}, Hyoyin and {Halpern}, Mark and {Hill}, Alex S. and {Hinshaw}, Gary and {H{\"o}fer}, Carolin and {Landecker}, T.~L. and {Li}, Zack and {MacEachern}, Joshua and {Masui}, Kiyoshi and {Mena-Parra}, Juan and {Milutinovic}, Nikola and {Mirhosseini}, Arash and {Newburgh}, Laura and {Ordog}, Anna and {Paul}, Sourabh and {Pen}, Ue-Li and {Pinsonneault-Marotte}, Tristan and {Reda}, Alex and {Shaw}, J. Richard and {Siegel}, Seth R. and {Vanderlinde}, Keith and {Wang}, Haochen and {Wiebe}, D.~V. and {Wulf}, Dallas and {The Chime Collaboration}},
        title = "{A Detection of Cosmological 21 cm Emission from CHIME in Cross-correlation with eBOSS Measurements of the Ly{\ensuremath{\alpha}} Forest}",
      journal = {\apj},
         year = 2024,
        month = mar,
       volume = {963},
       number = {1},
          eid = {23},
        pages = {23},
          doi = {10.3847/1538-4357/ad0f1d},
archivePrefix = {arXiv},
       eprint = {2309.04404},
 primaryClass = {astro-ph.CO},
       adsurl = {https://ui.adsabs.harvard.edu/abs/2024ApJ...963...23A}
}

@ARTICLE{Bharadwaj_sethi_2001,
       author = {{Bharadwaj}, Somnath and {Sethi}, Shiv K.},
        title = "{HI Fluctuations at Large Redshifts: I--Visibility correlation}",
      journal = {Journal of Astrophysics and Astronomy},
         year = 2001,
        month = dec,
       volume = {22},
       number = {4},
        pages = {293-307},
          doi = {10.1007/BF02702273},
archivePrefix = {arXiv},
       eprint = {astro-ph/0203269},
 primaryClass = {astro-ph},
       adsurl = {https://ui.adsabs.harvard.edu/abs/2001JApA...22..293B}
}

@ARTICLE{Bharadwaj_2001,
       author = {{Bharadwaj}, Somnath and {Nath}, Biman B. and {Sethi}, Shiv K.},
        title = "{Using HI to probe large scale structures at z{\ensuremath{\sim}}3}",
      journal = {Journal of Astrophysics and Astronomy},
         year = 2001,
        month = mar,
       volume = {22},
       number = {1},
        pages = {21-34},
          doi = {10.1007/BF02933588},
archivePrefix = {arXiv},
       eprint = {astro-ph/0003200},
 primaryClass = {astro-ph},
       adsurl = {https://ui.adsabs.harvard.edu/abs/2001JApA...22...21B}
}

@ARTICLE{Bharadwaj_2020,
       author = {{Bharadwaj}, Somnath and {Mazumdar}, Arindam and {Sarkar}, Debanjan},
        title = "{Quantifying the redshift space distortion of the bispectrum I: primordial non-Gaussianity}",
      journal = {\mnras},
         year = 2020,
        month = mar,
       volume = {493},
       number = {1},
        pages = {594-602},
          doi = {10.1093/mnras/staa279},
archivePrefix = {arXiv},
       eprint = {2001.10243},
 primaryClass = {astro-ph.CO},
       adsurl = {https://ui.adsabs.harvard.edu/abs/2020MNRAS.493..594B}
}

@ARTICLE{Bernardeau_2002,
       author = {{Bernardeau}, F. and {Colombi}, S. and {Gazta{\~n}aga}, E. and {Scoccimarro}, R.},
        title = "{Large-scale structure of the Universe and cosmological perturbation theory}",
      journal = {\physrep},
         year = 2002,
        month = sep,
       volume = {367},
       number = {1-3},
        pages = {1-248},
          doi = {10.1016/S0370-1573(02)00135-7},
archivePrefix = {arXiv},
       eprint = {astro-ph/0112551},
 primaryClass = {astro-ph},
       adsurl = {https://ui.adsabs.harvard.edu/abs/2002PhR...367....1B}
}

@ARTICLE{Chang_2010,
       author = {{Chang}, Tzu-Ching and {Pen}, Ue-Li and {Bandura}, Kevin and {Peterson}, Jeffrey B.},
        title = "{Hydrogen 21-cm Intensity Mapping at redshift 0.8}",
      journal = {arXiv e-prints},
         year = 2010,
        month = jul,
          eid = {arXiv:1007.3709},
        pages = {arXiv:1007.3709},
          doi = {10.48550/arXiv.1007.3709},
archivePrefix = {arXiv},
       eprint = {1007.3709},
 primaryClass = {astro-ph.CO},
       adsurl = {https://ui.adsabs.harvard.edu/abs/2010arXiv1007.3709C}
}

@ARTICLE{Cunnington_Li_2023,
       author = {{Cunnington}, Steven and {Li}, Yichao and {Santos}, Mario G. and {Wang}, Jingying and {Carucci}, Isabella P. and {Irfan}, Melis O. and {Pourtsidou}, Alkistis and {Spinelli}, Marta and {Wolz}, Laura and {Soares}, Paula S. and {Blake}, Chris and {Bull}, Philip and {Engelbrecht}, Brandon and {Fonseca}, Jos{\'e} and {Grainge}, Keith and {Ma}, Yin-Zhe},
        title = "{H I intensity mapping with MeerKAT: power spectrum detection in cross-correlation with WiggleZ galaxies}",
      journal = {\mnras},
         year = 2023,
        month = feb,
       volume = {518},
       number = {4},
        pages = {6262-6272},
          doi = {10.1093/mnras/stac3060},
archivePrefix = {arXiv},
       eprint = {2206.01579},
 primaryClass = {astro-ph.CO},
       adsurl = {https://ui.adsabs.harvard.edu/abs/2023MNRAS.518.6262C}
}

@ARTICLE{Desjacques_2018,
       author = {{Desjacques}, Vincent and {Jeong}, Donghui and {Schmidt}, Fabian},
        title = "{Large-scale galaxy bias}",
      journal = {\physrep},
         year = 2018,
        month = feb,
       volume = {733},
        pages = {1-193},
          doi = {10.1016/j.physrep.2017.12.002},
archivePrefix = {arXiv},
       eprint = {1611.09787},
 primaryClass = {astro-ph.CO},
       adsurl = {https://ui.adsabs.harvard.edu/abs/2018PhR...733....1D}
}

@ARTICLE{emcee_2013,
       author = {{Foreman-Mackey}, Daniel and {Hogg}, David W. and {Lang}, Dustin and {Goodman}, Jonathan},
        title = "{emcee: The MCMC Hammer}",
      journal = {\pasp},
         year = 2013,
        month = mar,
       volume = {125},
       number = {925},
        pages = {306},
          doi = {10.1086/670067},
archivePrefix = {arXiv},
       eprint = {1202.3665},
 primaryClass = {astro-ph.IM},
       adsurl = {https://ui.adsabs.harvard.edu/abs/2013PASP..125..306F}
}

@article{Fontanot_2025,
    author = "Fontanot, Fabio and De Lucia, Gabriella and Xie, Lizhi and Hirschmann, Michaela and Baugh, Carlton and Helly, John C.",
    title = "{Galaxy Assembly and Evolution in the P-Millennium simulation: galaxy clustering}",
    eprint = "2409.02194",
    archivePrefix = "arXiv",
    primaryClass = "astro-ph.GA",
    doi = "10.1051/0004-6361/202452029",
    journal = "Astron. Astrophys.",
    volume = "699",
    pages = "A108",
    year = "2025"
}

@ARTICLE{Fry_1984,
       author = {{Fry}, J.~N.},
        title = "{The Galaxy correlation hierarchy in perturbation theory}",
      journal = {\apj},
         year = 1984,
        month = apr,
       volume = {279},
        pages = {499-510},
          doi = {10.1086/161913},
       adsurl = {https://ui.adsabs.harvard.edu/abs/1984ApJ...279..499F}
}

@ARTICLE{Guandalin_2022,
       author = {{Guandalin}, Caroline and {Carucci}, Isabella P. and {Alonso}, David and {Moodley}, Kavilan},
        title = "{Clustering redshifts with the 21cm-galaxy cross-bispectrum}",
      journal = {\mnras},
         year = 2022,
        month = oct,
       volume = {516},
       number = {2},
        pages = {3029-3048},
          doi = {10.1093/mnras/stac2343},
archivePrefix = {arXiv},
       eprint = {2112.05034},
 primaryClass = {astro-ph.CO},
       adsurl = {https://ui.adsabs.harvard.edu/abs/2022MNRAS.516.3029G}
}

@ARTICLE{Li_2021,
       author = {{Li}, Lin-Cheng and {Staveley-Smith}, Lister and {Rhee}, Jonghwan},
        title = "{An HI intensity mapping survey with a Phased Array Feed}",
      journal = {Research in Astronomy and Astrophysics},
         year = 2021,
        month = mar,
       volume = {21},
       number = {2},
          eid = {030},
        pages = {030},
          doi = {10.1088/1674-4527/21/2/30},
archivePrefix = {arXiv},
       eprint = {2008.04081},
 primaryClass = {astro-ph.CO},
       adsurl = {https://ui.adsabs.harvard.edu/abs/2021RAA....21...30L}
}

@ARTICLE{Navarro_1996,
       author = {{Navarro}, Julio F. and {Frenk}, Carlos S. and {White}, Simon D.~M.},
        title = "{The Structure of Cold Dark Matter Halos}",
      journal = {\apj},
         year = 1996,
        month = may,
       volume = {462},
        pages = {563},
          doi = {10.1086/177173},
archivePrefix = {arXiv},
       eprint = {astro-ph/9508025},
 primaryClass = {astro-ph},
       adsurl = {https://ui.adsabs.harvard.edu/abs/1996ApJ...462..563N}
}

@ARTICLE{Majumdar_2020,
       author = {{Majumdar}, Suman and {Kamran}, Mohd and {Pritchard}, Jonathan R. and {Mondal}, Rajesh and {Mazumdar}, Arindam and {Bharadwaj}, Somnath and {Mellema}, Garrelt},
        title = "{Redshifted 21-cm bispectrum - I. Impact of the redshift space distortions on the signal from the Epoch of Reionization}",
      journal = {\mnras},
         year = 2020,
        month = dec,
       volume = {499},
       number = {4},
        pages = {5090-5106},
          doi = {10.1093/mnras/staa3168},
archivePrefix = {arXiv},
       eprint = {2007.06584},
 primaryClass = {astro-ph.CO},
       adsurl = {https://ui.adsabs.harvard.edu/abs/2020MNRAS.499.5090M}
}

@ARTICLE{Masui_2013,
       author = {{Masui}, K.~W. and {Switzer}, E.~R. and {Banavar}, N. and {Bandura}, K. and {Blake}, C. and {Calin}, L. -M. and {Chang}, T. -C. and {Chen}, X. and {Li}, Y. -C. and {Liao}, Y. -W. and {Natarajan}, A. and {Pen}, U. -L. and {Peterson}, J.~B. and {Shaw}, J.~R. and {Voytek}, T.~C.},
        title = "{Measurement of 21 cm Brightness Fluctuations at z \raisebox{-0.5ex}\textasciitilde 0.8 in Cross-correlation}",
      journal = {\apjl},
         year = 2013,
        month = jan,
       volume = {763},
       number = {1},
          eid = {L20},
        pages = {L20},
          doi = {10.1088/2041-8205/763/1/L20},
archivePrefix = {arXiv},
       eprint = {1208.0331},
 primaryClass = {astro-ph.CO},
       adsurl = {https://ui.adsabs.harvard.edu/abs/2013ApJ...763L..20M}
}

@ARTICLE{Matarrese_1997,
       author = {{Matarrese}, S. and {Verde}, L. and {Heavens}, A.~F.},
        title = "{Large-scale bias in the Universe: bispectrum method}",
      journal = {\mnras},
         year = 1997,
        month = oct,
       volume = {290},
       number = {4},
        pages = {651-662},
          doi = {10.1093/mnras/290.4.651},
archivePrefix = {arXiv},
       eprint = {astro-ph/9706059},
 primaryClass = {astro-ph},
       adsurl = {https://ui.adsabs.harvard.edu/abs/1997MNRAS.290..651M}
}

@ARTICLE{Marin_2012,
       author = {{Gil-Mar{\'\i}n}, H{\'e}ctor and {Wagner}, Christian and {Fragkoudi}, Frantzeska and {Jimenez}, Raul and {Verde}, Licia},
        title = "{An improved fitting formula for the dark matter bispectrum}",
      journal = {\jcap},
         year = 2012,
        month = feb,
       volume = {2012},
       number = {2},
          eid = {047},
        pages = {047},
          doi = {10.1088/1475-7516/2012/02/047},
archivePrefix = {arXiv},
       eprint = {1111.4477},
 primaryClass = {astro-ph.CO},
       adsurl = {https://ui.adsabs.harvard.edu/abs/2012JCAP...02..047G}
}

@ARTICLE{Paul_2023,
       author = {{Paul}, Sourabh and {Santos}, Mario G. and {Chen}, Zhaoting and {Wolz}, Laura},
        title = "{A first detection of neutral hydrogen intensity mapping on Mpc scales at $z\approx 0.32$ and $z\approx 0.44$}",
      journal = {arXiv e-prints},
         year = 2023,
        month = jan,
          eid = {arXiv:2301.11943},
        pages = {arXiv:2301.11943},
          doi = {10.48550/arXiv.2301.11943},
archivePrefix = {arXiv},
       eprint = {2301.11943},
 primaryClass = {astro-ph.CO},
       adsurl = {https://ui.adsabs.harvard.edu/abs/2023arXiv230111943P}
}

@ARTICLE{Sarkar_2019,
       author = {{Sarkar}, Debanjan and {Majumdar}, Suman and {Bharadwaj}, Somnath},
        title = "{Modelling the post-reionization neutral hydrogen (H I) 21-cm bispectrum}",
      journal = {\mnras},
         year = 2019,
        month = dec,
       volume = {490},
       number = {2},
        pages = {2880-2889},
          doi = {10.1093/mnras/stz2799},
archivePrefix = {arXiv},
       eprint = {1907.01819},
 primaryClass = {astro-ph.CO},
       adsurl = {https://ui.adsabs.harvard.edu/abs/2019MNRAS.490.2880S}
}

@ARTICLE{Shaw_2021,
       author = {{Shaw}, Abinash Kumar and {Bharadwaj}, Somnath and {Sarkar}, Debanjan and {Mazumdar}, Arindam and {Singh}, Sukhdeep and {Majumdar}, Suman},
        title = "{A fast estimator for quantifying the shape dependence of the 3D bispectrum}",
      journal = {\jcap},
         year = 2021,
        month = dec,
       volume = {2021},
       number = {12},
          eid = {024},
        pages = {024},
          doi = {10.1088/1475-7516/2021/12/024},
archivePrefix = {arXiv},
       eprint = {2107.14564},
 primaryClass = {astro-ph.CO},
       adsurl = {https://ui.adsabs.harvard.edu/abs/2021JCAP...12..024S}
}

@ARTICLE{Sarkar_2013,
       author = {{Guha Sarkar}, Tapomoy and {Hazra}, Dhiraj Kumar},
        title = "{Probing primordial non-Gaussianity: the 3D Bispectrum of Ly-{\ensuremath{\alpha}} forest and the redshifted 21-cm signal from the post reionization epoch}",
      journal = {\jcap},
         year = 2013,
        month = apr,
       volume = {2013},
       number = {4},
          eid = {002},
        pages = {002},
          doi = {10.1088/1475-7516/2013/04/002},
archivePrefix = {arXiv},
       eprint = {1211.4756},
 primaryClass = {astro-ph.CO},
       adsurl = {https://ui.adsabs.harvard.edu/abs/2013JCAP...04..002G}
}

@ARTICLE{berti24,
       author = {{Berti}, Maria and {Spinelli}, Marta and {Viel}, Matteo},
        title = "{21 cm intensity mapping cross-correlation with galaxy surveys: Current and forecasted cosmological parameters estimation for the SKAO}",
      journal = {\mnras},
         year = 2024,
        month = apr,
       volume = {529},
       number = {4},
        pages = {4803-4817},
          doi = {10.1093/mnras/stae755},
archivePrefix = {arXiv},
       eprint = {2309.00710},
 primaryClass = {astro-ph.CO},
       adsurl = {https://ui.adsabs.harvard.edu/abs/2024MNRAS.529.4803B}
}

@ARTICLE{mishra26,
       author = {{Mishra}, Satvik and {Trotta}, Roberto and {Viel}, Matteo},
        title = "{Large, fast, and accurate H I intensity maps with latent overlap diffusion}",
      journal = {\mnras},
         year = 2026,
        month = jan,
       volume = {545},
       number = {3},
          eid = {staf2071},
        pages = {staf2071},
          doi = {10.1093/mnras/staf2071},
archivePrefix = {arXiv},
       eprint = {2506.08086},
 primaryClass = {astro-ph.CO},
       adsurl = {https://ui.adsabs.harvard.edu/abs/2026MNRAS.545f2071M}
}

@ARTICLE{autieri26,
       author = {{Autieri}, G. and {Berti}, M. and {Spinelli}, M. and {Haridasu}, B.~S. and {Viel}, M.},
        title = "{Weighing neutrinos with 21cm intensity mapping at the SKAO}",
      journal = {\jcap},
         year = 2026,
        month = jan,
       volume = {2026},
       number = {1},
          eid = {050},
        pages = {050},
          doi = {10.1088/1475-7516/2026/01/050},
archivePrefix = {arXiv},
       eprint = {2504.18625},
 primaryClass = {astro-ph.CO},
       adsurl = {https://ui.adsabs.harvard.edu/abs/2026JCAP...01..050A}
}

@ARTICLE{villa15,
       author = {{Villaescusa-Navarro}, Francisco and {Viel}, Matteo and {Alonso}, David and {Datta}, Kanan K. and {Bull}, Philip and {Santos}, M{\'a}rio G.},
        title = "{Cross-correlating 21cm intensity maps with Lyman Break Galaxies in the post-reionization era}",
      journal = {\jcap},
         year = 2015,
        month = mar,
       volume = {2015},
       number = {3},
        pages = {034-034},
          doi = {10.1088/1475-7516/2015/03/034},
archivePrefix = {arXiv},
       eprint = {1410.7393},
 primaryClass = {astro-ph.CO},
       adsurl = {https://ui.adsabs.harvard.edu/abs/2015JCAP...03..034V}
}

@ARTICLE{Spinelli_2020,
       author = {{Spinelli}, Marta and {Zoldan}, Anna and {De Lucia}, Gabriella and {Xie}, Lizhi and {Viel}, Matteo},
        title = "{The atomic hydrogen content of the post-reionization Universe}",
      journal = {\mnras},
         year = 2020,
        month = apr,
       volume = {493},
       number = {4},
        pages = {5434-5455},
          doi = {10.1093/mnras/staa604},
archivePrefix = {arXiv},
       eprint = {1909.02242},
 primaryClass = {astro-ph.CO},
       adsurl = {https://ui.adsabs.harvard.edu/abs/2020MNRAS.493.5434S}
}

@ARTICLE{Switzer_2013,
       author = {{Switzer}, E.~R. and {Masui}, K.~W. and {Bandura}, K. and {Calin}, L. -M. and {Chang}, T. -C. and {Chen}, X. -L. and {Li}, Y. -C. and {Liao}, Y. -W. and {Natarajan}, A. and {Pen}, U. -L. and {Peterson}, J.~B. and {Shaw}, J.~R. and {Voytek}, T.~C.},
        title = "{Determination of z \raisebox{-0.5ex}\textasciitilde 0.8 neutral hydrogen fluctuations using the 21cm  intensity mapping autocorrelation.}",
      journal = {\mnras},
         year = 2013,
        month = jul,
       volume = {434},
        pages = {L46-L50},
          doi = {10.1093/mnrasl/slt074},
archivePrefix = {arXiv},
       eprint = {1304.3712},
 primaryClass = {astro-ph.CO},
       adsurl = {https://ui.adsabs.harvard.edu/abs/2013MNRAS.434L..46S}
}

@ARTICLE{Zheng_2007,
       author = {{Zheng}, Zheng and {Coil}, Alison L. and {Zehavi}, Idit},
        title = "{Galaxy Evolution from Halo Occupation Distribution Modeling of DEEP2 and SDSS Galaxy Clustering}",
      journal = {\apj},
         year = 2007,
        month = oct,
       volume = {667},
       number = {2},
        pages = {760-779},
          doi = {10.1086/521074},
archivePrefix = {arXiv},
       eprint = {astro-ph/0703457},
 primaryClass = {astro-ph},
       adsurl = {https://ui.adsabs.harvard.edu/abs/2007ApJ...667..760Z}
}

@ARTICLE{Claude_2019,
       author = {{Schmit}, Claude J. and {Heavens}, Alan F. and {Pritchard}, Jonathan R.},
        title = "{The gravitational and lensing-ISW bispectrum of 21 cm radiation}",
      journal = {\mnras},
         year = 2019,
        month = mar,
       volume = {483},
       number = {3},
        pages = {4259-4275},
          doi = {10.1093/mnras/sty3400},
archivePrefix = {arXiv},
       eprint = {1810.00973},
 primaryClass = {astro-ph.CO},
       adsurl = {https://ui.adsabs.harvard.edu/abs/2019MNRAS.483.4259S}
}

@ARTICLE{Cunnington_2020,
       author = {{Cunnington}, Steven and {Watkinson}, Catherine and {Pourtsidou}, Alkistis},
        title = "{The H I intensity mapping bispectrum including observational effects}",
      journal = {\mnras},
         year = 2021,
        month = oct,
       volume = {507},
       number = {2},
        pages = {1623-1639},
          doi = {10.1093/mnras/stab2200},
archivePrefix = {arXiv},
       eprint = {2102.11153},
 primaryClass = {astro-ph.CO},
       adsurl = {https://ui.adsabs.harvard.edu/abs/2021MNRAS.507.1623C}
}

@ARTICLE{Jolicouer_2021,
       author = {{Jolicoeur}, Sheean and {Maartens}, Roy and {De Weerd}, Eline M. and {Umeh}, Obinna and {Clarkson}, Chris and {Camera}, Stefano},
        title = "{Detecting the relativistic bispectrum in 21cm intensity maps}",
      journal = {\jcap},
         year = 2021,
        month = jun,
       volume = {2021},
       number = {6},
          eid = {039},
        pages = {039},
          doi = {10.1088/1475-7516/2021/06/039},
archivePrefix = {arXiv},
       eprint = {2009.06197},
 primaryClass = {astro-ph.CO},
       adsurl = {https://ui.adsabs.harvard.edu/abs/2021JCAP...06..039J}
}

@ARTICLE{Durrer_2020,
       author = {{Durrer}, Ruth and {Jalilvand}, Mona and {Kothari}, Rahul and {Maartens}, Roy and {Montanari}, Francesco},
        title = "{Full-sky bispectrum in redshift space for 21cm intensity maps}",
      journal = {\jcap},
         year = 2020,
        month = dec,
       volume = {2020},
       number = {12},
          eid = {003},
        pages = {003},
          doi = {10.1088/1475-7516/2020/12/003},
archivePrefix = {arXiv},
       eprint = {2008.02266},
 primaryClass = {astro-ph.CO},
       adsurl = {https://ui.adsabs.harvard.edu/abs/2020JCAP...12..003D}
}

@article{Karagiannis_2020,
    author = "Karagiannis, Dionysios and Slosar, An{\v{z}}e and Liguori, Michele",
    title = "{Forecasts on Primordial non-Gaussianity from 21 cm Intensity Mapping experiments}",
    eprint = "1911.03964",
    archivePrefix = "arXiv",
    primaryClass = "astro-ph.CO",
    doi = "10.1088/1475-7516/2020/11/052",
    journal = "JCAP",
    volume = "11",
    pages = "052",
    year = "2020"
}

@article{Karagiannis_2020b,
    author = "Karagiannis, Dionysios and Fonseca, Jos{\'e} and Maartens, Roy and Camera, Stefano",
    title = "{Probing primordial non-Gaussianity with the power spectrum and bispectrum of future 21 cm intensity maps}",
    eprint = "2010.07034",
    archivePrefix = "arXiv",
    primaryClass = "astro-ph.CO",
    doi = "10.1016/j.dark.2021.100821",
    journal = "Phys. Dark Univ.",
    volume = "32",
    pages = "100821",
    year = "2021"
}

@article{Karagiannis_2022,
    author = "Karagiannis, Dionysios and Maartens, Roy and Randrianjanahary, Liantsoa F.",
    title = "{Cosmological constraints from the power spectrum and bispectrum of 21cm intensity maps}",
    eprint = "2206.07747",
    archivePrefix = "arXiv",
    primaryClass = "astro-ph.CO",
    doi = "10.1088/1475-7516/2022/11/003",
    journal = "JCAP",
    volume = "11",
    pages = "003",
    year = "2022"
}

@ARTICLE{Chhabra_2025,
       author = {{Chhabra}, Minal and {Bharadwaj}, Somnath},
        title = "{Probing the HI distribution at small scales using 21-cm Intensity Mapping at large scales}",
      journal = {arXiv e-prints},
         year = 2025,
        month = aug,
          eid = {arXiv:2508.19126},
        pages = {arXiv:2508.19126},
          doi = {10.48550/arXiv.2508.19126},
archivePrefix = {arXiv},
       eprint = {2508.19126},
 primaryClass = {astro-ph.CO},
       adsurl = {https://ui.adsabs.harvard.edu/abs/2025arXiv250819126C}
}

@ARTICLE{Randrianjanahary_2024,
       author = {{Randrianjanahary}, Liantsoa F. and {Karagiannis}, Dionysios and {Maartens}, Roy},
        title = "{Cosmological constraints from the EFT power spectrum and tree-level bispectrum of 21 cm intensity maps}",
      journal = {Physics of the Dark Universe},
         year = 2024,
        month = jul,
       volume = {45},
          eid = {101530},
        pages = {101530},
          doi = {10.1016/j.dark.2024.101530},
archivePrefix = {arXiv},
       eprint = {2312.02511},
 primaryClass = {astro-ph.CO},
       adsurl = {https://ui.adsabs.harvard.edu/abs/2024PDU....4501530R}
}

@ARTICLE{Wang_2006,
       author = {{Wang}, Xiaomin and {Tegmark}, Max and {Santos}, M{\'a}rio G. and {Knox}, Lloyd},
        title = "{21 cm Tomography with Foregrounds}",
      journal = {\apj},
         year = 2006,
        month = oct,
       volume = {650},
       number = {2},
        pages = {529-537},
          doi = {10.1086/506597},
archivePrefix = {arXiv},
       eprint = {astro-ph/0501081},
 primaryClass = {astro-ph},
       adsurl = {https://ui.adsabs.harvard.edu/abs/2006ApJ...650..529W}
}

@ARTICLE{Alonso_2015,
       author = {{Alonso}, David and {Bull}, Philip and {Ferreira}, Pedro G. and {Santos}, M{\'a}rio G.},
        title = "{Blind foreground subtraction for intensity mapping experiments}",
      journal = {\mnras},
         year = 2015,
        month = feb,
       volume = {447},
       number = {1},
        pages = {400-416},
          doi = {10.1093/mnras/stu2474},
archivePrefix = {arXiv},
       eprint = {1409.8667},
 primaryClass = {astro-ph.CO},
       adsurl = {https://ui.adsabs.harvard.edu/abs/2015MNRAS.447..400A}
}

@ARTICLE{Zuo_2019,
       author = {{Zuo}, Shifan and {Chen}, Xuelei and {Ansari}, Reza and {Lu}, Youjun},
        title = "{21 cm Signal Recovery via Robust Principal Component Analysis}",
      journal = {\aj},
         year = 2019,
        month = jan,
       volume = {157},
       number = {1},
          eid = {4},
        pages = {4},
          doi = {10.3847/1538-3881/aaef3b},
archivePrefix = {arXiv},
       eprint = {1801.04082},
 primaryClass = {astro-ph.CO},
       adsurl = {https://ui.adsabs.harvard.edu/abs/2019AJ....157....4Z}
}

@ARTICLE{2012MNRAS.423.2518C,
       author = {{Chapman}, Emma and {Abdalla}, Filipe B. and {Harker}, Geraint and {Jeli{\'c}}, Vibor and {Labropoulos}, Panagiotis and {Zaroubi}, Saleem and {Brentjens}, Michiel A. and {de Bruyn}, A.~G. and {Koopmans}, L.~V.~E.},
        title = "{Foreground removal using FASTICA: a showcase of LOFAR-EoR}",
      journal = {\mnras},
         year = 2012,
        month = jul,
       volume = {423},
       number = {3},
        pages = {2518-2532},
          doi = {10.1111/j.1365-2966.2012.21065.x},
archivePrefix = {arXiv},
       eprint = {1201.2190},
 primaryClass = {astro-ph.CO},
       adsurl = {https://ui.adsabs.harvard.edu/abs/2012MNRAS.423.2518C}
}

@ARTICLE{Chapman_2013,
       author = {{Chapman}, Emma and {Abdalla}, Filipe B. and {Bobin}, J. and {Starck}, J.-L. and {Harker}, Geraint and {Jeli{\'c}}, Vibor and {Labropoulos}, Panagiotis and {Zaroubi}, Saleem and {Brentjens}, Michiel A. and {de Bruyn}, A.~G. and {Koopmans}, L.~V.~E.},
        title = "{The scale of the problem: recovering images of reionization with Generalized Morphological Component Analysis}",
      journal = {\mnras},
         year = 2013,
        month = feb,
       volume = {429},
       number = {1},
        pages = {165-176},
          doi = {10.1093/mnras/sts333},
archivePrefix = {arXiv},
       eprint = {1209.4769},
 primaryClass = {astro-ph.CO},
       adsurl = {https://ui.adsabs.harvard.edu/abs/2013MNRAS.429..165C}
}

@ARTICLE{Carucci_2020,
       author = {{Carucci}, Isabella P. and {Irfan}, Melis O. and {Bobin}, J{\'e}r{\^o}me},
        title = "{Recovery of 21-cm intensity maps with sparse component separation}",
      journal = {\mnras},
         year = 2020,
        month = nov,
       volume = {499},
       number = {1},
        pages = {304-319},
          doi = {10.1093/mnras/staa2854},
archivePrefix = {arXiv},
       eprint = {2006.05996},
 primaryClass = {astro-ph.CO},
       adsurl = {https://ui.adsabs.harvard.edu/abs/2020MNRAS.499..304C}
}

@ARTICLE{Olivari_2016,
       author = {{Olivari}, L.~C. and {Remazeilles}, M. and {Dickinson}, C.},
        title = "{Extracting H I cosmological signal with generalized needlet internal linear combination}",
      journal = {\mnras},
         year = 2016,
        month = mar,
       volume = {456},
       number = {3},
        pages = {2749-2765},
          doi = {10.1093/mnras/stv2884},
archivePrefix = {arXiv},
       eprint = {1509.00742},
 primaryClass = {astro-ph.CO},
       adsurl = {https://ui.adsabs.harvard.edu/abs/2016MNRAS.456.2749O}
}

@ARTICLE{Carucci_2025,
       author = {{Carucci}, I.~P. and {Bernal}, J.~L. and {Cunnington}, S. and {Santos}, M.~G. and {Wang}, J. and {Fonseca}, J. and {Grainge}, K. and {Irfan}, M.~O. and {Li}, Y. and {Pourtsidou}, A. and {Spinelli}, M. and {Wolz}, L.},
        title = "{Hydrogen intensity mapping with MeerKAT: Preserving cosmological signal by optimising contaminant separation}",
      journal = {\aap},
         year = 2025,
        month = nov,
       volume = {703},
          eid = {A222},
        pages = {A222},
          doi = {10.1051/0004-6361/202453461},
archivePrefix = {arXiv},
       eprint = {2412.06750},
 primaryClass = {astro-ph.CO},
       adsurl = {https://ui.adsabs.harvard.edu/abs/2025A&A...703A.222C}
}

@article{Breysse:2016szq,
    author = "Breysse, Patrick C. and Kovetz, Ely D. and Behroozi, Peter S. and Dai, Liang and Kamionkowski, Marc",
    title = "{Insights from probability distribution functions of intensity maps}",
    eprint = "1609.01728",
    archivePrefix = "arXiv",
    primaryClass = "astro-ph.CO",
    doi = "10.1093/mnras/stx203",
    journal = "Mon. Not. Roy. Astron. Soc.",
    volume = "467",
    number = "3",
    pages = "2996--3010",
    year = "2017"
}

@article{Bernal:2023ovz,
    author = "Bernal, Jos\'e Luis",
    title = "{Toward accurate modeling of line-intensity mapping one-point statistics: Including extended profiles}",
    eprint = "2309.06481",
    archivePrefix = "arXiv",
    primaryClass = "astro-ph.CO",
    doi = "10.1103/PhysRevD.109.043517",
    journal = "Phys. Rev. D",
    volume = "109",
    number = "4",
    pages = "043517",
    year = "2024"
}

@ARTICLE{2025JCAP...07..054K,
       author = {{Kamran}, Mohd and {Sahl{\'e}n}, Martin and {Sarkar}, Debanjan and {Majumdar}, Suman},
        title = "{The re-markable 21-cm power spectrum. Part I. Probing the HI distribution in the post-reionization era using marked statistics}",
      journal = {\jcap},
         year = 2025,
        month = jul,
       volume = {2025},
       number = {7},
          eid = {054},
        pages = {054},
          doi = {10.1088/1475-7516/2025/07/054},
archivePrefix = {arXiv},
       eprint = {2409.05187},
 primaryClass = {astro-ph.CO},
       adsurl = {https://ui.adsabs.harvard.edu/abs/2025JCAP...07..054K}
}

@article{massara_cosmological_2023,
	title = {Cosmological {Information} in the {Marked} {Power} {Spectrum} of the {Galaxy} {Field}},
	volume = {951},
	issn = {0004-637X},
	url = {https://dx.doi.org/10.3847/1538-4357/acd44d},
	doi = {10.3847/1538-4357/acd44d},
	language = {en},
	number = {1},
	urldate = {2025-07-22},
	journal = {The Astrophysical Journal},
	author = {Massara, Elena and Villaescusa-Navarro, Francisco and Hahn, ChangHoon and Abidi, Muntazir M. and Eickenberg, Michael and Ho, Shirley and Lemos, Pablo and Dizgah, Azadeh Moradinezhad and Blancard, Bruno Régaldo-Saint},
	month = jul,
	year = {2023},
	note = {Publisher: The American Astronomical Society},
	pages = {70},
}

@article{bag_shape_2018,
	title = {The shape and size distribution of {H} ii regions near the percolation transition},
	volume = {477},
	issn = {0035-8711, 1365-2966},
	url = {https://academic.oup.com/mnras/article/477/2/1984/4944231},
	doi = {10.1093/mnras/sty714},
	language = {en},
	number = {2},
	urldate = {2024-02-20},
	journal = {Monthly Notices of the Royal Astronomical Society},
	author = {Bag, Satadru and Mondal, Rajesh and Sarkar, Prakash and Bharadwaj, Somnath and Sahni, Varun},
	month = jun,
	year = {2018},
	pages = {1984--1992},
}

@article{pathak_distinguishing_2022,
	title = {Distinguishing reionization models using the largest cluster statistics of the 21-cm maps},
	volume = {2022},
	issn = {1475-7516},
	url = {https://iopscience.iop.org/article/10.1088/1475-7516/2022/11/027},
	doi = {10.1088/1475-7516/2022/11/027},
	language = {en},
	number = {11},
	urldate = {2024-02-20},
	journal = {Journal of Cosmology and Astroparticle Physics},
	author = {Pathak, Aadarsh and Bag, Satadru and Dasgupta, Saswata and Majumdar, Suman and Mondal, Rajesh and Kamran, Mohd and Sarkar, Prakash},
	month = nov,
	year = {2022},
	pages = {027},
}

@article{dosibhatla_2025_lss-morphology,
    author = "Dosibhatla, Manas Mohit and Majumdar, Suman and Murmu, Chandra Shekhar and Pal, Samit Kumar and Dasgupta, Saswata and Bag, Satadru and Datta, Abhirup",
    title = "{Tracing large-scale structure morphology with multiwavelength line intensity maps}",
    eprint = "2508.09112",
    archivePrefix = "arXiv",
    primaryClass = "astro-ph.CO",
    doi = "10.1088/1475-7516/2026/06/014",
    journal = "JCAP",
    volume = "06",
    pages = "014",
    year = "2026"
}

@ARTICLE{bingo,
       author = {{Abdalla}, Elcio and {Ferreira}, Elisa G.~M. and {Landim}, Ricardo G. and {Costa}, Andre A. and {Fornazier}, Karin S.~F. and {Abdalla}, Filipe B. and {Barosi}, Luciano and {Brito}, Francisco A. and {Queiroz}, Amilcar R. and {Villela}, Thyrso and {Wang}, Bin and {Wuensche}, Carlos A. and {Marins}, Alessandro and {Novaes}, Camila P. and {Liccardo}, Vincenzo and {Shan}, Chenxi and {Zhang}, Jiajun and {Zhang}, Zhongli and {Zhu}, Zhenghao and {Browne}, Ian and {Delabrouille}, Jacques and {Santos}, Larissa and {dos Santos}, Marcelo V. and {Xu}, Haiguang and {Anton}, Sonia and {Battye}, Richard and {Chen}, Tianyue and {Dickinson}, Clive and {Ma}, Yin-Zhe and {Maffei}, Bruno and {de Mericia}, Eduardo J. and {Motta}, Pablo and {Otobone}, Carlos H.~N. and {Peel}, Michael W. and {Roychowdhury}, Sambit and {Remazeilles}, Mathieu and {Ribeiro}, Rafael M. and {Sang}, Yu and {Santos}, Joao R.~L. and {dos Santos}, Juliana F.~R. and {Silva}, Gustavo B. and {Vieira}, Frederico and {Vieira}, Jordany and {Xiao}, Linfeng and {Zhang}, Xue and {Zhu}, Yongkai},
        title = "{The BINGO project. I. Baryon acoustic oscillations from integrated neutral gas observations}",
      journal = {\aap},
         year = 2022,
        month = aug,
       volume = {664},
          eid = {A14},
        pages = {A14},
          doi = {10.1051/0004-6361/202140883},
archivePrefix = {arXiv},
       eprint = {2107.01633},
 primaryClass = {astro-ph.CO},
       adsurl = {https://ui.adsabs.harvard.edu/abs/2022A&A...664A..14A}
}

@INPROCEEDINGS{chord,
       author = {{Vanderlinde}, Keith and {Liu}, Adrian and {Gaensler}, Bryan and {Bond}, Dick and {Hinshaw}, Gary and {Ng}, Cherry and {Chiang}, Cynthia and {Stairs}, Ingrid and {Brown}, Jo-Anne and {Sievers}, Jonathan and {Mena}, Juan and {Smith}, Kendrick and {Bandura}, Kevin and {Masui}, Kiyoshi and {Spekkens}, Kristine and {Belostotski}, Leo and {Dobbs}, Matt and {Turok}, Neil and {Boyle}, Patrick and {Rupen}, Michael and {Landecker}, Tom and {Pen}, Ue-Li and {Kaspi}, Victoria},
        title = "{The Canadian Hydrogen Observatory and Radio-transient Detector (CHORD)}",
    booktitle = {Canadian Long Range Plan for Astronomy and Astrophysics White Papers},
         year = 2019,
       volume = {2020},
        month = oct,
          eid = {28},
        pages = {28},
          doi = {10.5281/zenodo.3765414},
archivePrefix = {arXiv},
       eprint = {1911.01777},
 primaryClass = {astro-ph.IM},
       adsurl = {https://ui.adsabs.harvard.edu/abs/2019clrp.2020...28V}
}

@ARTICLE{fast,
       author = {{Nan}, Rendong and {Li}, Di and {Jin}, Chengjin and {Wang}, Qiming and {Zhu}, Lichun and {Zhu}, Wenbai and {Zhang}, Haiyan and {Yue}, Youling and {Qian}, Lei},
        title = "{The Five-Hundred Aperture Spherical Radio Telescope (fast) Project}",
      journal = {International Journal of Modern Physics D},
         year = 2011,
        month = jan,
       volume = {20},
       number = {6},
        pages = {989-1024},
          doi = {10.1142/S0218271811019335},
archivePrefix = {arXiv},
       eprint = {1105.3794},
 primaryClass = {astro-ph.IM},
       adsurl = {https://ui.adsabs.harvard.edu/abs/2011IJMPD..20..989N}
}

@INPROCEEDINGS{hirax,
       author = {{Newburgh}, L.~B. and {Bandura}, K. and {Bucher}, M.~A. and {Chang}, T. -C. and {Chiang}, H.~C. and {Cliche}, J.~F. and {Dav{\'e}}, R. and {Dobbs}, M. and {Clarkson}, C. and {Ganga}, K.~M. and {Gogo}, T. and {Gumba}, A. and {Gupta}, N. and {Hilton}, M. and {Johnstone}, B. and {Karastergiou}, A. and {Kunz}, M. and {Lokhorst}, D. and {Maartens}, R. and {Macpherson}, S. and {Mdlalose}, M. and {Moodley}, K. and {Ngwenya}, L. and {Parra}, J.~M. and {Peterson}, J. and {Recnik}, O. and {Saliwanchik}, B. and {Santos}, M.~G. and {Sievers}, J.~L. and {Smirnov}, O. and {Stronkhorst}, P. and {Taylor}, R. and {Vanderlinde}, K. and {Van Vuuren}, G. and {Weltman}, A. and {Witzemann}, A.},
        title = "{HIRAX: a probe of dark energy and radio transients}",
    booktitle = {Ground-based and Airborne Telescopes VI},
         year = 2016,
       editor = {{Hall}, Helen J. and {Gilmozzi}, Roberto and {Marshall}, Heather K.},
       series = {Society of Photo-Optical Instrumentation Engineers (SPIE) Conference Series},
       volume = {9906},
        month = aug,
          eid = {99065X},
        pages = {99065X},
          doi = {10.1117/12.2234286},
archivePrefix = {arXiv},
       eprint = {1607.02059},
 primaryClass = {astro-ph.IM},
       adsurl = {https://ui.adsabs.harvard.edu/abs/2016SPIE.9906E..5XN}
}

@ARTICLE{Chime_detection_2025,
       author = {{CHIME Collaboration} and {Amiri}, Mandana and {Bandura}, Kevin and {Chakraborty}, Arnab and {Cliche}, Jean-Fran{\c{c}}ois and {Dobbs}, Matt and {Foreman}, Simon and {Gray}, Liam and {Halpern}, Mark and {Hill}, Alex S and {Hinshaw}, Gary and {H{\"o}fer}, Carolin and {Joseph}, Albin and {Kruger}, Nolan and {Landecker}, T.~L. and {van Lieshout}, Rik and {MacEachern}, Joshua and {Masui}, Kiyoshi W. and {Mena-Parra}, Juan and {Miller}, Kyle and {Milutinovic}, Nikola and {Mirhosseini}, Arash and {Newburgh}, Laura and {Ordog}, Anna and {Pen}, Ue-Li and {Pinsonneault-Marotte}, Tristan and {Reda}, Alex and {Renard}, Andre and {Sakaguri}, Kana and {Shaw}, J. Richard and {Shaikh}, Shabbir and {Siegel}, Seth R. and {Singh}, Saurabh and {Spear}, David and {Uchibori}, Yukari and {Vanderlinde}, Keith and {Wang}, Haochen and {Wiebe}, Donald V. and {Wulf}, Dallas},
        title = "{Detection of the Cosmological 21 cm Signal in Auto-correlation at z \raisebox{-0.5ex}\textasciitilde 1 with the Canadian Hydrogen Intensity Mapping Experiment}",
      journal = {arXiv e-prints},
         year = 2025,
        month = nov,
          eid = {arXiv:2511.19620},
        pages = {arXiv:2511.19620},
          doi = {10.48550/arXiv.2511.19620},
archivePrefix = {arXiv},
       eprint = {2511.19620},
 primaryClass = {astro-ph.CO},
       adsurl = {https://ui.adsabs.harvard.edu/abs/2025arXiv251119620C}
}

@ARTICLE{GMRT_2017,
       author = {{Gupta}, Y. and {Ajithkumar}, B. and {Kale}, H.~S. and {Nayak}, S. and {Sabhapathy}, S. and {Sureshkumar}, S. and {Swami}, R.~V. and {Chengalur}, J.~N. and {Ghosh}, S.~K. and {Ishwara-Chandra}, C.~H. and {Joshi}, B.~C. and {Kanekar}, N. and {Lal}, D.~V. and {Roy}, S.},
        title = "{The upgraded GMRT: opening new windows on the radio Universe}",
      journal = {Current Science},
         year = 2017,
        month = aug,
       volume = {113},
       number = {4},
        pages = {707-714},
          doi = {10.18520/cs/v113/i04/707-714},
       adsurl = {https://ui.adsabs.harvard.edu/abs/2017CSci..113..707G}
}

@INPROCEEDINGS{Meerkat_2017,
       author = {{Santos}, M. and {Bull}, P. and {Camera}, S. and {Chen}, S. and {Fonseca}, J. and {Heywood}, I. and {Hilton}, M. and {Jarvis}, M. and {Jozsa}, G.~I.~G. and {Knowles}, K. and {Leeuw}, L. and {Maartens}, R. and {Malefahlo}, E. and {McAlpine}, K. and {Moodley}, K. and {Patel}, P. and {Pourtsidou}, A. and {Prescott}, M. and {Spekkens}, K. and {Taylor}, R. and {Witzemann}, A. and {Whittam}, I.~H.},
        title = "{A Large Sky Survey with MeerKAT}",
    booktitle = {MeerKAT Science: On the Pathway to the SKA},
         year = 2016,
        month = jan,
          eid = {32},
        pages = {32},
          doi = {10.22323/1.277.0032},
archivePrefix = {arXiv},
       eprint = {1709.06099},
 primaryClass = {astro-ph.CO},
       adsurl = {https://ui.adsabs.harvard.edu/abs/2016mks..confE..32S}
}

@ARTICLE{DeLucia_2014,
       author = {{De Lucia}, Gabriella and {Tornatore}, Luca and {Frenk}, Carlos S. and {Helmi}, Amina and {Navarro}, Julio F. and {White}, Simon D.~M.},
        title = "{Elemental abundances in Milky Way-like galaxies from a hierarchical galaxy formation model}",
      journal = {\mnras},
         year = 2014,
        month = nov,
       volume = {445},
       number = {1},
        pages = {970-987},
          doi = {10.1093/mnras/stu1752},
archivePrefix = {arXiv},
       eprint = {1407.7867},
 primaryClass = {astro-ph.GA},
       adsurl = {https://ui.adsabs.harvard.edu/abs/2014MNRAS.445..970D}
}

@ARTICLE{Hirschmann_2016,
       author = {{Hirschmann}, Michaela and {De Lucia}, Gabriella and {Fontanot}, Fabio},
        title = "{Galaxy assembly, stellar feedback and metal enrichment: the view from the GAEA model}",
      journal = {\mnras},
         year = 2016,
        month = sep,
       volume = {461},
       number = {2},
        pages = {1760-1785},
          doi = {10.1093/mnras/stw1318},
archivePrefix = {arXiv},
       eprint = {1512.04531},
 primaryClass = {astro-ph.GA},
       adsurl = {https://ui.adsabs.harvard.edu/abs/2016MNRAS.461.1760H}
}

@ARTICLE{Xie_2017,
       author = {{Xie}, Lizhi and {De Lucia}, Gabriella and {Hirschmann}, Michaela and {Fontanot}, Fabio and {Zoldan}, Anna},
        title = "{H$_{2}$-based star formation laws in hierarchical models of galaxy formation}",
      journal = {\mnras},
         year = 2017,
        month = jul,
       volume = {469},
       number = {1},
        pages = {968-993},
          doi = {10.1093/mnras/stx889},
archivePrefix = {arXiv},
       eprint = {1611.09372},
 primaryClass = {astro-ph.GA},
       adsurl = {https://ui.adsabs.harvard.edu/abs/2017MNRAS.469..968X}
}

@ARTICLE{Fontanot_2020,
       author = {{Fontanot}, Fabio and {De Lucia}, Gabriella and {Hirschmann}, Michaela and {Xie}, Lizhi and {Monaco}, Pierluigi and {Menci}, Nicola and {Fiore}, Fabrizio and {Feruglio}, Chiara and {Cristiani}, Stefano and {Shankar}, Francesco},
        title = "{The rise of active galactic nuclei in the galaxy evolution and assembly semi-analytic model}",
      journal = {\mnras},
         year = 2020,
        month = aug,
       volume = {496},
       number = {3},
        pages = {3943-3960},
          doi = {10.1093/mnras/staa1716},
archivePrefix = {arXiv},
       eprint = {2002.10576},
 primaryClass = {astro-ph.CO},
       adsurl = {https://ui.adsabs.harvard.edu/abs/2020MNRAS.496.3943F}
}

@ARTICLE{Xie_2020,
       author = {{Xie}, Lizhi and {De Lucia}, Gabriella and {Hirschmann}, Michaela and {Fontanot}, Fabio},
        title = "{The influence of environment on satellite galaxies in the GAEA semi-analytic model}",
      journal = {\mnras},
         year = 2020,
        month = nov,
       volume = {498},
       number = {3},
        pages = {4327-4344},
          doi = {10.1093/mnras/staa2370},
archivePrefix = {arXiv},
       eprint = {2003.12757},
 primaryClass = {astro-ph.GA},
       adsurl = {https://ui.adsabs.harvard.edu/abs/2020MNRAS.498.4327X}
}

@ARTICLE{DeLucia_2024b,
       author = {{De Lucia}, Gabriella and {Fontanot}, Fabio and {Xie}, Lizhi and {Hirschmann}, Michaela},
        title = "{Tracing the quenching journey across cosmic time}",
      journal = {\aap},
         year = 2024,
        month = jul,
       volume = {687},
          eid = {A68},
        pages = {A68},
          doi = {10.1051/0004-6361/202349045},
archivePrefix = {arXiv},
       eprint = {2401.06211},
 primaryClass = {astro-ph.GA},
       adsurl = {https://ui.adsabs.harvard.edu/abs/2024A&A...687A..68D}
}

@ARTICLE{castander_2025,
       author = {{Euclid Collaboration} and {Castander}, F.~J. and {Fosalba}, P. and {Stadel}, J. and {Potter}, D. and {Carretero}, J. and {Tallada-Cresp{\'\i}}, P. and {Pozzetti}, L. and {Bolzonella}, M. and {Mamon}, G.~A. and {Blot}, L. and {Hoffmann}, K. and {Huertas-Company}, M. and {Monaco}, P. and {Gonzalez}, E.~J. and {De Lucia}, G. and {Scarlata}, C. and {Breton}, M.-A. and {Linke}, L. and {Viglione}, C. and {Li}, S.-S. and {Zhai}, Z. and {Baghkhani}, Z. and {Pardede}, K. and {Neissner}, C. and {Teyssier}, R. and {Crocce}, M. and {Tutusaus}, I. and {Miller}, L. and {Congedo}, G. and {Biviano}, A. and {Hirschmann}, M. and {Pezzotta}, A. and {Aussel}, H. and {Hoekstra}, H. and {Kitching}, T. and {Percival}, W.~J. and {Guzzo}, L. and {Mellier}, Y. and {Oesch}, P.~A. and {Bowler}, R.~A.~A. and {Bruton}, S. and {Allevato}, V. and {Gonzalez-Perez}, V. and {Manera}, M. and {Avila}, S. and {Kov{\'a}cs}, A. and {Aghanim}, N. and {Altieri}, B. and {Amara}, A. and {Amendola}, L. and {Andreon}, S. and {Auricchio}, N. and {Baccigalupi}, C. and {Baldi}, M. and {Balestra}, A. and {Bardelli}, S. and {Bender}, R. and {Bernardeau}, F. and {Bodendorf}, C. and {Bonino}, D. and {Branchini}, E. and {Brescia}, M. and {Brinchmann}, J. and {Camera}, S. and {Capobianco}, V. and {Carbone}, C. and {Casas}, S. and {Castellano}, M. and {Castignani}, G. and {Cavuoti}, S. and {Cimatti}, A. and {Colodro-Conde}, C. and {Conselice}, C.~J. and {Conversi}, L. and {Copin}, Y. and {Corcione}, L. and {Courbin}, F. and {Courtois}, H.~M. and {Da Silva}, A. and {Degaudenzi}, H. and {Di Giorgio}, A.~M. and {Dinis}, J. and {Douspis}, M. and {Dubath}, F. and {Duncan}, C.~A.~J. and {Dupac}, X. and {Dusini}, S. and {Ealet}, A. and {Farina}, M. and {Farrens}, S. and {Ferriol}, S. and {Fotopoulou}, S. and {Fourmanoit}, N. and {Frailis}, M. and {Franceschi}, E. and {Franzetti}, P. and {Galeotta}, S. and {Gillard}, W. and {Gillis}, B. and {Giocoli}, C. and {G{\'o}mez-Alvarez}, P. and {Granett}, B.~R. and {Grazian}, A. and {Grupp}, F. and {Haugan}, S.~V.~H. and {Holliman}, M.~S. and {Holmes}, W. and {Hook}, I. and {Hormuth}, F. and {Hornstrup}, A. and {Hudelot}, P. and {Ili{\'c}}, S. and {Jahnke}, K. and {Jhabvala}, M. and {Joachimi}, B. and {Keih{\"a}nen}, E. and {Kermiche}, S. and {Kiessling}, A. and {Kilbinger}, M. and {Kohley}, R. and {Kubik}, B. and {K{\"u}mmel}, M. and {Kunz}, M. and {Kurki-Suonio}, H. and {Lahav}, O. and {Laureijs}, R. and {Le Mignant}, D. and {Liebing}, P. and {Ligori}, S. and {Lilje}, P.~B. and {Lindholm}, V. and {Lloro}, I. and {Maino}, D. and {Maiorano}, E. and {Mansutti}, O. and {Marcin}, S. and {Marggraf}, O. and {Markovic}, K. and {Martinelli}, M. and {Martinet}, N. and {Marulli}, F. and {Massey}, R. and {Masters}, D.~C. and {Maurogordato}, S. and {McCracken}, H.~J. and {Medinaceli}, E. and {Mei}, S. and {Melchior}, M. and {Meneghetti}, M. and {Merlin}, E. and {Meylan}, G. and {Mohr}, J.~J. and {Moresco}, M. and {Moscardini}, L. and {Munari}, E. and {Nakajima}, R. and {Nichol}, R.~C. and {Niemi}, S.-M. and {Padilla}, C. and {Paech}, K. and {Paltani}, S. and {Pasian}, F. and {Peacock}, J.~A. and {Pedersen}, K. and {Pettorino}, V. and {Pires}, S. and {Polenta}, G. and {Poncet}, M. and {Popa}, L.~A. and {Raison}, F. and {Rebolo}, R. and {Renzi}, A. and {Rhodes}, J. and {Riccio}, G. and {Romelli}, E. and {Roncarelli}, M. and {Rosset}, C. and {Rossetti}, E. and {Rusholme}, B. and {Saglia}, R. and {Sakr}, Z. and {S{\'a}nchez}, A.~G. and {Sapone}, D. and {Schewtschenko}, J.~A. and {Schirmer}, M. and {Schneider}, P. and {Schrabback}, T. and {Scodeggio}, M. and {Secroun}, A. and {Sefusatti}, E. and {Seidel}, G. and {Serrano}, S. and {Sirignano}, C. and {Sirri}, G. and {Stanco}, L. and {Starck}, J.-L. and {Steinwagner}, J. and {Taylor}, A.~N. and {Teplitz}, H.~I.},
        title = "{Euclid: V. The Flagship galaxy mock catalogue: A comprehensive simulation for the Euclid mission}",
      journal = {\aap},
         year = 2025,
        month = may,
       volume = {697},
          eid = {A5},
        pages = {A5},
          doi = {10.1051/0004-6361/202450853},
archivePrefix = {arXiv},
       eprint = {2405.13495},
 primaryClass = {astro-ph.CO},
       adsurl = {https://ui.adsabs.harvard.edu/abs/2025A&A...697A...5E}
}

@ARTICLE{Pezzotta_2024,
       author = {{Euclid Collaboration} and {Pezzotta}, A. and {Moretti}, C. and {Zennaro}, M. and {Moradinezhad Dizgah}, A. and {Crocce}, M. and {Sefusatti}, E. and {Ferrero}, I. and {Pardede}, K. and {Eggemeier}, A. and {Barreira}, A. and {Angulo}, R.~E. and {Marinucci}, M. and {Camacho Quevedo}, B. and {de la Torre}, S. and {Alkhanishvili}, D. and {Biagetti}, M. and {Breton}, M.-A. and {Castorina}, E. and {D'Amico}, G. and {Desjacques}, V. and {Guidi}, M. and {K{\"a}rcher}, M. and {Oddo}, A. and {Pellejero Ibanez}, M. and {Porciani}, C. and {Pugno}, A. and {Salvalaggio}, J. and {Sarpa}, E. and {Veropalumbo}, A. and {Vlah}, Z. and {Amara}, A. and {Andreon}, S. and {Auricchio}, N. and {Baldi}, M. and {Bardelli}, S. and {Bender}, R. and {Bodendorf}, C. and {Bonino}, D. and {Branchini}, E. and {Brescia}, M. and {Brinchmann}, J. and {Camera}, S. and {Capobianco}, V. and {Carbone}, C. and {Cardone}, V.~F. and {Carretero}, J. and {Casas}, S. and {Castander}, F.~J. and {Castellano}, M. and {Cavuoti}, S. and {Cimatti}, A. and {Congedo}, G. and {Conselice}, C.~J. and {Conversi}, L. and {Copin}, Y. and {Corcione}, L. and {Courbin}, F. and {Courtois}, H.~M. and {Da Silva}, A. and {Degaudenzi}, H. and {Di Giorgio}, A.~M. and {Dinis}, J. and {Dupac}, X. and {Dusini}, S. and {Ealet}, A. and {Farina}, M. and {Farrens}, S. and {Fosalba}, P. and {Frailis}, M. and {Franceschi}, E. and {Galeotta}, S. and {Gillis}, B. and {Giocoli}, C. and {Granett}, B.~R. and {Grazian}, A. and {Grupp}, F. and {Guzzo}, L. and {Haugan}, S.~V.~H. and {Hormuth}, F. and {Hornstrup}, A. and {Jahnke}, K. and {Joachimi}, B. and {Keih{\"a}nen}, E. and {Kermiche}, S. and {Kiessling}, A. and {Kilbinger}, M. and {Kitching}, T. and {Kubik}, B. and {Kunz}, M. and {Kurki-Suonio}, H. and {Ligori}, S. and {Lilje}, P.~B. and {Lindholm}, V. and {Lloro}, I. and {Maiorano}, E. and {Mansutti}, O. and {Marggraf}, O. and {Markovic}, K. and {Martinet}, N. and {Marulli}, F. and {Massey}, R. and {Medinaceli}, E. and {Mellier}, Y. and {Meneghetti}, M. and {Merlin}, E. and {Meylan}, G. and {Moresco}, M. and {Moscardini}, L. and {Munari}, E. and {Niemi}, S.-M. and {Padilla}, C. and {Paltani}, S. and {Pasian}, F. and {Pedersen}, K. and {Percival}, W.~J. and {Pettorino}, V. and {Pires}, S. and {Polenta}, G. and {Pollack}, J.~E. and {Poncet}, M. and {Popa}, L.~A. and {Pozzetti}, L. and {Raison}, F. and {Renzi}, A. and {Rhodes}, J. and {Riccio}, G. and {Romelli}, E. and {Roncarelli}, M. and {Rossetti}, E. and {Saglia}, R. and {Sapone}, D. and {Sartoris}, B. and {Schneider}, P. and {Schrabback}, T. and {Secroun}, A. and {Seidel}, G. and {Seiffert}, M. and {Serrano}, S. and {Sirignano}, C. and {Sirri}, G. and {Stanco}, L. and {Surace}, C. and {Tallada-Cresp{\'\i}}, P. and {Taylor}, A.~N. and {Tereno}, I. and {Toledo-Moreo}, R. and {Torradeflot}, F. and {Tutusaus}, I. and {Valentijn}, E.~A. and {Valenziano}, L. and {Vassallo}, T. and {Wang}, Y. and {Weller}, J. and {Zamorani}, G. and {Zoubian}, J. and {Zucca}, E. and {Biviano}, A. and {Bozzo}, E. and {Burigana}, C. and {Colodro-Conde}, C. and {Di Ferdinando}, D. and {Mainetti}, G. and {Martinelli}, M. and {Mauri}, N. and {Sakr}, Z. and {Scottez}, V. and {Tenti}, M. and {Viel}, M. and {Wiesmann}, M. and {Akrami}, Y. and {Allevato}, V. and {Anselmi}, S. and {Baccigalupi}, C. and {Ballardini}, M. and {Bernardeau}, F. and {Blanchard}, A. and {Borgani}, S. and {Bruton}, S. and {Cabanac}, R. and {Cappi}, A. and {Carvalho}, C.~S. and {Castignani}, G. and {Castro}, T. and {Ca{\~n}as-Herrera}, G. and {Chambers}, K.~C. and {Contarini}, S. and {Cooray}, A.~R. and {Coupon}, J. and {Davini}, S. and {De Lucia}, G. and {Desprez}, G. and {Di Domizio}, S. and {Dole}, H. and {D{\'\i}az-S{\'a}nchez}, A. and {Escartin Vigo}, J.~A. and {Escoffier}, S. and {Ferreira}, P.~G. and {Finelli}, F. and {Gabarra}, L.},
        title = "{Euclid preparation. XLI. Galaxy power spectrum modelling in real space}",
      journal = {\aap},
         year = 2024,
        month = jul,
       volume = {687},
          eid = {A216},
        pages = {A216},
          doi = {10.1051/0004-6361/202348939},
archivePrefix = {arXiv},
       eprint = {2312.00679},
 primaryClass = {astro-ph.CO},
       adsurl = {https://ui.adsabs.harvard.edu/abs/2024A&A...687A.216E}
}

@ARTICLE{Pozzetti_2016,
       author = {{Pozzetti}, L. and {Hirata}, C.~M. and {Geach}, J.~E. and {Cimatti}, A. and {Baugh}, C. and {Cucciati}, O. and {Merson}, A. and {Norberg}, P. and {Shi}, D.},
        title = "{Modelling the number density of H{\ensuremath{\alpha}} emitters for future spectroscopic near-IR space missions}",
      journal = {\aap},
         year = 2016,
        month = may,
       volume = {590},
          eid = {A3},
        pages = {A3},
          doi = {10.1051/0004-6361/201527081},
archivePrefix = {arXiv},
       eprint = {1603.01453},
 primaryClass = {astro-ph.GA},
       adsurl = {https://ui.adsabs.harvard.edu/abs/2016A&A...590A...3P}
}

@ARTICLE{Baugh_2019,
       author = {{Baugh}, C.~M. and {Gonzalez-Perez}, Violeta and {Lagos}, Claudia D.~P. and {Lacey}, Cedric G. and {Helly}, John C. and {Jenkins}, Adrian and {Frenk}, Carlos S. and {Benson}, Andrew J. and {Bower}, Richard G. and {Cole}, Shaun},
        title = "{Galaxy formation in the Planck Millennium: the atomic hydrogen content of dark matter haloes}",
      journal = {\mnras},
         year = 2019,
        month = mar,
       volume = {483},
       number = {4},
        pages = {4922-4937},
          doi = {10.1093/mnras/sty3427},
archivePrefix = {arXiv},
       eprint = {1808.08276},
 primaryClass = {astro-ph.GA},
       adsurl = {https://ui.adsabs.harvard.edu/abs/2019MNRAS.483.4922B}
}

@ARTICLE{Springel_2001,
       author = {{Springel}, Volker and {White}, Simon D.~M. and {Tormen}, Giuseppe and {Kauffmann}, Guinevere},
        title = "{Populating a cluster of galaxies - I. Results at z=0}",
      journal = {\mnras},
         year = 2001,
        month = dec,
       volume = {328},
       number = {3},
        pages = {726-750},
          doi = {10.1046/j.1365-8711.2001.04912.x},
archivePrefix = {arXiv},
       eprint = {astro-ph/0012055},
 primaryClass = {astro-ph},
       adsurl = {https://ui.adsabs.harvard.edu/abs/2001MNRAS.328..726S}
}

@ARTICLE{Bharadwaj_Srikant_2004,
       author = {{Bharadwaj}, Somnath and {Srikant}, Pennathur Sridharan},
        title = "{HI fluctuations at large redshifts: III {\textemdash} Simulating the signal expected at GMRT}",
      journal = {Journal of Astrophysics and Astronomy},
         year = 2004,
        month = mar,
       volume = {25},
       number = {1-2},
        pages = {67-80},
          doi = {10.1007/BF02702289},
archivePrefix = {arXiv},
       eprint = {astro-ph/0402262},
 primaryClass = {astro-ph},
       adsurl = {https://ui.adsabs.harvard.edu/abs/2004JApA...25...67B}
}

@BOOK{Peebles_1980,
       author = {{Peebles}, P.~J.~E.},
        title = "{The large-scale structure of the universe}",
         year = 1980,
       adsurl = {https://ui.adsabs.harvard.edu/abs/1980lssu.book.....P}
}
\bibliographystyle{aasjournalv7}



\end{document}